\documentclass[12pt,preprint]{aastex}

\begin{document}

\title{Twisted flux tube emergence from the convection zone to the corona}

\author{Juan Mart\'inez-Sykora}
\and
\author{Viggo Hansteen} 
\and 
\author{Mats Carlsson }
\affil{Institute of Theoretical Astrophysics, University of Oslo, P.O. Box 1029 Blindern, N-0315 Oslo, Norway}

\newcommand{\myemail}{juanms@astro.uio.no}
\newcommand{\viscous}{\underline{\underline{\tau}}}
\newcommand{\resistive}{\underline{\underline{\eta}}}

\begin{abstract}

Three dimensional numerical simulations of the emergence of a horizontal magnetic flux tube 
with different twist levels are carried out in a computational domain spanning the upper 
layers of the convection zone to the lower corona. We use the {\em Oslo Staggered Code} (OSC) 
to solve the full MHD equations with non-grey and non-LTE radiative transfer and thermal 
conduction along the magnetic field lines. The emergence of the
magnetic flux tube input at the bottom boundary into a weakly magnetized atmosphere is presented. 
The photospheric and chromospheric response is described with magnetograms, synthetic continuum 
images at different wavelengths, as well as synthetic Ca~{\sc ii} H-line images, and velocity field 
distributions. 

The emergence of a magnetic flux tube into such an atmosphere
results in varied atmospheric responses. 
In the photosphere the granular size increases when the flux tube approaches from 
below, as has been reported previously in the literature. In the convective overshoot region some
200~km above the photosphere adiabatic expansion produces cooling, darker regions with 
the structure of granulation cells. We also find collapsed granulation in the boundaries 
of the rising flux tube. Once the flux tube has crossed the photosphere, bright points related 
with concentrated magnetic field, vorticity, high vertical velocities and heating by compressed 
material are found at heights up to 500~km above the photosphere. At greater heights in the 
magnetized chromosphere, we find that the rising flux tube produces a large, cool, dim, magnetized
bubble that tends to expel the usual chromospheric oscillations. In addition the rising flux tube
dramatically increases the chromospheric scale height, pushing the transition region and corona
aside such that the chromosphere extends up to 6~Mm above the photosphere in the center of the
rising flux tube. The emergence of magnetic flux tubes through the photosphere and subsequent
rise through the chromosphere up to the bottom of the lower corona is a relatively slow process,
taking of order 1~hour to complete. In later papers we intend to discuss the merging of the 
newly emerged field with the pre-existing coronal field and its diagnostic consequences.

\end{abstract}


\section{Introduction}

One of the stated goals of the newly launched Hinode satellite is to follow
the evolution of magnetic flux from the moment of emergence through the photosphere and
into the chromosphere and corona. Such observations come at a timely moment as they
should prove a fertile testing ground for the 3D numerical models of flux emergence that
are now becoming available due to the technological development of massively 
parallel computers and algorithms to utilize these.

A great variety of simplified simulations have been developed in order
to study the different phases of flux emergence. Originally, early simulations 
were performed in 1D due to computational limitations. 
Studies of rising flux tubes were carried out to gain an understanding of 
the emergence latitudes of active regions, their tilt angles as well as asymmetry between 
leading and following polarities and magnetic forces such as tension 
\citep{moreno1986,fan1993,Caligari1995}.  The early 80's saw the appearance
of 2D simulations where \citet{Shussler1979} and
\citet{longcope1996} found that an emerging magnetic flux tube without twist 
is fragmented. \citet{moreno1996} investigated the effect
of the surrounding flows on the flux tube's rise to the surface and
demonstrated that a minimum twist in the tube was required in order to
suppress the conversion of the tube into vortex pairs \citep{lites1995}.
\citet{emonet2001} studied the Von K{\'a}rm{\'a}n vortex street 
generated in the boundary layer of the tube \citep{Dorch+Nordlund1998}.
In the 90's 2D simulations were extended to 3D. \citet{fan1998}
investigated flux tubes moving through an adiabatic stratification 
found
that the degree of twist must be large enough to avoid fragmentation yet small 
enough to avoid the kink instability.  There are also other 3D emergence 
models that while not including radiative transfer, nor any detailed 
treatment of convection cells below the photosphere, have added 
schematic descriptions of the chromosphere and corona
\citep{matsumoto1998,fan2001,magara2004,Magara:2006lr,manchester2004,
archontis2004} in order to gain a handle on the effects to be expected.

On the other hand, in the last decade more realistic 3D MHD models of active
region flux emergence have been developed up to the photosphere. 
\citet{cheung2007} consider flux emergence through a convective layer 
including detailed photospheric radiative transfer, but excluding the layers above
the photosphere. Other simulations involve solar surface magnetoconvection
\citep{Stein:2002fk,Stein:2006qy,Stein:2003uq,Vogler:2005fj}, but do not consider
flux emergence from below.

The description of the outer layers of the solar atmosphere are also growing
in sophistication. \citet{Gudiksen:2005lr} recently performed realistic 
simulations of the solar corona driven by a stochastic approximation of 
granulation. \citet{Hansteen2004,Hansteen+Carlsson+Gudiksen2007} and \citet{abbett2007} 
studied 3D MHD simulations with a 
convective layer below the photosphere and included upper atmosphere (chromosphere,
transition region and lower-corona) with the radiative losses in the
quiet Sun and conduction along the magnetic field treated in detail.

In this paper our goal is to study horizontal flux tube emergence from the 
upper-convection zone to the lower corona in a realistic 3D MHD model. The model is 
designed to have a corona driven by magnetoconvection. The radiative
losses from the photosphere and lower chromosphere are computed in a 
non-grey manner and scattering in the chromosphere is handled to a certain 
approximation \citep{Nordlund1982,Skartlien2000}.  For the
upper-chromosphere and corona we assume effectively thin radiative losses.
A flux tube is injected into the bottom boundary of an atmosphere that contains 
a preexisting but weak magnetic field. We follow the evolution of the flux tube as
it rises through the various atmospheric layers: How much flux enters
the differing regions? When do we see observational signatures of flux emergence?
Where and how does the reconnection process happen? How does this affect chromospheric and 
coronal dynamics and structure? are questions we will discuss.

In section~\ref{sec:equations} we summarize how the MHD equations with
radiative transfer, thermal conduction, viscosity and resistivity 
are solved by the {\em Oslo Staggered Code}.  
Section~\ref{sec:condition} shows the initial and boundary
conditions of our numerical box for the different simulations. We
describe in more detail how the magnetic flux tube is input in the bottom
boundary. The results of our simulations are described in section~\ref{sec:results}
centered on the different events observed in the different layers of the
simulation box. We start with the photosphere, then treat the reverse granulation layer,
the end of the overshooting and beginning of the chromosphere and 
finally the upper chromosphere/transition region and lower corona. 

\section{Equations and Numerical Method}\label{sec:equations}

In order to model the rise of magnetic flux tubes through the upper convection layer and their 
emergence through the photosphere and into the chromosphere and corona we solve the equations 
of MHD using the {\it Oslo Stagger Code} (OSC):

\begin{equation}
\frac{\partial \rho}{\partial t} + \nabla ({\rho \bf u})=0 \label{cont}
\end{equation}
\begin{equation}
\frac{\partial {\bf u}}{\partial t}  + ({\bf u} \nabla) {\bf u} + \frac{1}{\rho} \nabla (P+\viscous) =
	+ \frac{{\bf J} \times {\bf B}}{\rho}+{\bf g},\label{moment} 
\end{equation}
\begin{equation}
\frac{\partial (e)}{\partial t}  + \nabla (e{\bf u})  + {\bf u} \nabla P=
	 \nabla {\bf F}_r+\nabla {\bf F}_c+Q_{Joule}+Q_{visc},\label{ener} 
\end{equation}
\begin{equation}
\frac{\partial{\bf B}}{\partial t}=\nabla \times ({\bf u} \times {\bf B})-\nabla\times (\resistive {\bf J}), \label{induction}
\end{equation}

\noindent where $\rho$ represents the mass density, ${\bf u}$ the fluid velocity, $P$ the
gas pressure, ${\bf J}$ the current density, ${\bf B}$ the magnetic field, ${\bf g}$ 
gravitational acceleration, and $e$ the internal energy. The viscous stress tensor is 
written $\viscous$ and the resistivity $\resistive$. ${\bf F}_r$ represents the radiative flux, 
${\bf F}_c$ represents the conductive flux, and $Q_{Joule}$ and $Q_{visc}$ are the joule 
heating and viscous heating respectively. 

These equations are solved using an extended version of the numerical code described
in \citet{Dorch+Nordlund1998,Mackay+Galsgaard2001} and in more detail
by Nordlund \& Galsgaard at http://www.astro.ku.dk/$\sim$kg and 
\citet{Hansteen+Carlsson+Gudiksen2007}.
Extended in this case means the inclusion of
thermal conduction along the magnetic field, non-grey, non-LTE radiative transfer and 
characteristic boundary conditions on the upper and lower boundaries. The latter 
allow the passage of waves out of the computational box while at the same time allowing 
the specification of an input magnetic field and entropy at the lower boundary.

In short, the code functions as follows: The variables are represented
on staggered meshes, such that the density and the internal energy
are volume centered, the magnetic field components and the momentum densities
${\bf p}=\rho{\bf u}$ are face centered, while the electric field ${\bf E}$
and the current density are edge centered. A sixth order accurate 
method involving the three nearest neighbor points on each side is
used for determining the partial spatial derivatives. In the cases where
variables are needed at positions other than their defined positions 
a fifth order interpolation scheme is used. The equations are stepped 
forward in time using the explicit 3rd order predictor-corrector 
procedure defined by \citet{Hyman1979}, modified for variable time steps. 
In order to suppress numerical noise, high-order artificial diffusion is added both
in the forms of a viscosity and in the form of a magnetic diffusivity.

The radiative flux divergence from the photosphere and lower chromosphere is obtained
by angle and wavelength integration of the transport equation. Assuming isotropic opacities 
$\chi_\lambda$ and emissivities we find
\begin{equation}
\nabla{\bf F}_r=4\pi\int_\lambda\epsilon_\lambda\chi_\lambda(B_\lambda-J_\lambda)d\lambda,
\end{equation}

\noindent where $\epsilon_\lambda$ is the photon destruction probability. 
The method of solving the transport equation assumes opacities are in LTE and coherent
scattering. \citet{Nordlund1982} devised a technique of solution based on group mean opacities, 
in which the spectrum is divided into four bins representing strong, medium and weak lines 
in addition to the continuum. The transfer equation is formulated for these bins and a group 
mean source function is calculated for each bin. These
source functions contain an approximate scattering term and an exact contribution from thermal
emissivity. The resulting 3D scattering problems are solved by iteration based on a one-ray
approximation in the angle integral for the mean intensity as developed by \citet{Skartlien2000}.

For the upper chromosphere and corona we assume optically thin radiative losses such that:

\begin{equation}
\nabla{\bf F}_r\propto \rho^2f(T)e^{-\tau}.
\end{equation}

\noindent where $f(T)$ is the optically thin radiative loss function based on the coronal
approximation and atomic data collected in the HAO spectral diagnostics package 
\citep{HAO_Diaper1994}; $f(T)$ is based on the elements hydrogen, helium, carbon, oxygen, neon 
and iron. The $e^{-\tau}$ term prevents this term from having an effect in the
deep photosphere, where we assume an ``optical depth'', $\tau$, proportional to the gas pressure.

In the mid and upper chromosphere we include non-LTE radiative losses from hydrogen continua, hydrogen
lines and lines from singly ionized calcium. These losses are calculated from

\begin{equation}
\nabla{\bf F}_r\propto N_e\rho f_i(T)\epsilon_i(m_c)
\end{equation}

\noindent where $N_e$ is the electron density, $f_i(T)$ is a pre-calculated function based on collisional
excitation and radiative de-excitation, separate for hydrogen and calcium, and $\epsilon_i(m_c)$ is
the non-LTE escape probability as a function of column mass ($m_c$). This escape probability 
function is calculated from 1D dynamical chromospheric models in which the radiative losses
are computed in detail (Carlsson \& Stein 1992, 1995, 1997, 2002).
\nocite{Carlsson+Stein1992}
\nocite{Carlsson+Stein1995}
\nocite{Carlsson+Stein1997}
\nocite{Carlsson+Stein2002}
In addition to these optically thin losses in the
upper atmosphere we have added an {\it ad hoc} heating term to prevent the 
atmosphere from cooling much below $2000$~K in the upper chromosphere.

The Courant condition for a diffusive operator such as that describing thermal conduction
scales with the square of the grid size $\Delta z^2$ instead of with $\Delta z$ as applies
to the magneto-hydrodynamic operator. 
This severely limits the time step $\Delta t$ the code can be stably run at. Our 
solution to this problem is to proceed by operator splitting such that the operator advancing 
the variables in time is $L=L_{hydro}+L_{conduction}$. Thus the conductive part of the energy 
equation is solved by discretizing.

\begin{equation}
{\partial e\over\partial t}=\nabla{\bf F}_c={-\nabla\kappa_\parallel\nabla_\parallel T}
\end{equation} 

\noindent using the Crank-Nicholson method and solving the resulting implicit problem using
a multi-grid solver. 
The formulation used in the code described here is based on the method used by Malagoli, 
Dubey, Cattaneo as shown at 
{\tt http://astro.uchicago.edu/Computing/On\_ Line/cfd95/camelse.html}, but extended to 
3D and to confining conduction to follow the magnetic field lines.

The units employed in this works are in SI. Sometimes we refer the distance in Mm ($10^6$m) and the time in hs ($10^2$~s).

\section{Initial and boundary conditions}\label{sec:condition}

The five models described here are run on a grid of $256\times 128\times 160$
points spanning $8\times 4\times 16$~Mm$^3$ and $16\times 8\times 16$~Mm$^3$.  
At these resolutions the models have been run for roughly one hour solar time.

The average temperature at the bottom boundary is maintained by setting 
the entropy of the fluid entering the computational domain. 
The bottom boundary is otherwise open, allowing fluid to enter and leave as required. 
The upper boundary is set so that the temperature gradient is zero; no
conductive heat flux enters or leaves the computational domain through the top 
boundary. 

Of course, without coronal heating the corona would cool on a time scale of roughly 
an hour (depending on the magnetic field topology), when no heat flux enters the model
through the upper boundary. We have therefore seeded the 
initial model with a magnetic field in which sufficient stresses can be built up
to maintain coronal temperatures in the upper part of the computational domain, as 
previously shown to be feasible by \citet{Gudiksen+Nordlund2004}. 
The initial field was obtained by semi-randomly spreading some $20-30$ positive and negative 
patches of vertical field at the bottom boundary, then calculating the potential field
that arises from this distribution in the remainder of the domain. Stresses sufficient 
to maintain a minimal corona are built up by photospheric motions after roughly $20$~minutes solar 
time.

At the top boundary the hydrodynamic variables (aside from the temperature)
and the magnetic field are set by characteristic extrapolations. This method hinders 
most of the reflections that are due to the presence of the upper boundary. 
No joule heating is added in the top five computational zones.

\subsection{Initial model}
\label{sec:initial-model}

\begin{figure}
  \includegraphics[width=9.5cm]{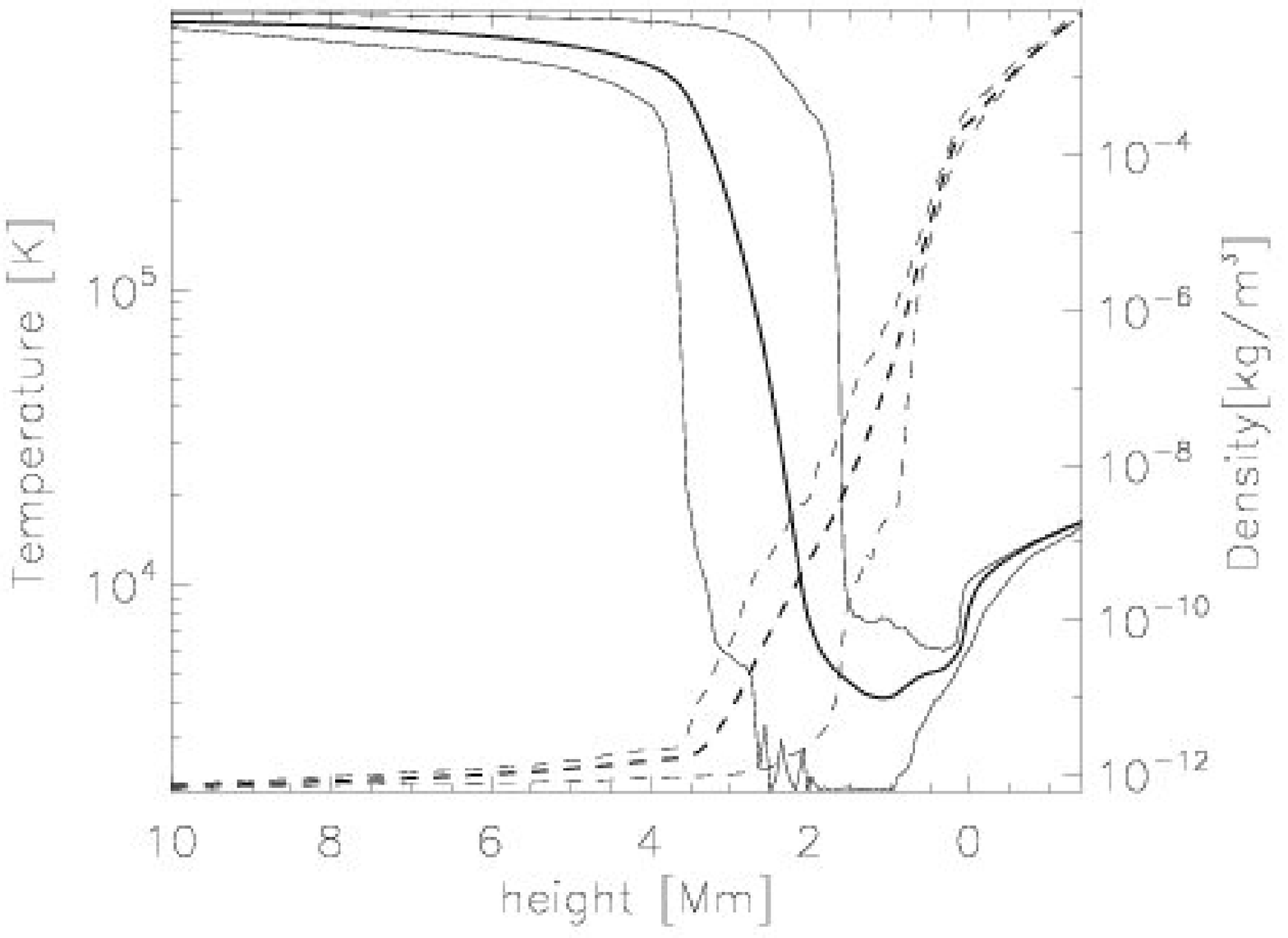}
  \includegraphics[width=9.5cm]{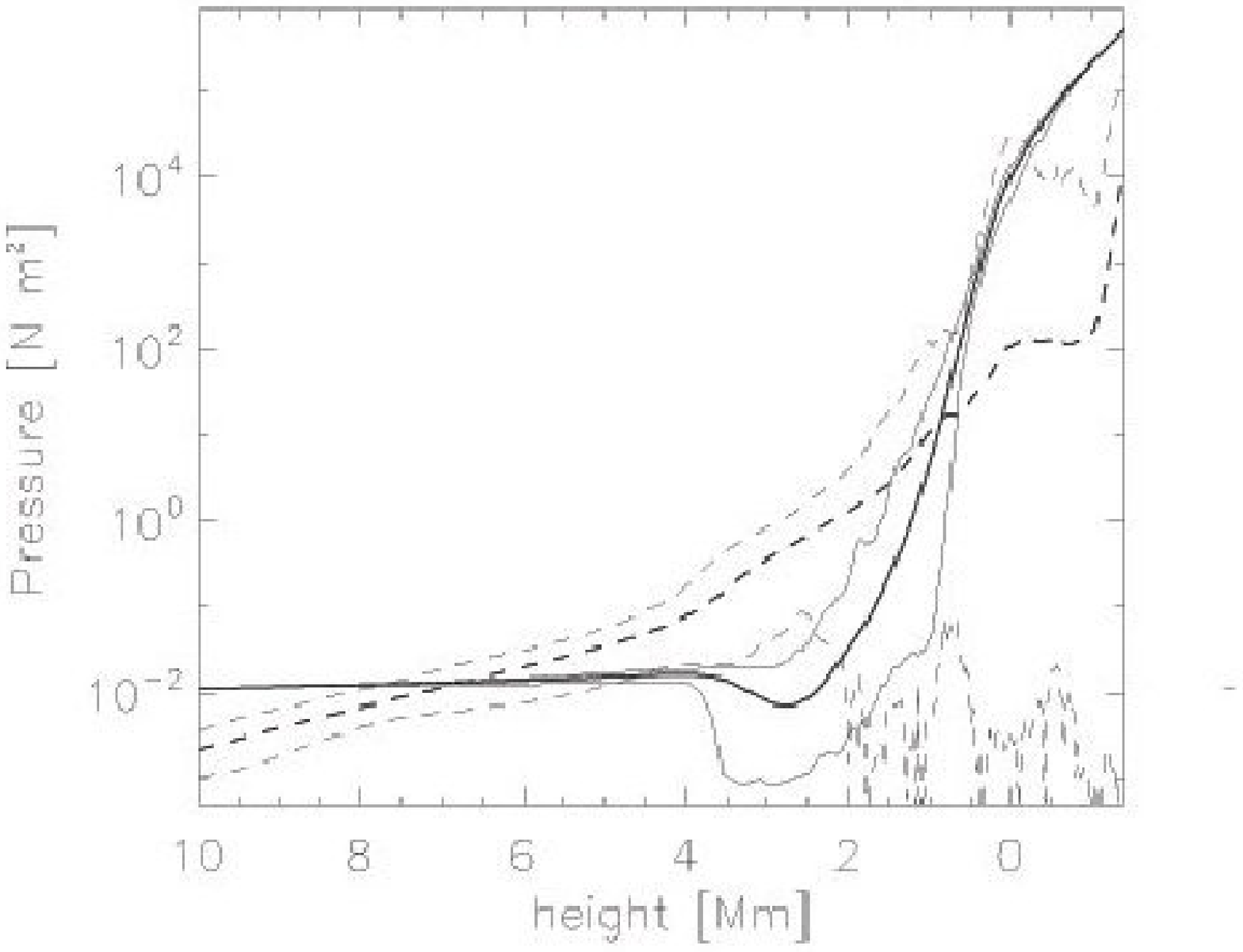}
 \caption{\label{fig:initsetup}Left panel: Temperatures (solid) and densities (dashed). Right panel: Gas (solid) and 
          magnetic (dashed) pressures. All as a function of height in the initial model. 
          Minimum and maximum values are shown in grey, average values are shown in black. The
          photosphere is situated at $z\approx0$~Mm. The models extent in height is from $-1.4$ to 
          $14$~Mm (top 4 Mm not shown here).}
 \end{figure}

The initial atmosphere, into which magnetic field is injected, includes the 
upper convection zone, the photosphere, chromosphere, transition region, 
and corona. In the left panel of figure~\ref{fig:initsetup} we show the lower 12~Mm (of 16~Mm) 
minimum, maximum, and horizontally averaged temperatures and densities as a function of
height $z$. 
At the base of the model, $1.4$~Mm below the photosphere, the average temperature 
is some $16\,200$~K and average density $6.7\times10^{-3}$~kg m$^{-3}$. With increasing
height the temperature falls close to adiabatically up to the photosphere, which is located 
at $z\approx0$~Mm, where the temperature gradient becomes much steeper and the average 
temperature falls to $6\,000$~K. The average density falls roughly $1.5$ orders of magnitude
to $2.5\times 10^{-4}$~kg m$^{-3}$ in the same height range. 
Quantities vary much more in the chromosphere, which extends from the top of the photosphere 
up to a height of some 4~Mm in some locations. The chromospheric temperature varies from 
a low of $2000$~K up to almost $10\,000$~K in the strongest shocks and in regions of strong 
magnetic field.
The density also has a high horizontal contrast in the chromosphere while at the same time 
falling by $7-8$ orders of magnitude vertically. The location of the transition region varies 
from $1.5$ to $4$~Mm above the 
photosphere. In this region the temperature rises rapidly to coronal values of roughly 1~MK, 
while coronal densities are of order $10^{-12}$~kg m$^{-3}$ in this model. 

In the right panel of 
figure~\ref{fig:initsetup} we plot the minimum, maximum, and horizontally averaged gas and
magnetic pressures as a function of height. 
The average gas pressure is $8.7\times 10^5$~Pa at the bottom boundary while the average 
magnetic pressure at the same depth is much smaller at $1.6\times 10^3$~Pa. The ratio between
these is still large in the photosphere, but note that the maximum magnetic pressure is
as large as the maximum gas pressure here; photospheric motions can concentrate the magnetic
field to pressures that equal or exceed photospheric pressure in intergranular lanes. 
The average plasma $\beta$ (ratio of gas pressure to magnetic pressure) remains greater 
than one up to some $600$~km above the 
photosphere, but the $\beta=1$ plane is quite corrugated and can extend up to $z=1000$~km. 
This means that the upper chromosphere, transition region and lower corona all are 
low $\beta$ regions. In the initial model, there is little net magnetic 
flux and the field is small at great heights, we therefore find some high $\beta$ regions 
at heights greater than $z=8$~Mm or so. 

\begin{figure}
  \includegraphics[width=9.5cm]{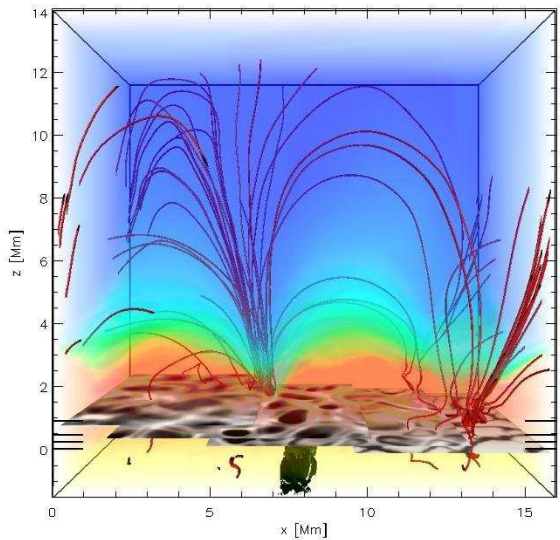}
  \caption{\label{fig:initbox}The computational box from the convection zone to the corona. 
           The color table is the temperature.
           The magnetic flux tube coming from the bottom 
           boundary (green lines). The initial (weak) background magnetic field
           lines (red lines). 
           The {\em z} axis runs from the corona ($z=14$~Mm top boundary) 
           to the convection zone ($z=-1.4$~Mm bottom boundary). We center
           our attention to 4 different heights shown in the figure as partial
           planes with the
           temperature (grey-scale coded); at heights $z=10$~km (photosphere),
           $z=234$~km (reverse granulation), $z=458$~km (photosphere-chromosphere)
           and $z=906$~km (mid-chromosphere).
           The black lines at the left and right side show the 
           heights of the four layers.}
\end{figure}

The numerical box and the axes used are shown in figure~\ref{fig:initbox}. The color scale 
represents the temperature. The red lines are the magnetic field lines in the initial background
field. Note that field lines extending into the corona are mainly rooted in regions that cover 
all $y$ values and around $x=7$~Mm and around $x=14$~Mm. In addition we show planes of temperature
at four heights: in the photosphere $z=10$~km, in the reverse granulation $z=234$~km, at
the layer $z=458$~km, and in the chromosphere at $z=900$~km. We will show details
of how the initial atmosphere reacts to flux emergence at these heights later in this paper.
The green lines show the magnetic field lines of the flux tube coming in from the bottom boundary.

\subsection{Injection of magnetic flux}
\label{seq:injection}

We introduce a magnetic sheet or magnetic flux tubes into the lower boundary of the model 
described in section~\ref{sec:initial-model}. The OSC uses a numerical method that, in principle,
does not change $\nabla \cdot{\bf B}$ in time. The solenoidal condition must be enforced on 
the boundary to ensure that no magnetic monopoles are introduced. This is implemented by applying 
the boundary condition to the {\it electric} field, the staggered mesh will then enforce 
$\nabla\cdot{\bf B}=0$ to the numerical accuracy of the operators at the boundary.

The magnetic field variation at the boundary is defined by:

\begin{equation}
\frac{\partial{\bf B}}{\partial t}=\nabla \times {\bf E},
\end{equation}
\noindent where, for example for the $x$ component of the electric field we have
\begin{equation}
E^n_x = E_x + \frac{\Delta (B_y)}{\tau}\Delta z
\end{equation}
\noindent where $E^n_x$ is the new {\it x} component of the electric field at the boundary,  
$\tau$ is the time step size, and $\Delta z$ is the vertical cell size, 
$\Delta (B_y) = B^n_y - B_y$ is the difference between the value of the magnetic field we would
like to impose at the boundary, $B^n_y$, and the current boundary field $B_y$. 

We have run models in which a constant horizontal field in the $y$ direction has been injected, 
and several models where a flux tube is introduced. The flux tube is horizontally rectilinear 
with twisted magnetic fields lines. The expression for the magnetic field 
has a structure given by

\begin{eqnarray}
{\bf B}_{long}&=& B_o\, \exp \left(-\frac{r^2}{R^2}\right)\, {\bf e}_{z}\label{eq:blong}\\
{\bf B}_{trans}&=&B_{long}\,r\,q\,{\bf e}_{\phi} ,\label{eq:btrans}
\end{eqnarray}

\noindent where  $r=\sqrt{(x-x_o)^2+(z-z_o)^2}$ is the radial distance to the center of the 
tube that has radius $R$. ${\bf B}_{long}$, ${\bf B}_{trans}$ are the longitudinal and 
transversal magnetic fields in cylindrical coordinates respectively. The 
parameter $q$ is used by \citet{linton1996} and \citet{fan1998} to define the twist of the 
magnetic field. Following \citet{mark2006}, we define a new twist parameter, $\lambda$, as

\begin{eqnarray}
\lambda\equiv q\,R. \label{lambda}
\end{eqnarray}

As the flux tube enters the computational box, the height of the center of the tube 
($z_o$) changes in time. The speed of flux tube, 
$({dz_o/dt})$, is set to the average of the velocity of plasma inflow at the boundary 
in the region where the magnetic flux tube is located each time step.

The field defined by equations~\ref{eq:blong} and \ref{eq:btrans} is easily seen to be 
a horizontal axisymmetric magnetic flux tube in which the longitudinal field has
a gaussian profile in the radial direction.

\section{Results}\label{sec:results}

We have carried out five simulations with different twist and magnetic field strength in order 
to study the effects of flux emergence in the photosphere and in the chromosphere. A summary of the 
runs completed is shown in table \ref{tab:runs}. In describing
the reaction of the atmosphere to the introduction of new magnetic flux we will specifically
concentrate on four layers placed at heights of $z=10, 235, 450$, and $900$~km represented
in the figure \ref{fig:initbox}. 

We find that all runs show a similar series of events after the emerging magnetic flux
pierces the photosphere. These events, that occur at roughly the same time in all models, 
are summarized in table \ref{tab:events}, which shows the processes observed in the 
simulations ordered in time. These processes are described in the following sections according 
to where they happen.

In the following sections all the figures shown correspond to
simulation A4 unless otherwise noted, both because it has run for a long solar 
time ($T=3200$~s) and as it has the most pronounced processes as a result of flux tube 
emergence. 

\subsection{Photosphere}\label{sec:photosphere}

\begin{figure}
\begin{center}
\includegraphics[width=7.5cm]{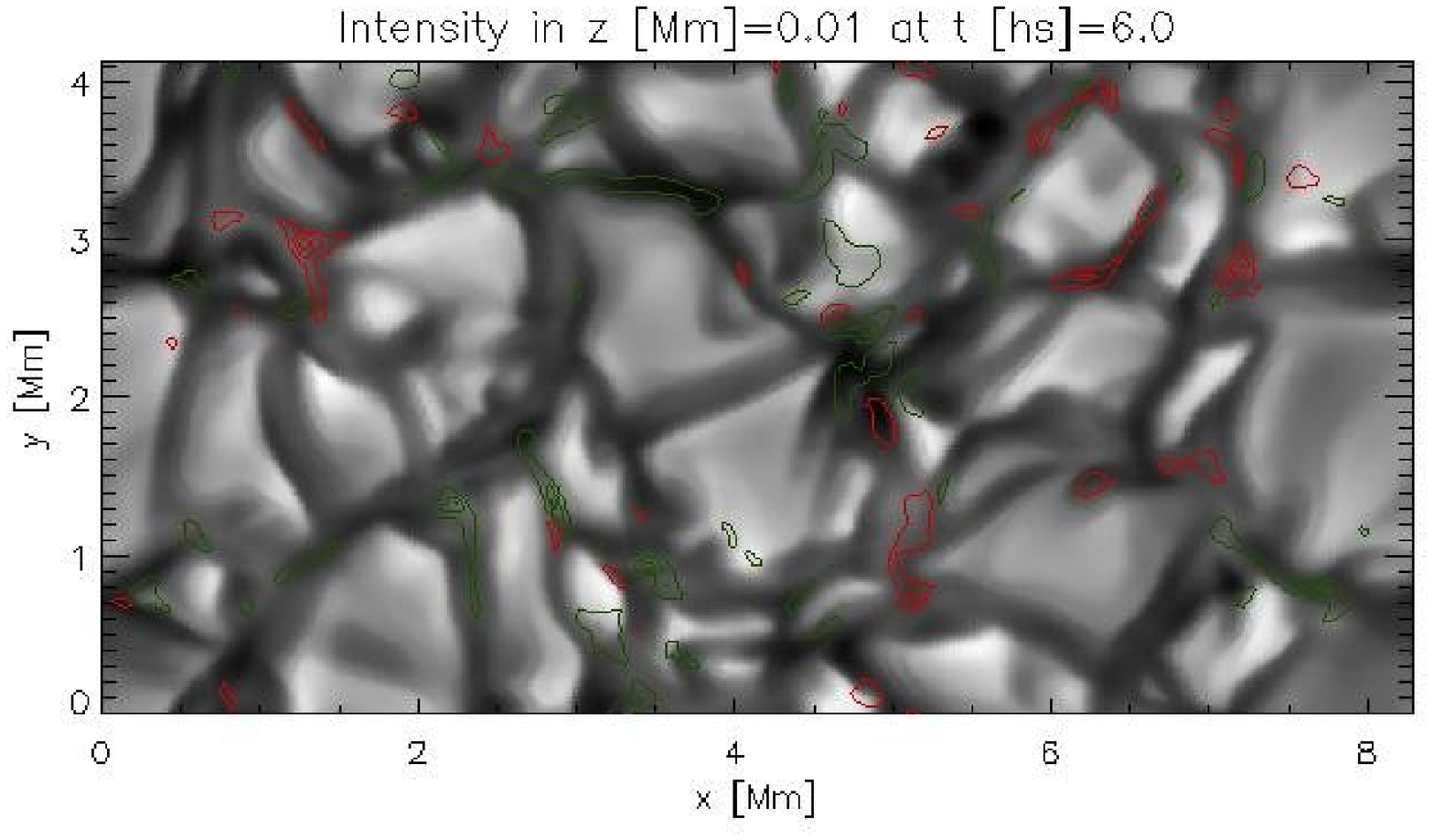}
\includegraphics[width=7.5cm]{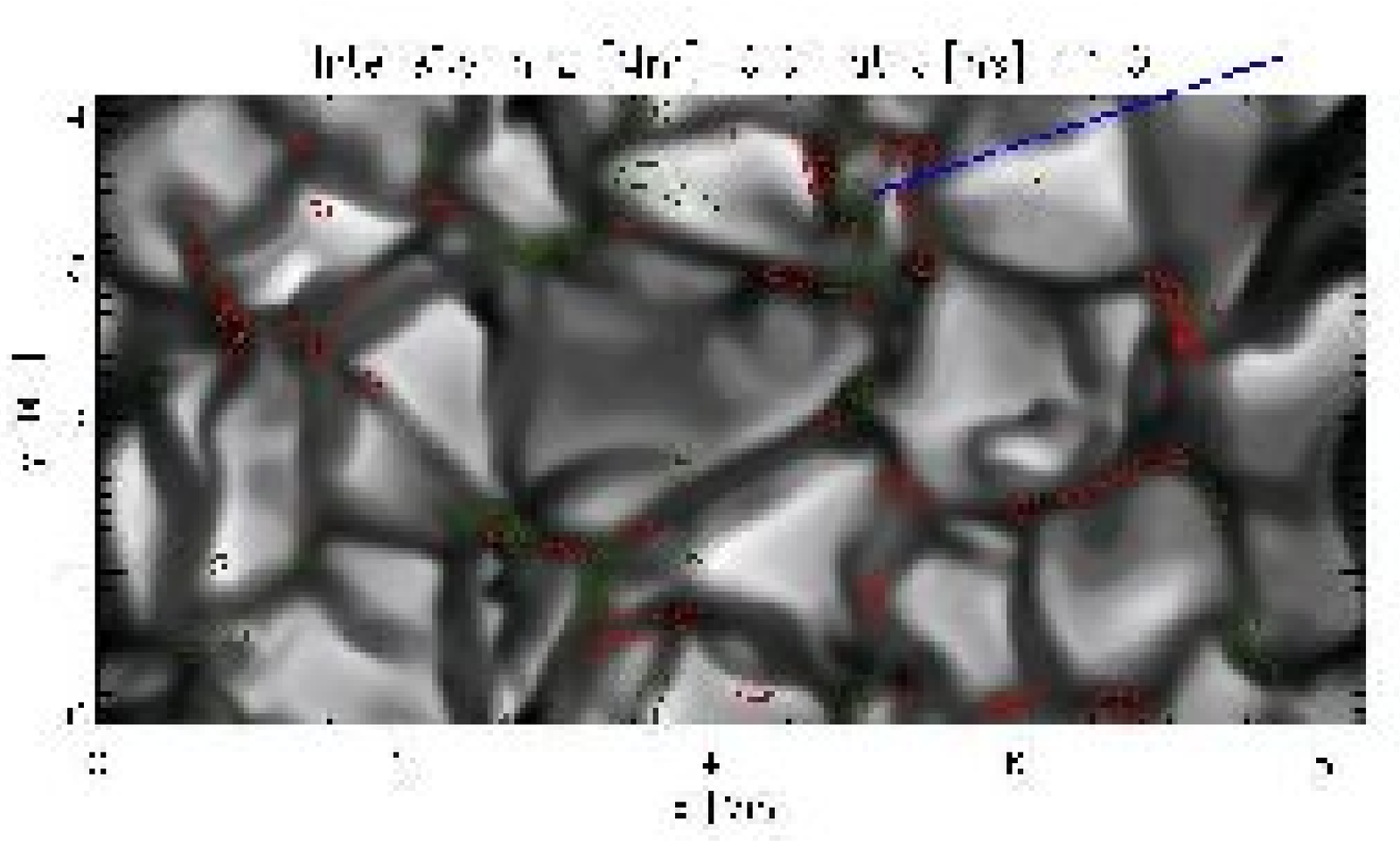}
\includegraphics[width=7.5cm]{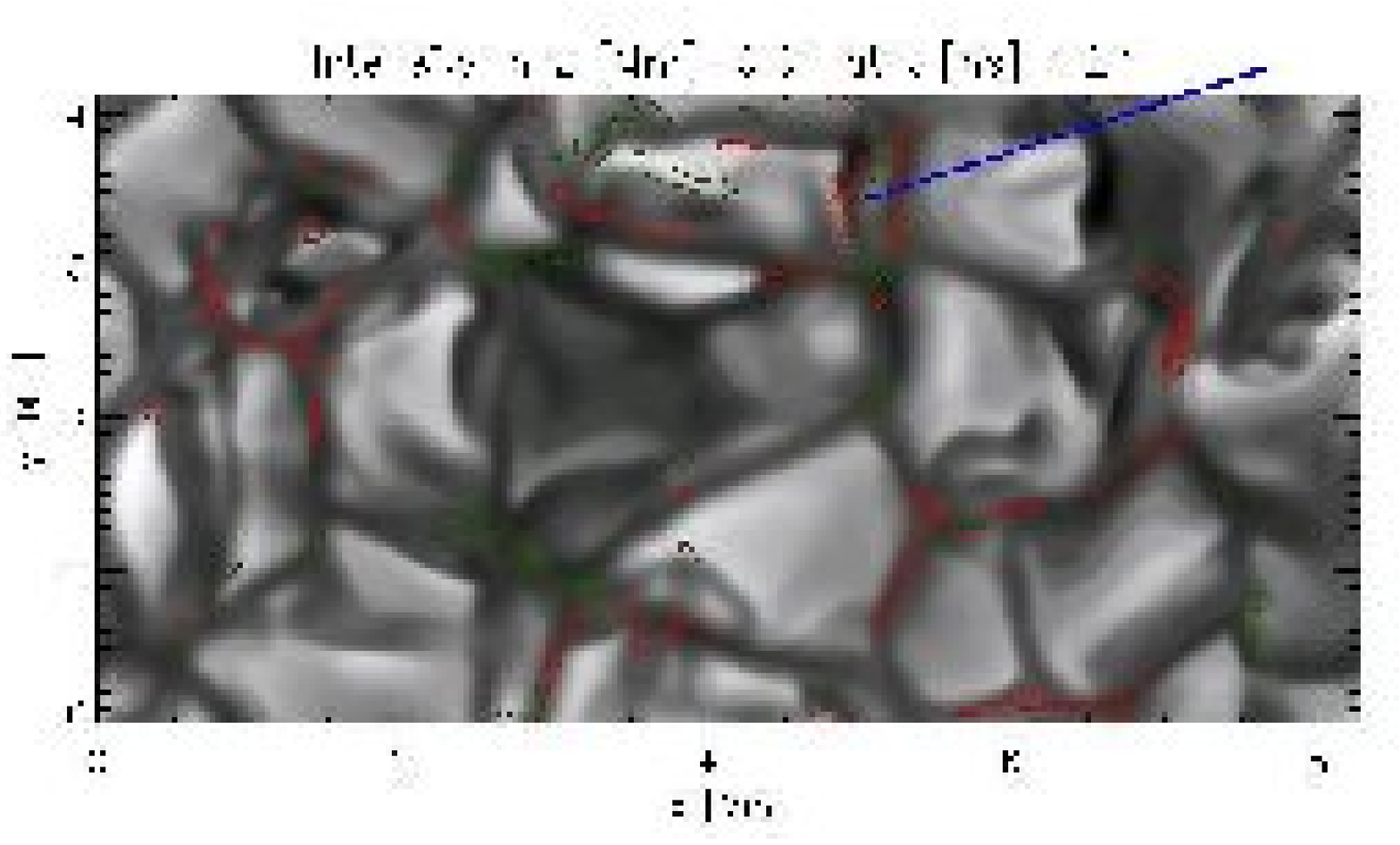}
\includegraphics[width=7.5cm]{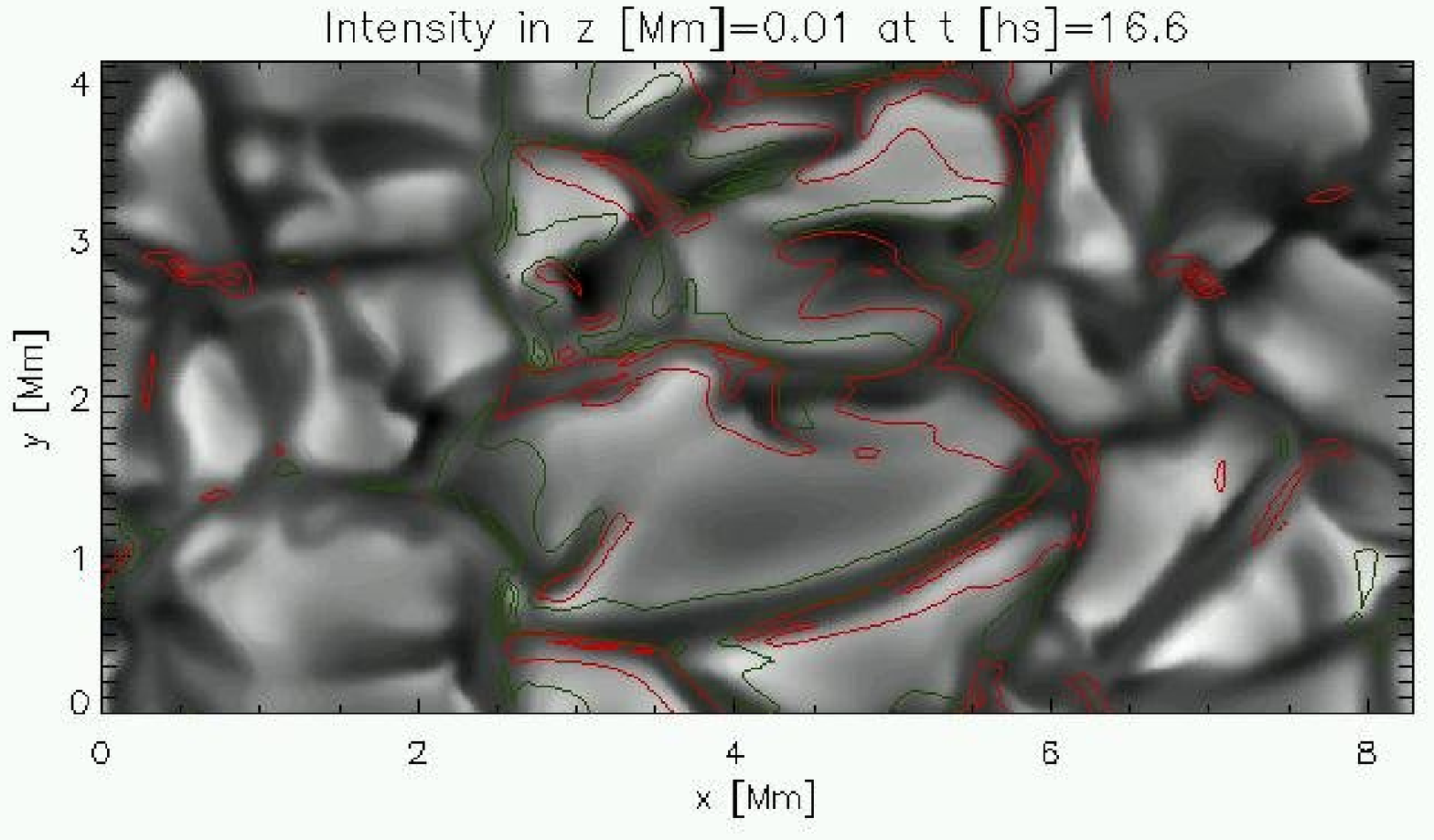}
\includegraphics[width=7.5cm]{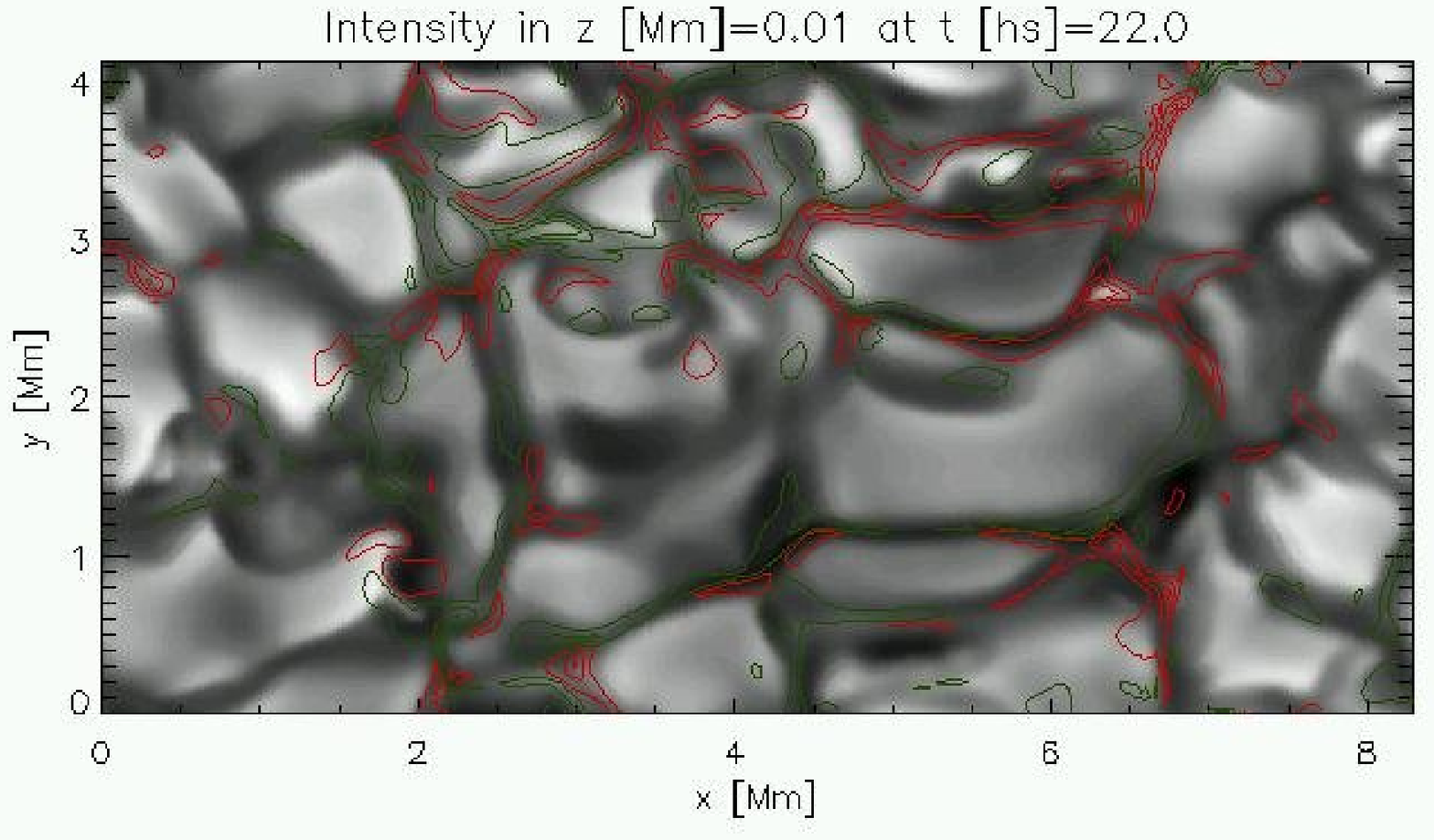}
\includegraphics[width=7.5cm]{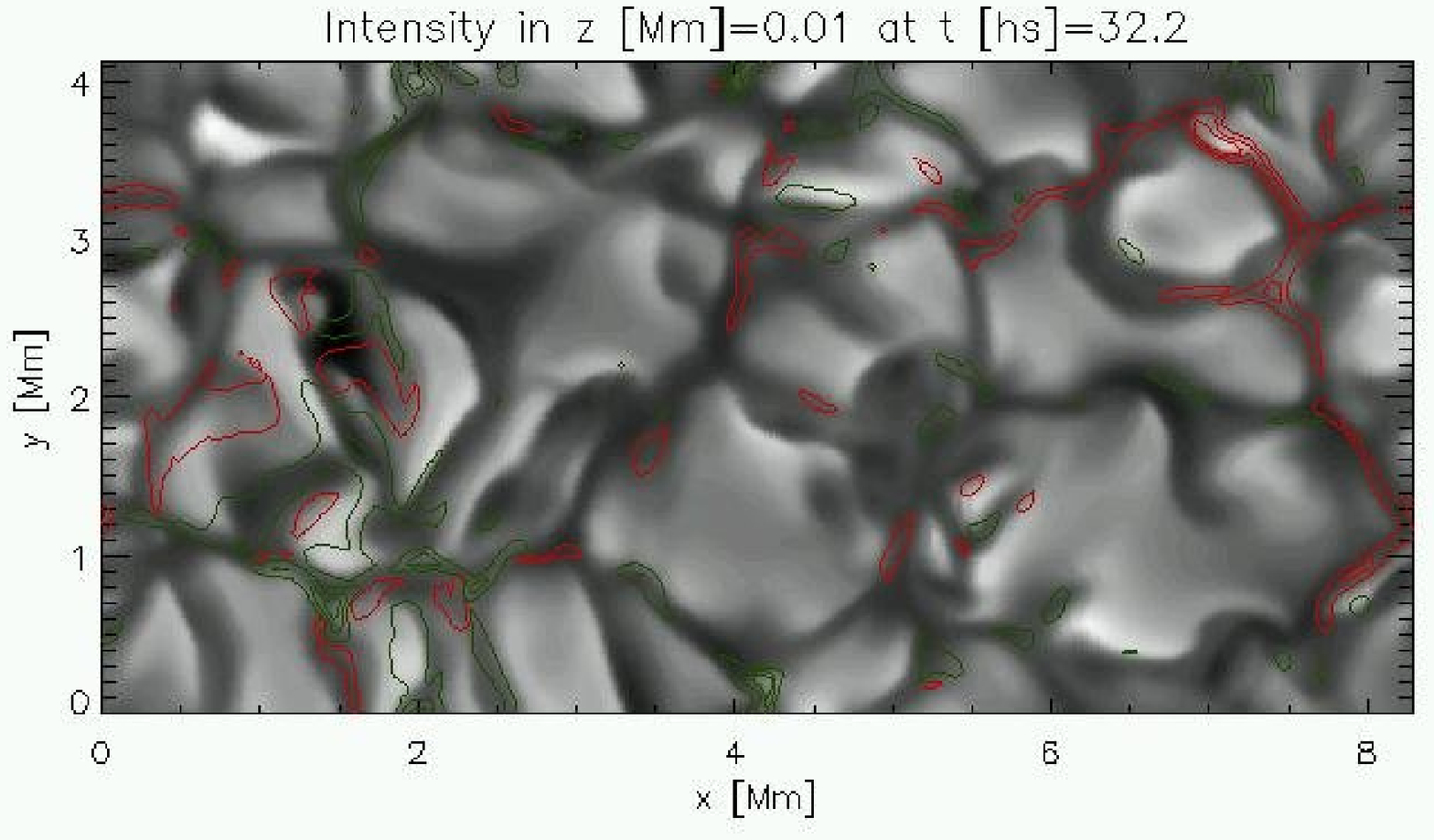}
\end{center}
\caption{\label{fig:intphot} The continuum intensity shown at 
times 600~s, before the flux tube goes through 
the photosphere, 
at time 1100~s, when the flux tube is close to the
photosphere, 
at time 1210~s, when the big cells are cooling by
expansion,
at time 1660~s, when the tube is crossing the photosphere,
 at time 2200~s, when the structure of the cells appears to have 
returned back to normal and at time 3220~s, from the left to
right and top to bottom respectively. The red contours correspond to
positive $B_z$ and green contours to negative $B_z$ at $z=10$~km.
An example of a collapsed granule is marked with a blue line.}
\end{figure}

\begin{figure}
\begin{center}
	\includegraphics[width=7.5cm]{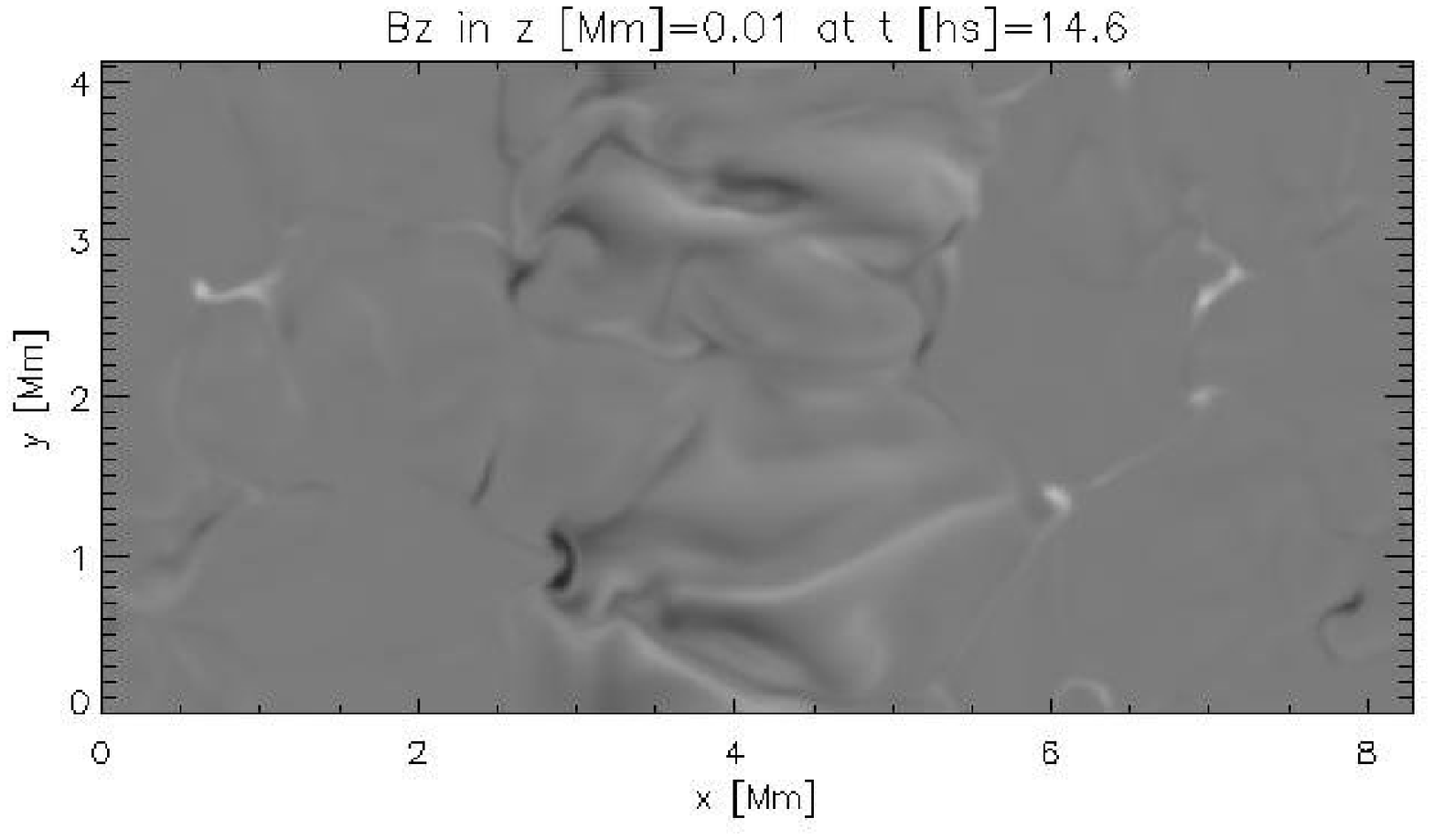}
	\includegraphics[width=7.5cm]{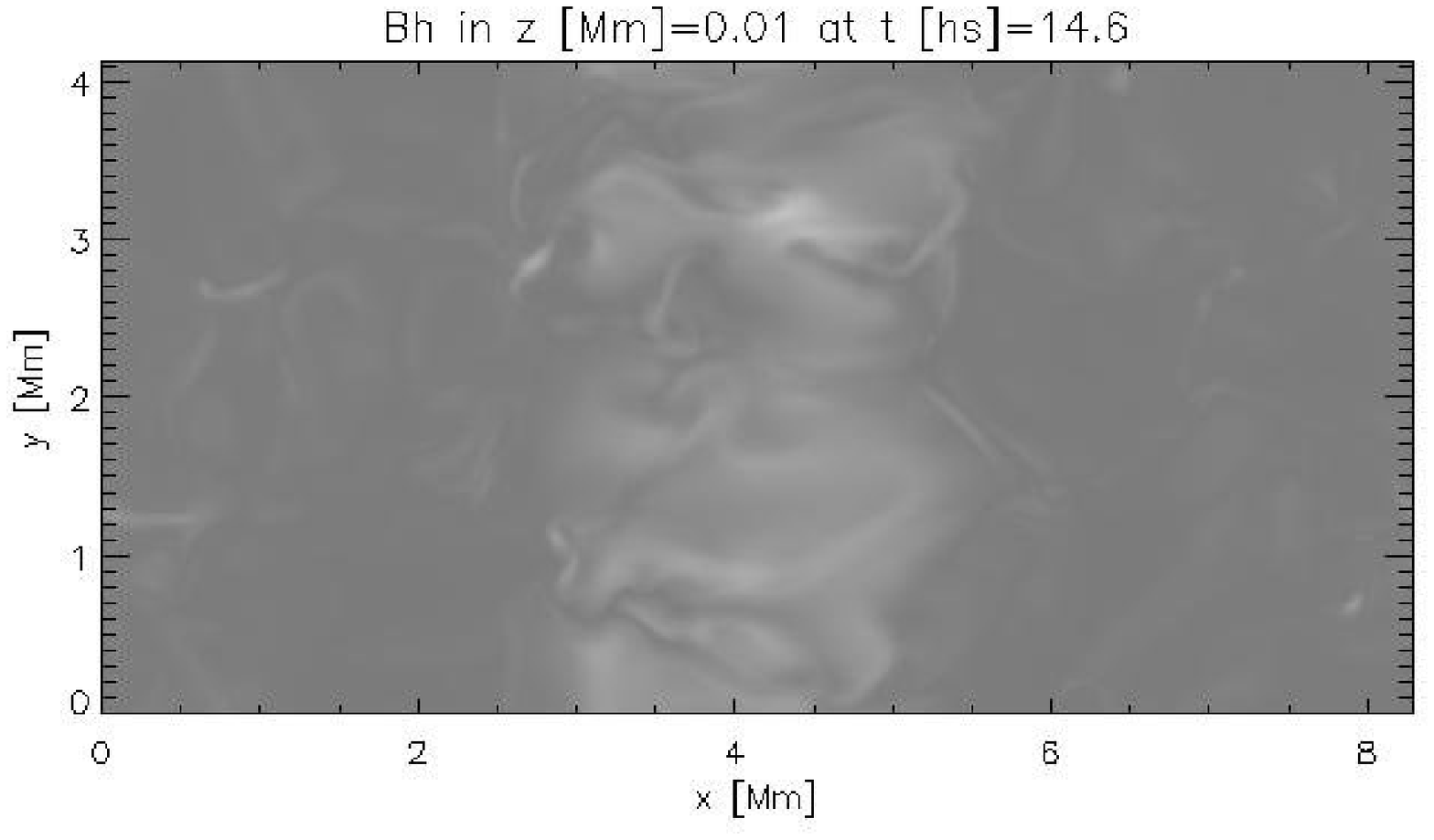}
	\includegraphics[width=7.5cm]{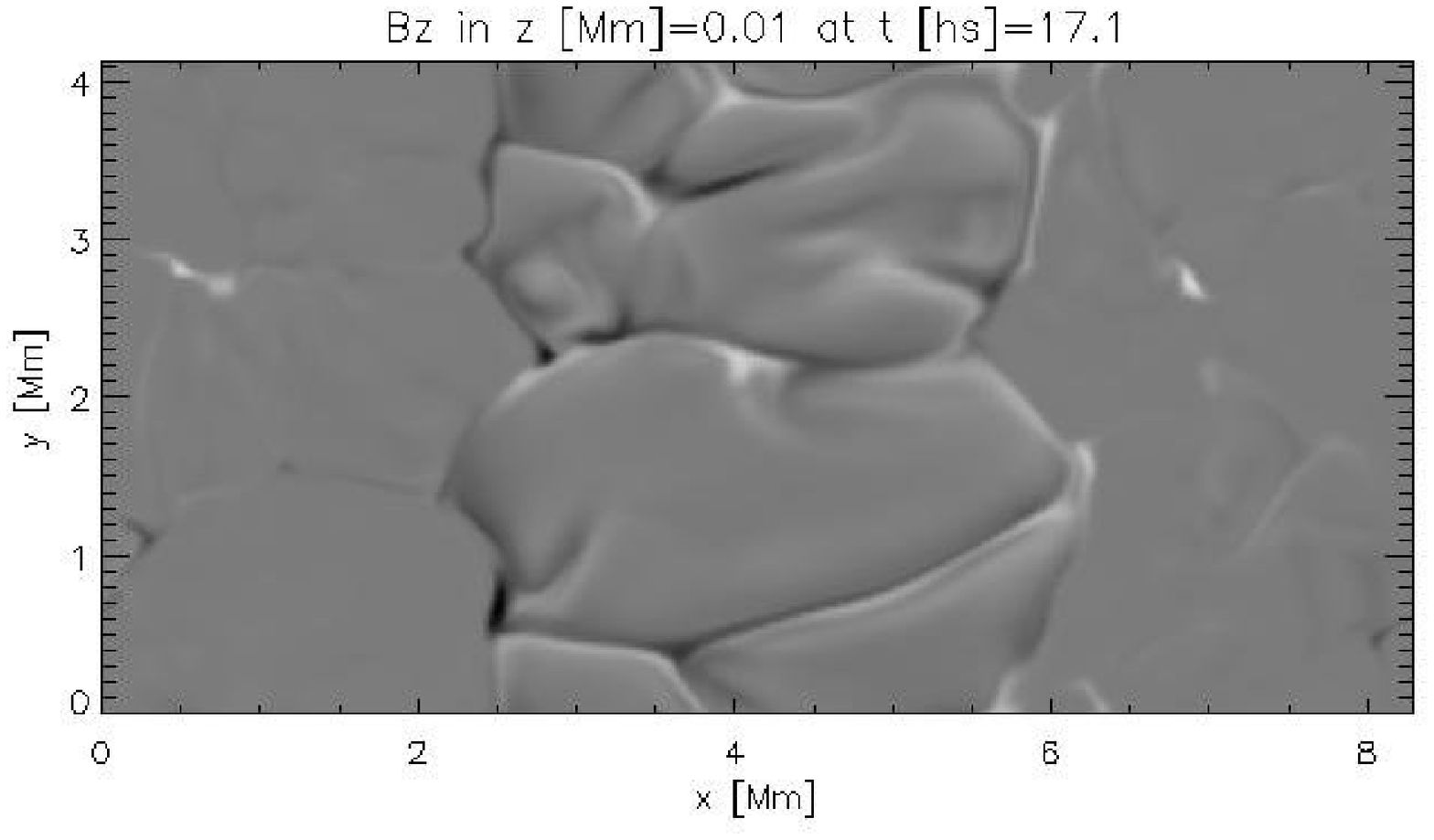}
	\includegraphics[width=7.5cm]{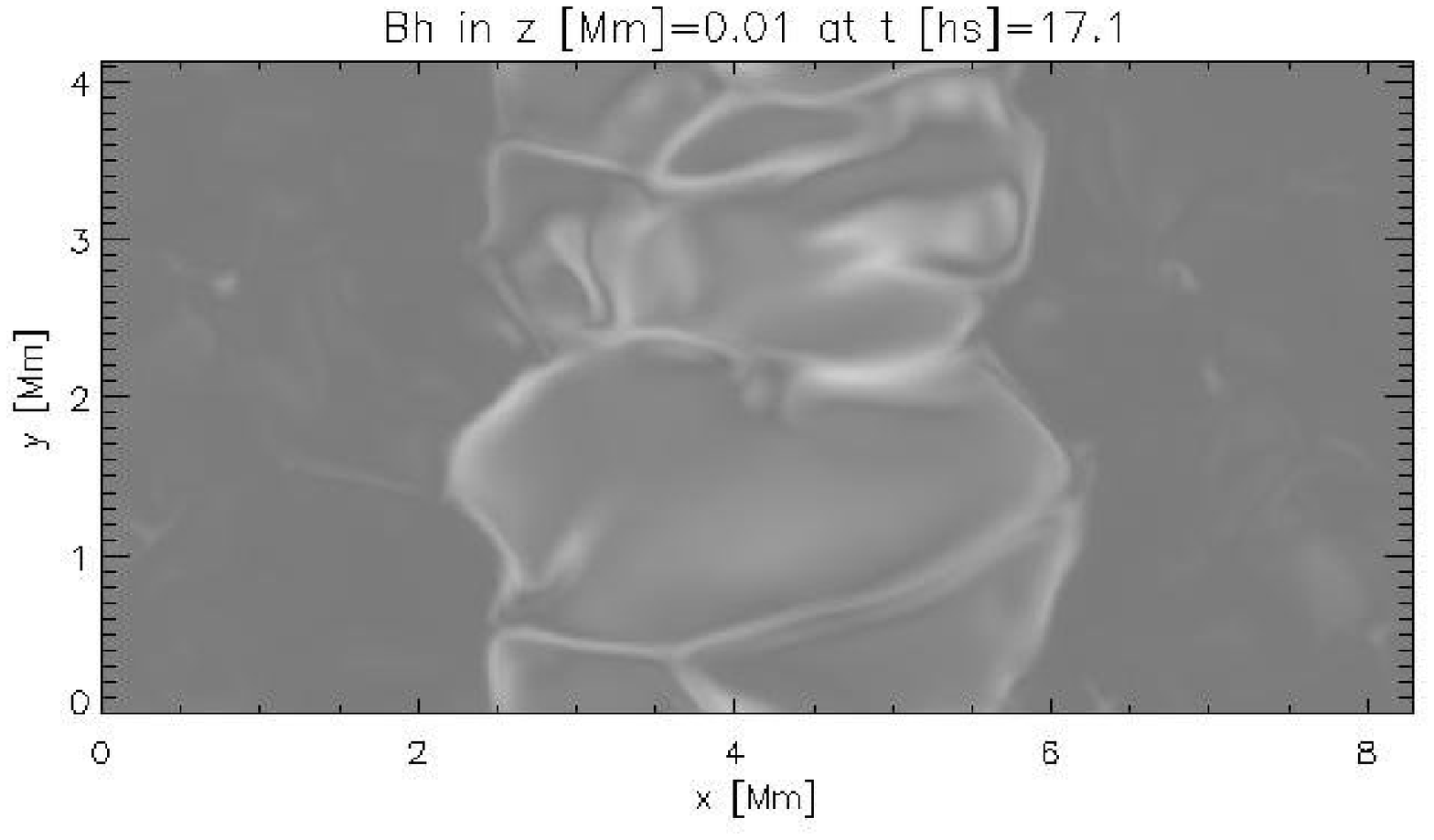}
	\includegraphics[width=7.5cm]{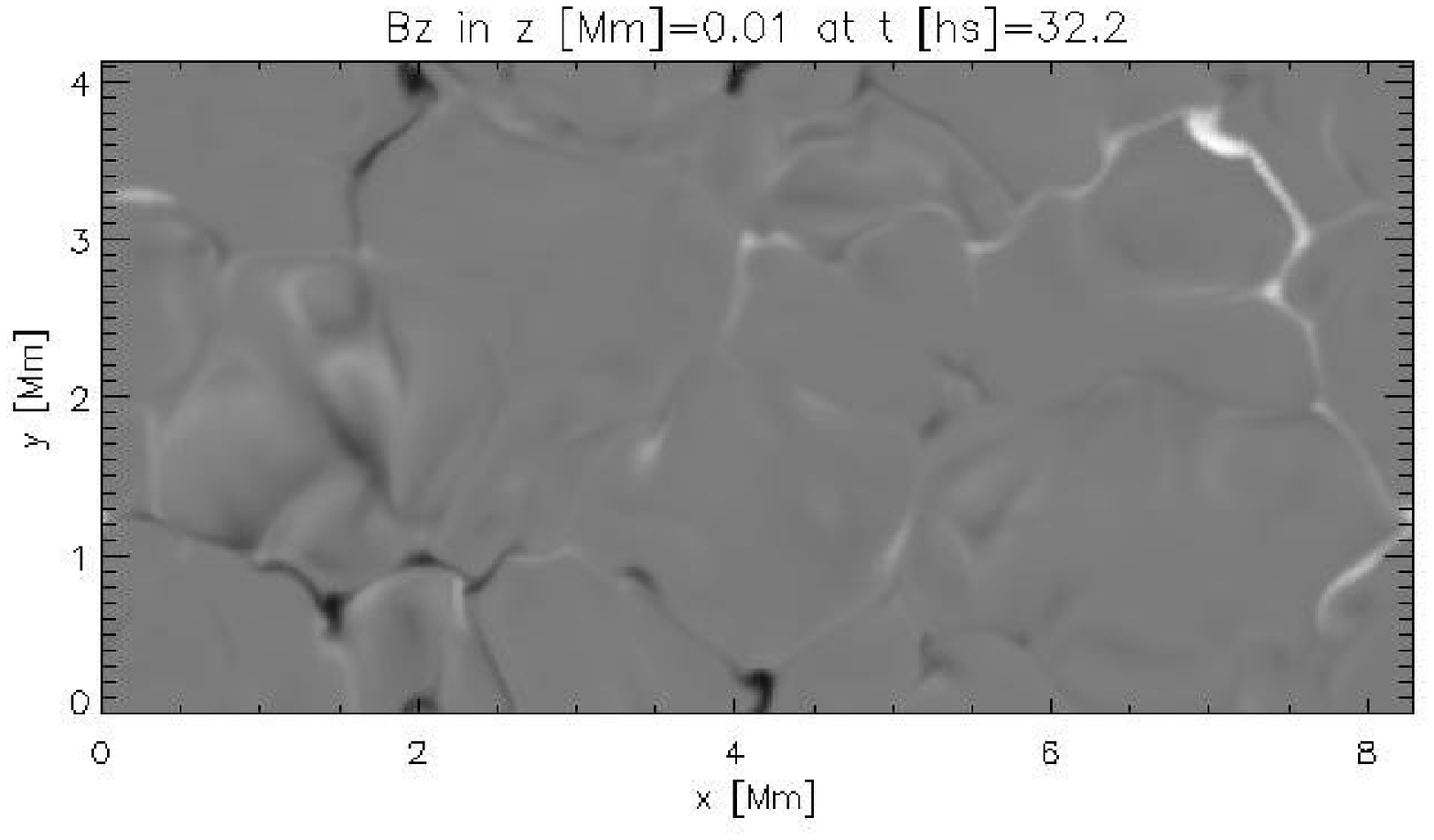}
	\includegraphics[width=7.5cm]{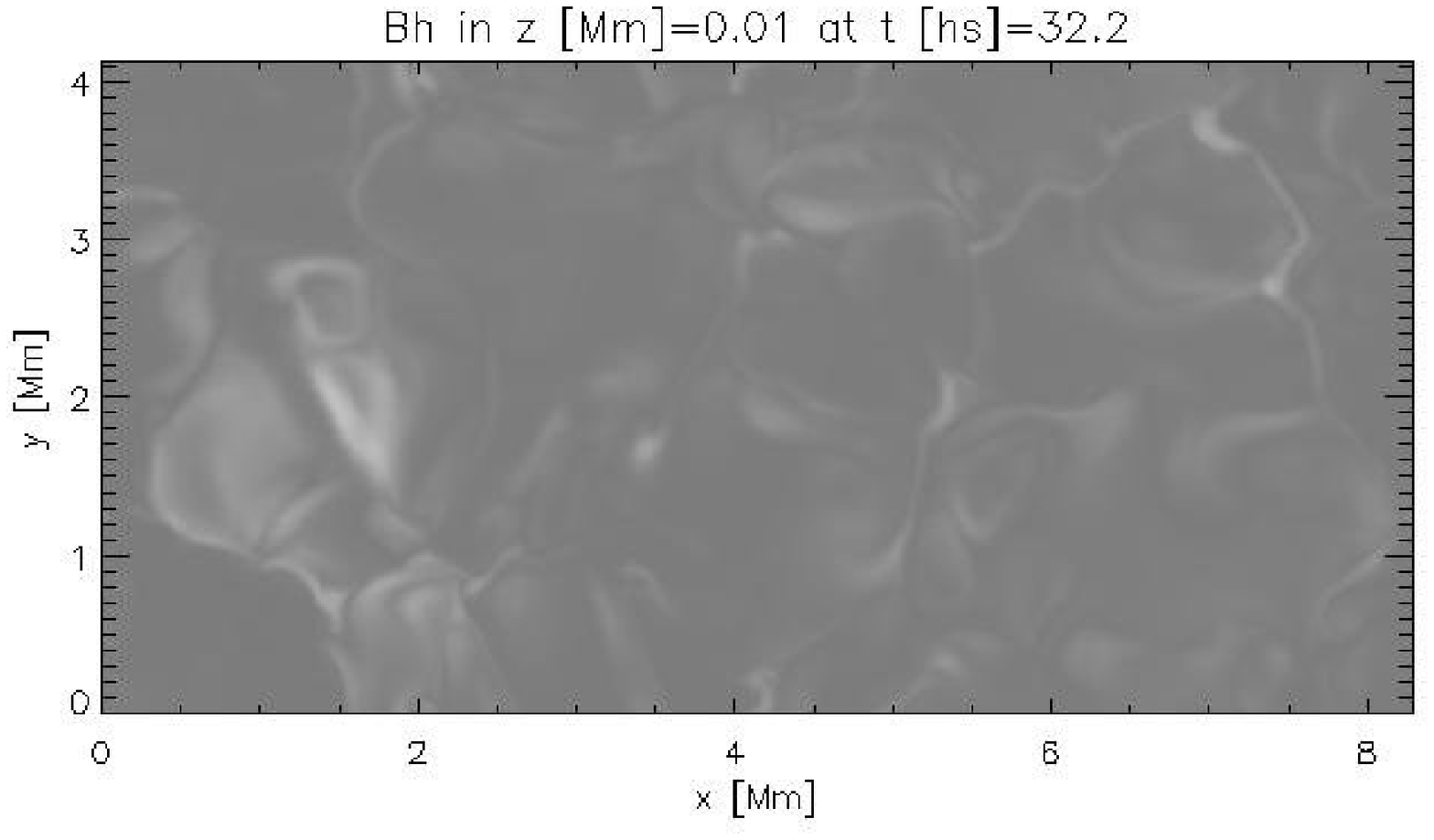}
 \end{center}
 \caption{\label{fig:fieldphot} 
 $B_z$ (longitudinal ``magnetograms'') in the photosphere, at $z=10$~km,
 shown with grey-scale intensity at three different times in the simulation in the left panels. 
 $B_h$ (horizontal magnetic field) in the photosphere at the same height and same instants shown in the 
 right panels. The grey-scale goes from -1111.5~G to 1376.3~G. The times chosen are 
  at time $1460$~s, when the tube just has started to cross the photosphere (top panels); 
 at time $1710$~s, as the tube is well into the photosphere (middle panels); 
 and at time $3220$~s (bottom panels).}
\end{figure}

The photosphere is located at $z\approx0$~Mm and is, up to the time when 
the emerging flux penetrates, as described in section~\ref{sec:initial-model},
with a cellular temperature structure similar to that shown in the upper left 
panel of figure~\ref{fig:intphot}. As the flux tube approaches the photosphere from 
below the cells immediately above the flux tube expand as seen in the upper 
right panel of figure~\ref{fig:intphot}. Eventually, horizontal field pierces the 
photospheric surface in the granular cell centers as seen in the second panel of the 
right column of figure~\ref{fig:fieldphot}. The horizontal field is rapidly moved to the
intergranular lanes as a result of the granular flow (figure~\ref{fig:fieldphot} third row). 
The granular cells where the field penetrates remain larger than in the undisturbed 
state as the flux tube passes through the photosphere. 
At some point, the expansion lowers the temperature of the
expanding cells through adiabatic cooling. At later stages a small amount of
the magnetic flux that pierces the photosphere is returned to the convection zone.
The photospheric evolution of emerging flux as described above has already been 
extensively studied previously and reported in the literature e.g. \citet{cheung2007} and references 
cited therein.

In all the simulations we find that 
approximately 2100~s passes from the time the magnetic flux tube crosses the 
photosphere until the time the photosphere again appears to regain its ``normal'' state, 
{\it i.e.} until the photospheric again is similar to the initial state before 
the flux tube passes through (see figure~\ref{fig:intphot}). Let us study the figures
in more detail.

\begin{figure}
\begin{center}
\includegraphics[width=7.5cm]{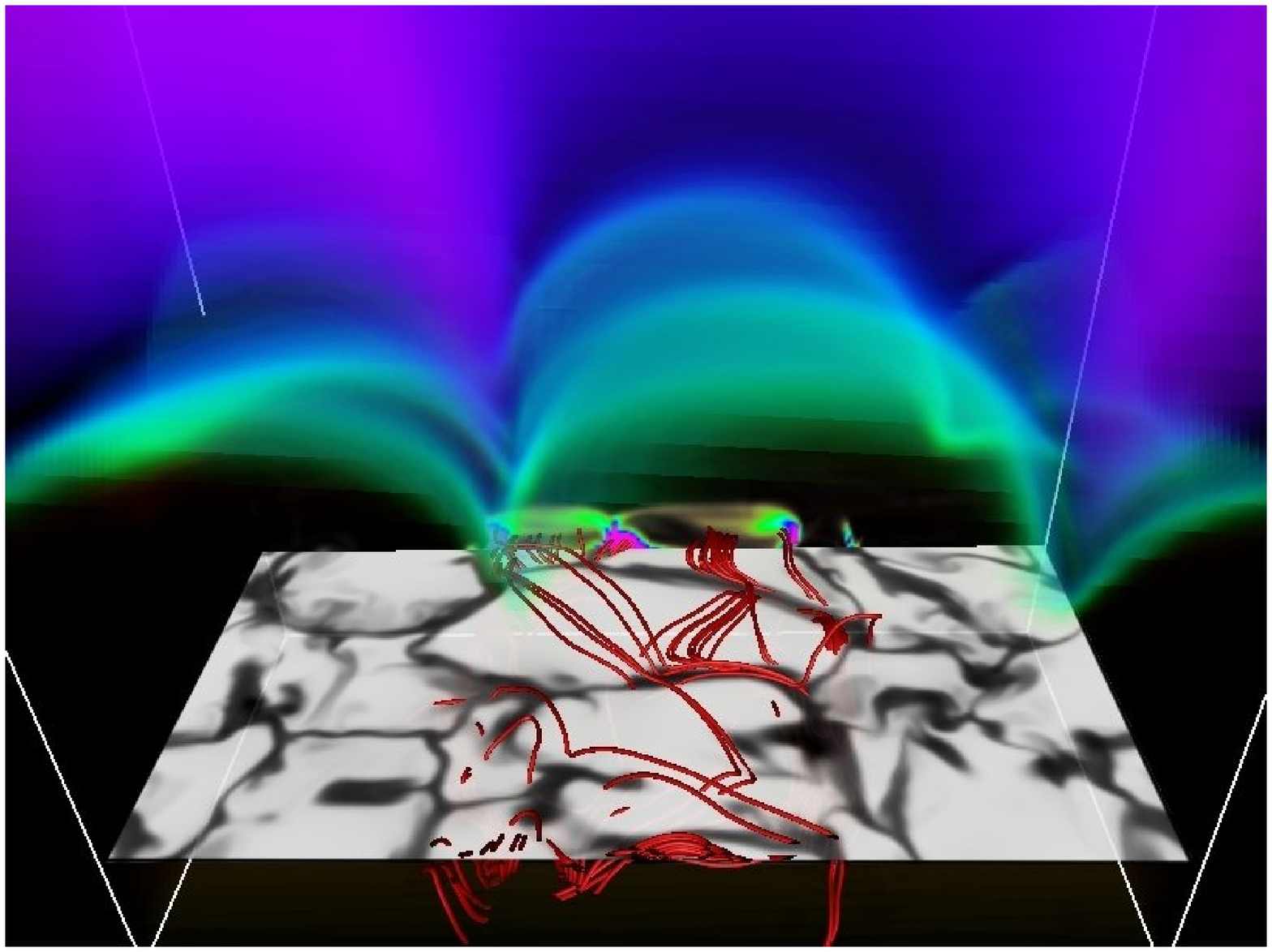}
\includegraphics[width=7.5cm]{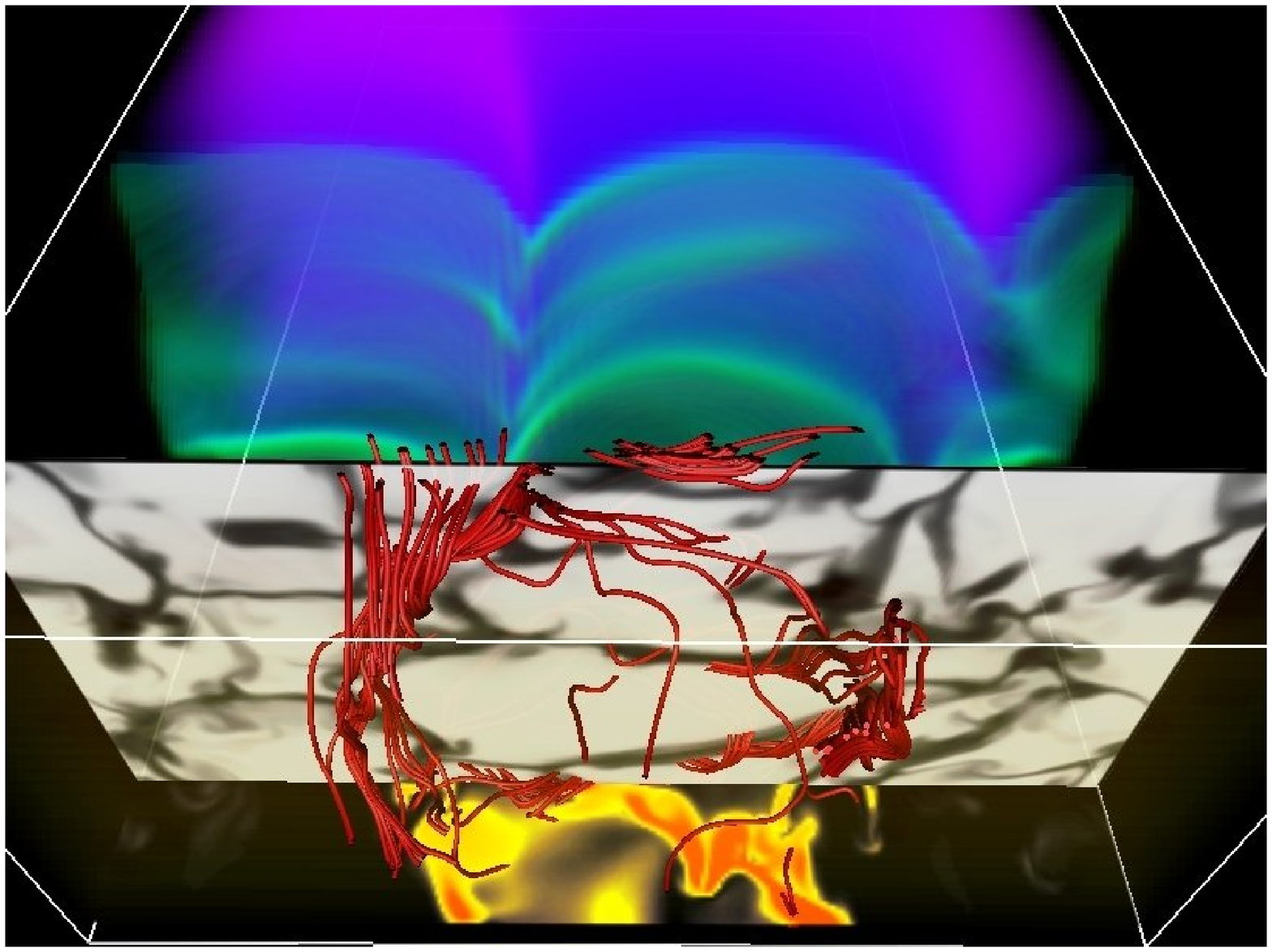}
  \end{center}
  \caption{\label{fig:3dpho} 3D image of the simulation at time $t=1700$~s.
  Temperature at the photosphere shown with grey-scale intensity, the red
  lines are the magnetic field lines of the tube, temperatures in 
  the 
  corona shown with green-violet scale color (green is the transition
  region) and with orange-scale is the magnetic field strength. 
  View from above (left panel) and from below (right panel).}
\end{figure}

In figure~\ref{fig:3dpho} we show two 3D images of the A4 simulation 
from two different viewpoints; from above and from below the photosphere as seen at 
time $t=1700$~s. The figure shows how the temperature structure of the granular cells 
has been changed by the presence of the flux tube, and in addition how the structure 
of the flux tube itself has been modified. The tube has suffered an expansion and 
splitting, gaining an irregular structure due to the movements of the convective cells. 
 
The magnetic field evolution in the photosphere is shown in the figure
\ref{fig:fieldphot}. The maximum magnetic field-strength in the photosphere
varies between 850~G and 1060~G. However, during the first hectosecond (hs)
after the flux tube enters the photosphere the maximum rises to between 
970~G and 1030~G. Later, the magnetic field is concentrated by the granular flow
and the maximum rises to some 1100~G . The increased maximum magnetic 
field-strength is observed 
from 1700~s until the simulation ends, in some instants rising to 1700~G 
during the last few hectoseconds of the simulation. The magnetic field is almost 
horizontal and centered in the cells when the flux tube starts to cross 
the photosphere. The magnetic field is confined by the fluid and 
displaced to cell edges at which time its orientation becomes vertical. 
After most of the original tube has crossed the photosphere,  some remaining 
magnetic flux that has remained below the photosphere appears. This remaining
flux is small compared to the main tube's emergence and appears far away on either side of 
the central tube axis. The emergence process for this remaining flux is similar 
to the main tube's emergence, in that almost horizontal field rises in the cell 
center to be subsequently transported to the intergranular lanes where it
becomes vertical as also reported earlier by 
\citet{cheung2007} (see the last row of figure~\ref{fig:fieldphot}).

The intensity in the opacity bin with smallest opacity, designed to mimic the continuum, 
is shown in figure \ref{fig:intphot} along with contours of the magnetic field strength 
at height $z=10$~km at 6 different instants in time; at 600, 1100, 1210, 1660, 
2200, and 3220~s.
The evolution of the intensity is very similar to that of the temperature:
Granular cells, located where the flux tube emerges, become larger and darker 
as they expand. These large cells cool as
their size increases due both to radiative looses and to adiabatic expansion. 
After the tube has crossed the photosphere the granulation pattern returns to
normal, roughly at times greater than 2100~s.

The temperature range before the tube crosses the photosphere is 
[6.0,10.2]~kK. While the upper value of the photospheric temperature remains 
the same, the lower limit varies: when the tube is just below the photosphere 
it rises to 6.2~kK, thereafter decreasing to 5.6~kK as large granulations cells 
cool by expansion (at time 1300~s). In the latter stages of the simulation
the lower temperature stabilizes at 5.8~kK.

The range in vertical velocity is roughly $[-6,5.5]$~km s$^{-1}$ where positive means
upwards flow. It is well known that downflows in the intergranules are
faster than the upward granular flow. 
However, it appears that 
the maximum and minimum vertical velocity 
increase after time 1100~s, {\it i.e.}
during and after the tube crosses the photosphere. 
The maximum upward velocity increases to 8~km s$^{-1}$ when the 
tube is in the photosphere, {\it i.e.} from 1100~s to 1600~s. 
Later the maximum returns to normal values, but still shows large variations in time. 
The maximum horizontal velocity oscillates between [6,7]~km s$^{-1}$ except at 
time 1350~s when there is a big jump to $10$~km s$^{-1}$, that lasts until
$t\approx 2000$~s. We do not see any remarkable patterns in the extrema of vertical
velocity before, during or after the tube crosses the photosphere.

It is worth remarking that we see several examples of collapsed cells 
\citep[see][]{Skartlien:2000lr} visible near the boundary of the flux tube as it 
crosses the photosphere (see the top-right at time 1100~s and bottom left at time 1210~s panels
at $x\approx 3$~Mm and at $x\approx 5$~Mm, the blue line of the figure 
\ref{fig:intphot}). 
Squeezing of the granule occurs as the adjacent plasma
expands during the rise of the flux tube and by the tendency for downflows 
surrounding a small granule to merge in the boundary layers. 
There are also bright points visible after the tube crosses the
photosphere, as seen in the lower left panel of the figure~\ref{fig:intphot}
at position $(x,y)\approx (6.5,2.7)$~Mm at time 2200~s or the bottom-right
panel at the position $7,3.8$~Mm. At that position the temperature is low
at photospheric height, but the magnetic field is strong and the decreased gas density 
(and thus photospheric opacity) 
allows us to see deeper into the atmosphere where the temperature is higher
\citep[see also][]{Shelyag:2004lr,Carlsson+Stein+Nordlund+Scharmer2004}.

In figure~\ref{fig:fluxret} we show the evolution of the mean of
magnetic flux of simulations A1, A2, A3 and A4 by area.  
We define the mean of the magnetic flux per unit of area between 2 heights by:

\begin{equation}
<\phi>=\int^L_0\left(\int^{L_h}_0 \int^u_l {\bf B} d{\bf S} \right)dn/(L |{\bf S}|) \label{eq:phi1}
\end{equation}

\noindent where $\phi$ is the mean of magnetic flux integrated in the
area in the vertical plane between $z=l$ and $z=u$  in the vertical component of ${\bf S}$,{\it i.e.}
$L_z=u-l$ and
along the whole horizontal component of ${\bf S}$, {\it i.e.} from $0$ to $L_h$. The mean is 
carried out
along the perpendicular component of ${\bf S}$, {\it i.e.}
it is defined as the integral over the line perpendicular to the 
surface ${\bf S}$ divided by the length in that direction of the box domain.
We are interested in measuring the flux in the $y$ direction, then the expression 
\ref{eq:phi1} simplifies to:

\begin{equation}
<\phi>=\int^{L{_y}}_0\left(\int_0^{L_{x}}\int^u_l  B_y dzdx\right) dy/(L_y L_x L_z)\label{eq:phi2} 
\end{equation}

\noindent where $L_x$  is the length of the box in the $x$ direction, $L_y$ is the length 
in the $y$ direction and the integration in $z$ goes from $z=l$ to $z=u$.
We consider two different integrated regions in the figure \ref{fig:fluxret}, one is below the 
photosphere ($z<0.01$~Mm), {\it i.e.} from $l=-1.5$~Mm to $u=0.01$~Mm, in red color, 
and the second from $l=0.01$~Mm 
to $u=245$~km, in black. 
The figure shows that the greater the twist, the greater the amount of magnetic flux that crosses 
the photosphere. On the other hand, the time needed for the tube
to cross the photosphere is similar in all simulations.
(It should be kept in mind that there is also magnetic flux being lost
through the bottom boundary of the simulation box in addition to that rising 
beyond the photosphere into the upper atmosphere.) 
The simulation A2 (with weaker injected field $B_o$) has a smaller total flux and 
the slope ($d\langle\phi\rangle/dt$) is much smaller than in the other simulations. 
When the tube has finished crossing the photosphere ($t>2000$~s)
the flux retained above the photosphere is in quasi-equilibrium and 
oscillates with a period of roughly 5 minutes.
This oscillation is due in part to the magnetic field that is returned 
to the photosphere and in part because the direction of the tube 
changes in time.
The oscillation in flux strength coincides with oscillations in maximum downward 
velocity variation with time, but there is a small delay of around 30~s between the 
flux and the velocity variation. We find the same phenomenon in upper layers (see section
\ref{sec:revgran} and \ref{upcrom}). 

The average amount of magnetic flux in the convection zone is greater than in the photosphere, 
which means not all of the flux tube crosses into the photosphere. 
The only portion of the magnetic tube that reaches the photosphere is the part that has a 
steeply decreasing field strength with height. \citet{cheung2007} have pointed out that the 
higher the level of twist in the initial flux tube, the larger the fraction of magnetic flux 
that can pass into the regions above the photosphere. This can be explained as a consequence
of the magnetic tension of the transverse fields tending to keep the tube coherent as shown
by \citet{moreno1996}. Following the linear stability analysis of \citet{Acheson:1979lr} we find
that the photosphere has a superadiabatic excess profile, defined 
by $\delta_T=\nabla-\nabla_{ad}$, that is shaped like a gaussian with 
height. In this case the magnetic buoyancy instability sees a barrier in the 
photosphere and the emerging tube velocity decreases as it comes closer to 
the photosphere. Only the portion of the flux tube where the gradient of the magnetic field
strength is larger than the superadiabatic excess in the photosphere can reach the layers
above the photosphere. The relation between the buoyancy instability and the superadiabatic
excess has been studied by several authors \citep{Acheson:1979lr,magara2001,archontis2004}.
 
The field left behind after the tube has passed is more vertical than it was before the 
tube passes through a given particular height. This behavior is more evident near the end 
of the simulation where the magnetic field is more vertical in the 
chromosphere. The ratio between 
$B_z$ and $|B|$ oscillates in time in much the same manner as does the maximum downward 
velocity in the photosphere between $z=0$~km and $z=458$~km.

\subsection{Reverse granulation}\label{sec:revgran}

Intensities originating a few hundred kilometers above the
granulation show a pattern that is reverse that of granulation,
with low intensity above granules and higher intensity above
intergranular lanes \citep{Evans:1972fk,
Suemoto:1987qy,Suemoto:1990uq,Rutten:2003fj}. 
The reason for the inverse contrast is the cooling from 
a divergent velocity field above granules and heating from
a convergent velocity field above the intergranular lanes.
The final intensity pattern from this layer may also
be influenced by internal gravity waves but does not
appear to be related to the magnetic field
\citep{Rutten:2004kx,Leenaarts:2005yq}.
However, as we discuss below, when the magnetic flux crosses this region, 
the contrast and the structure of the reverse granulation is modified. 

\begin{figure}
\begin{center}
\includegraphics[width=7.5cm]{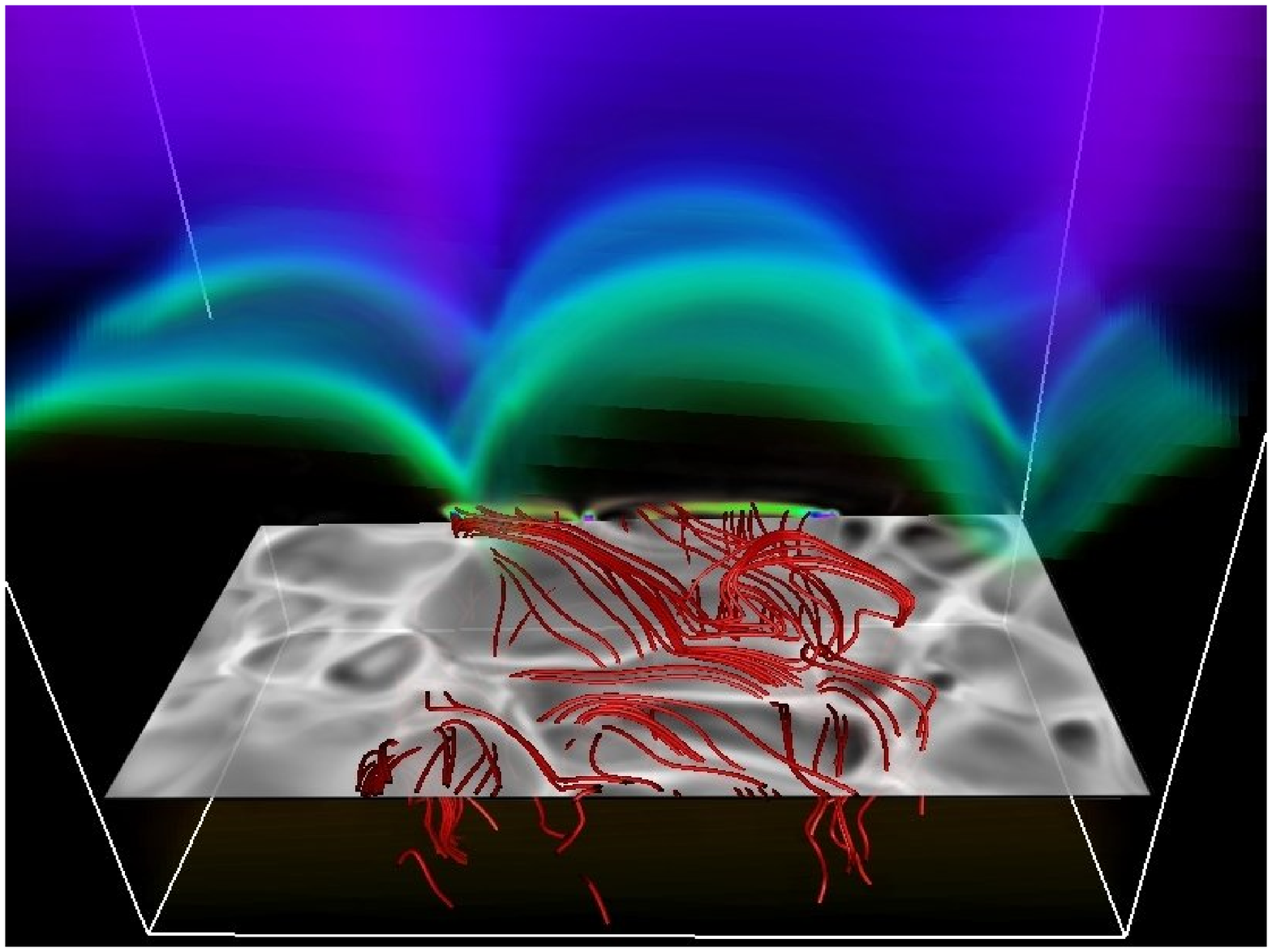}
\includegraphics[width=7.5cm]{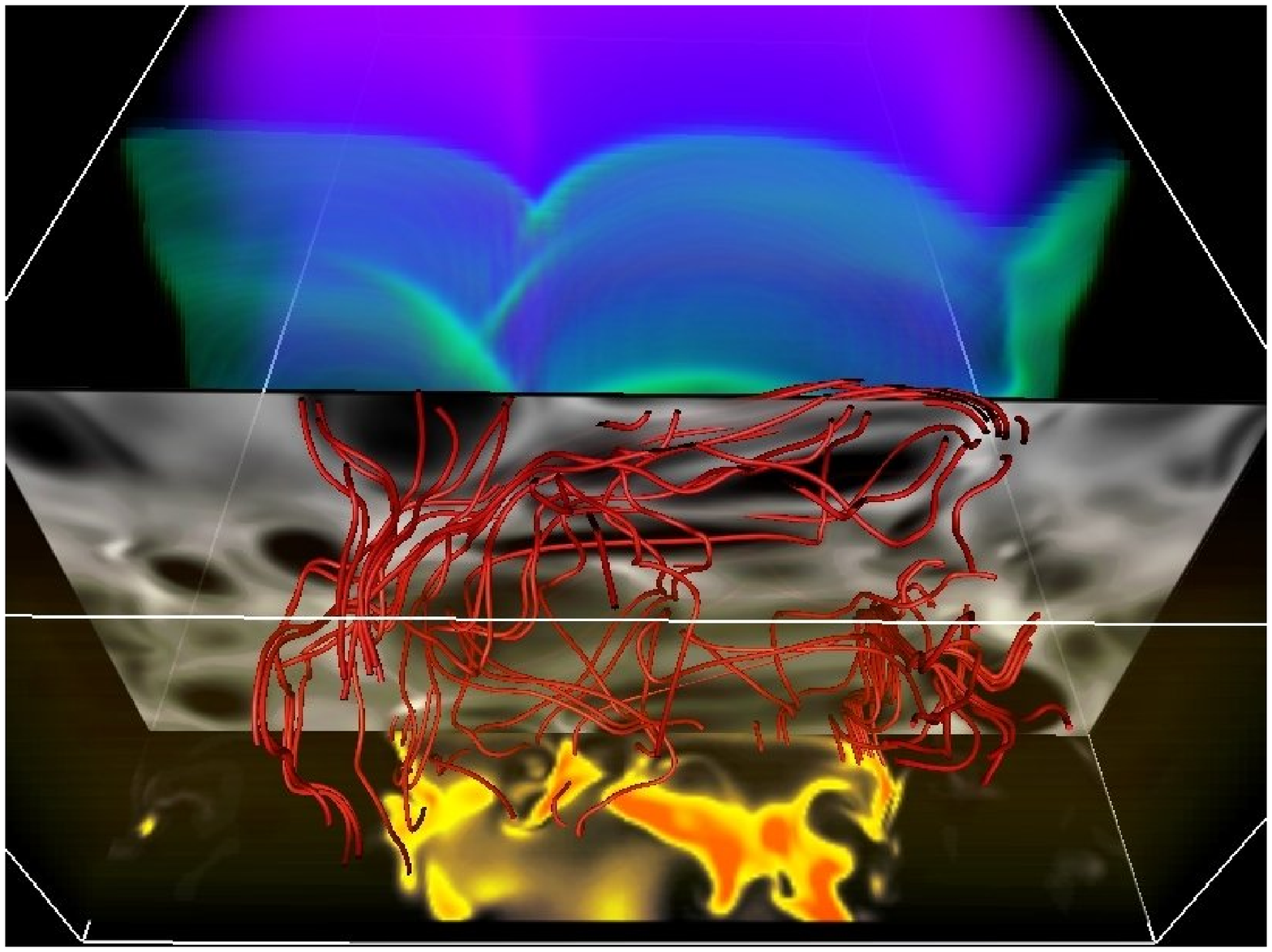}
  \end{center}
  \caption{\label{fig:3drev} 3D image of the simulation at time $t=1900$~s.
  Temperature at the height ($z=234$~km) shown
  with grey-scale intensity. The rest of the color scale is the
  same as in figure~\ref{fig:3dpho}. 
  View from above (left panel) and from below (right panel).}
\end{figure}

Figure~\ref{fig:3drev} shows 3D images in two
different views at $t=1900$~s; as seen from above and from below the reverse granulation layer. 
The figure shows how the reverse
granulation has been changed by the presence of the rising magnetic flux tube and 
is colder in the interior of the reverse granulation cells in regions where the tube is 
located than outside. Also at this height we can see how the structure of the tube
has been disrupted as it has passed through the atmosphere. 
In fact, the structure of the tube is more split and fragmented than that seen in 
figure~\ref{fig:3dpho}.

\begin{figure}
\begin{center}
\includegraphics[width=7.5cm]{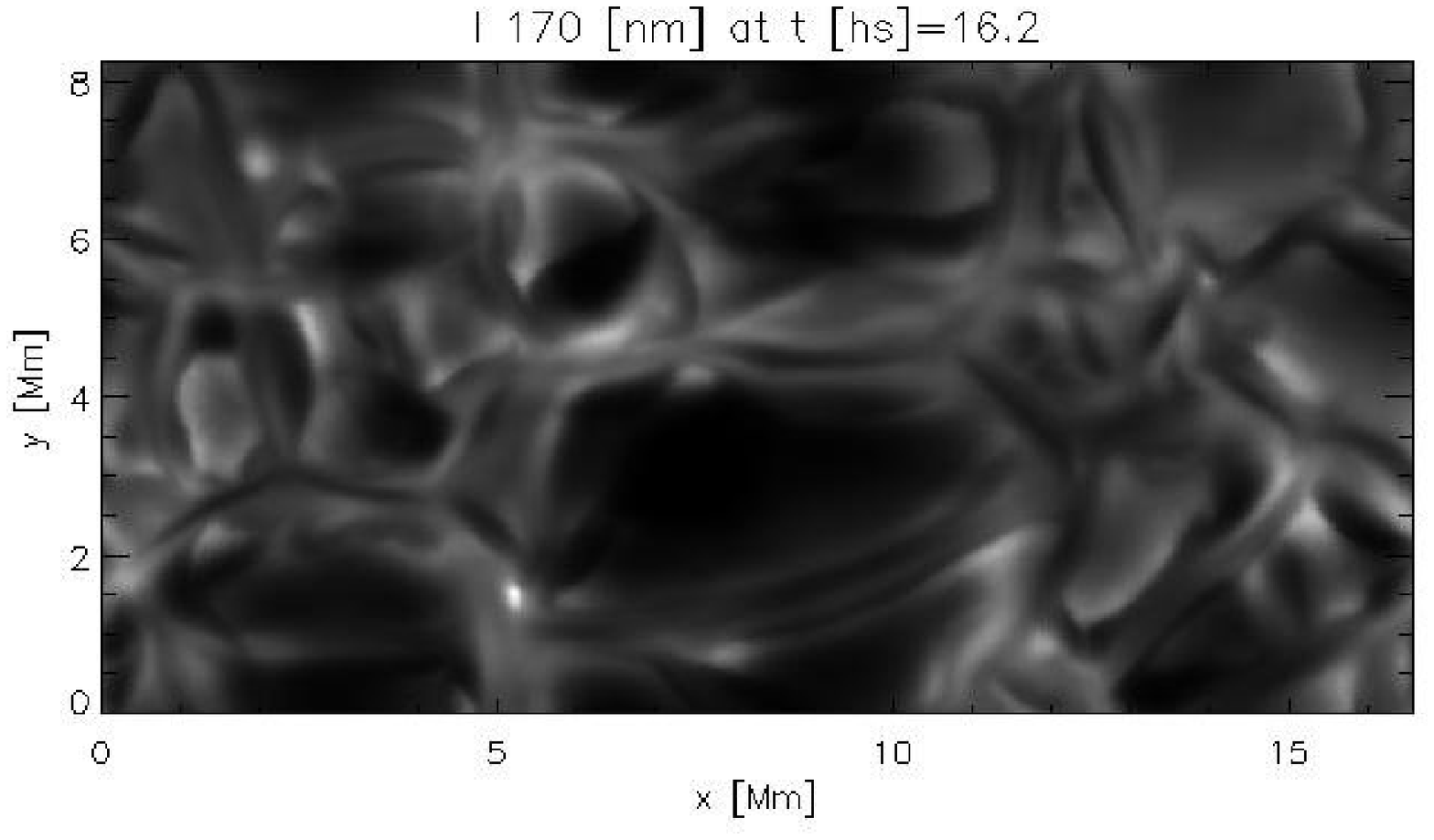}
\includegraphics[width=7.5cm]{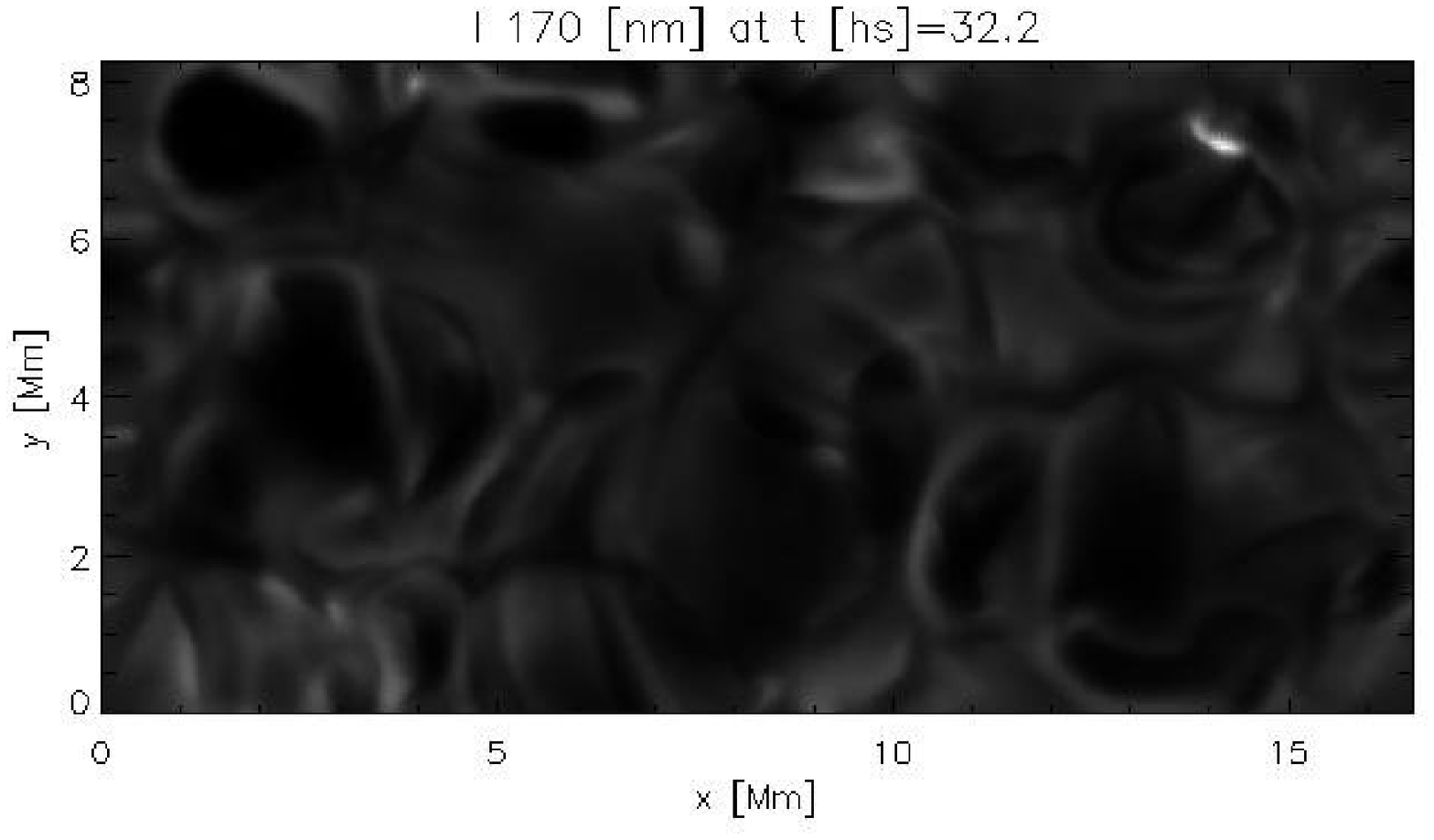}
\includegraphics[width=7.5cm]{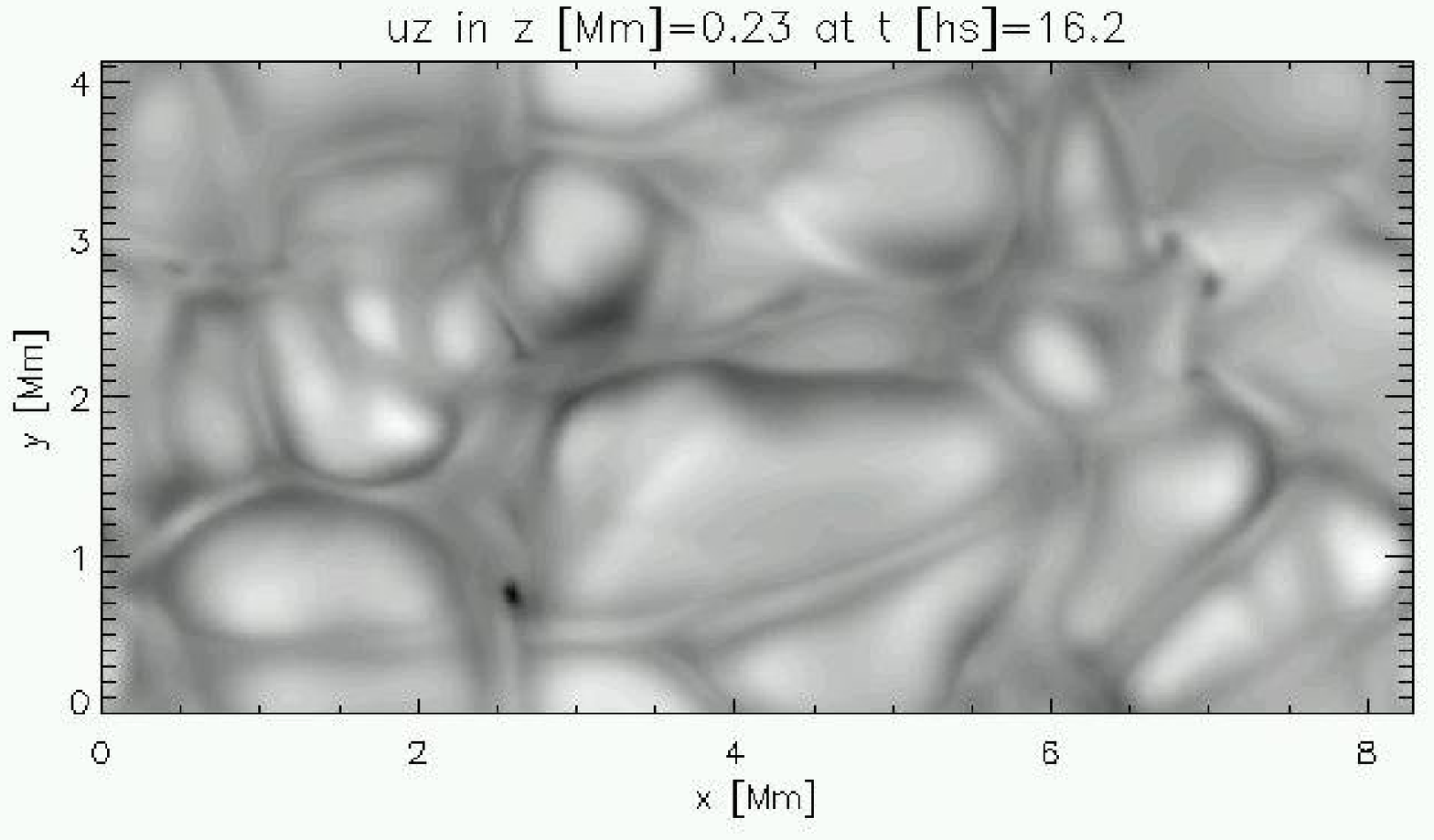}
\includegraphics[width=7.5cm]{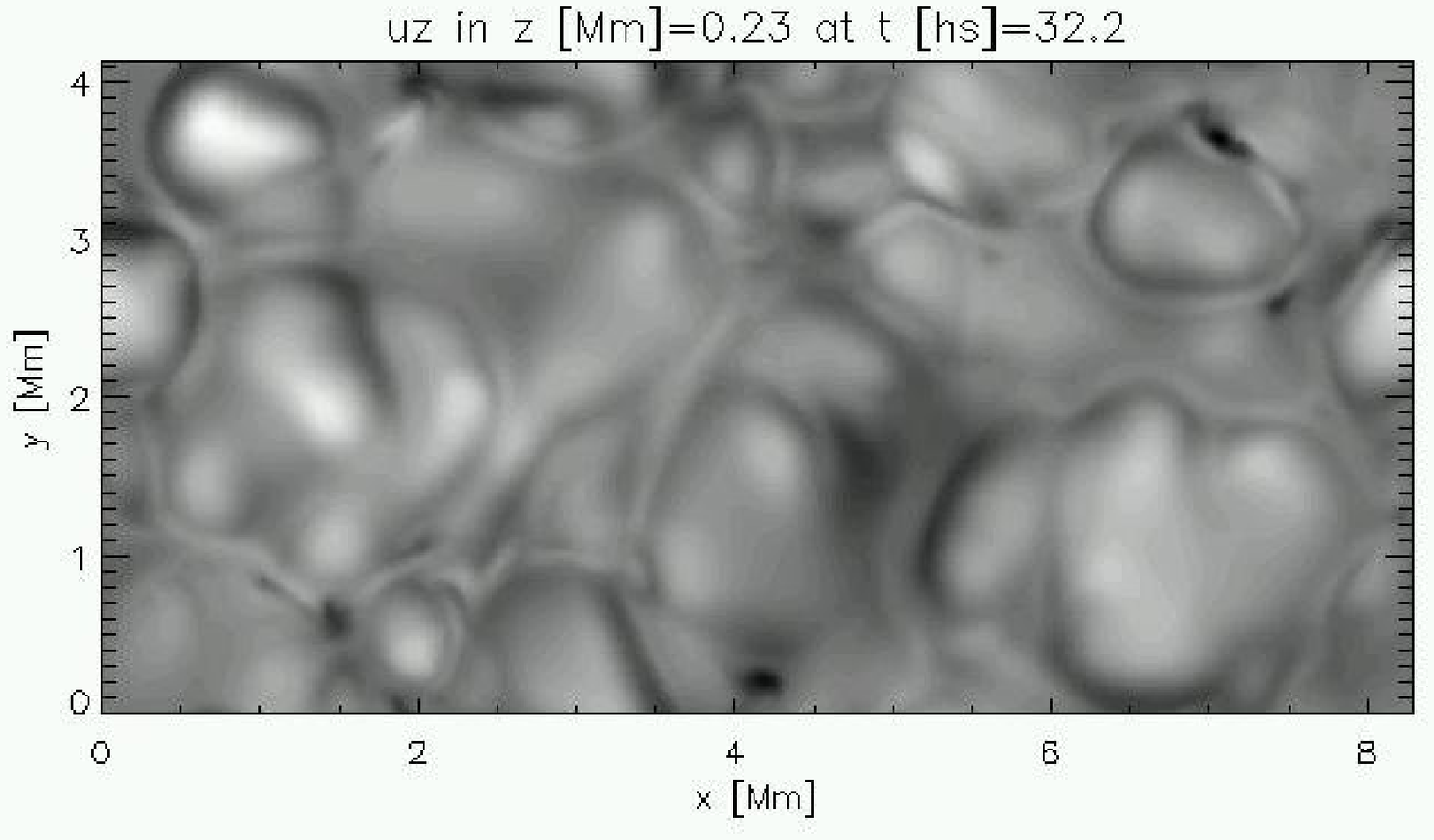}
\includegraphics[width=7.5cm]{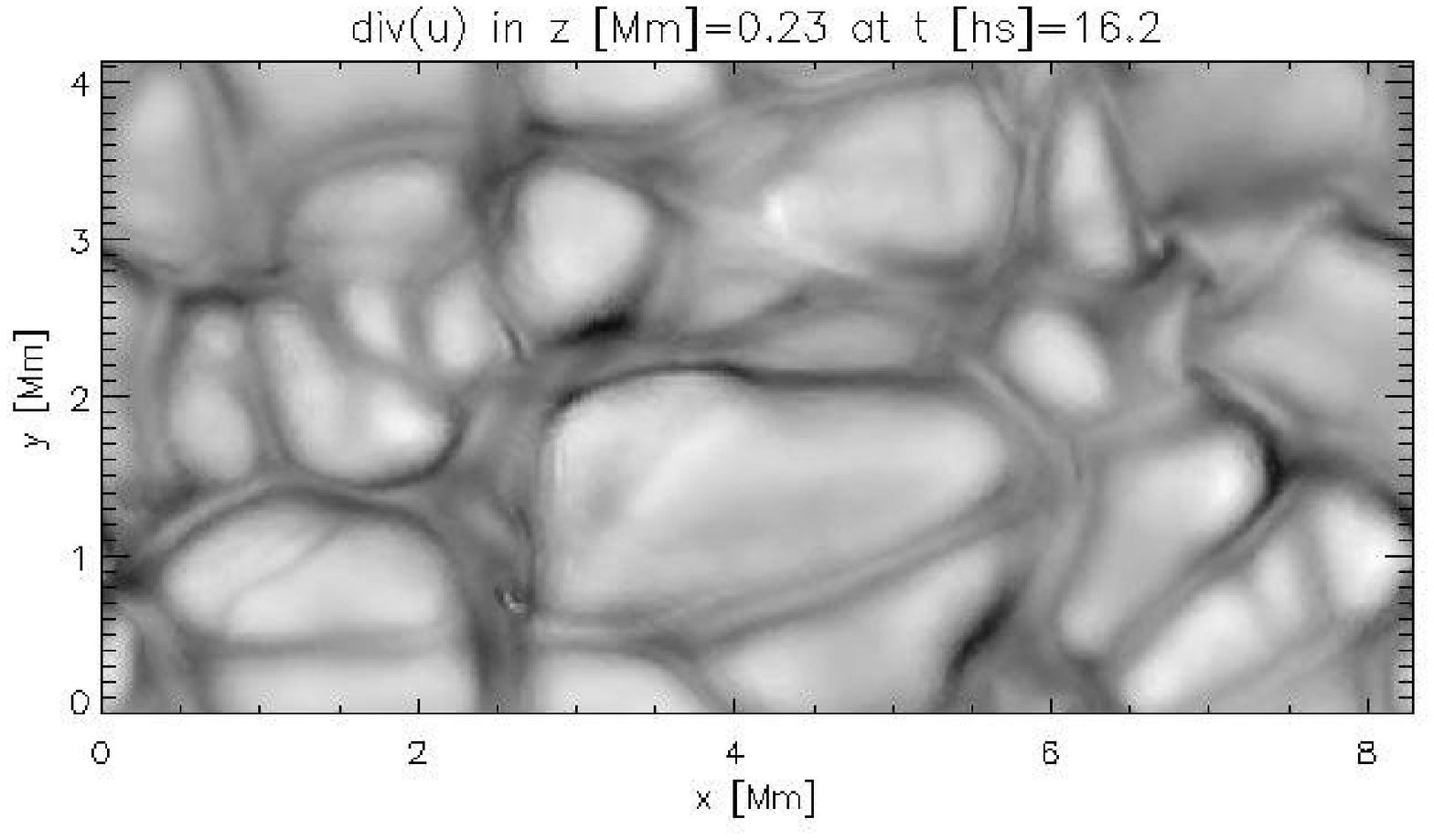}
\includegraphics[width=7.5cm]{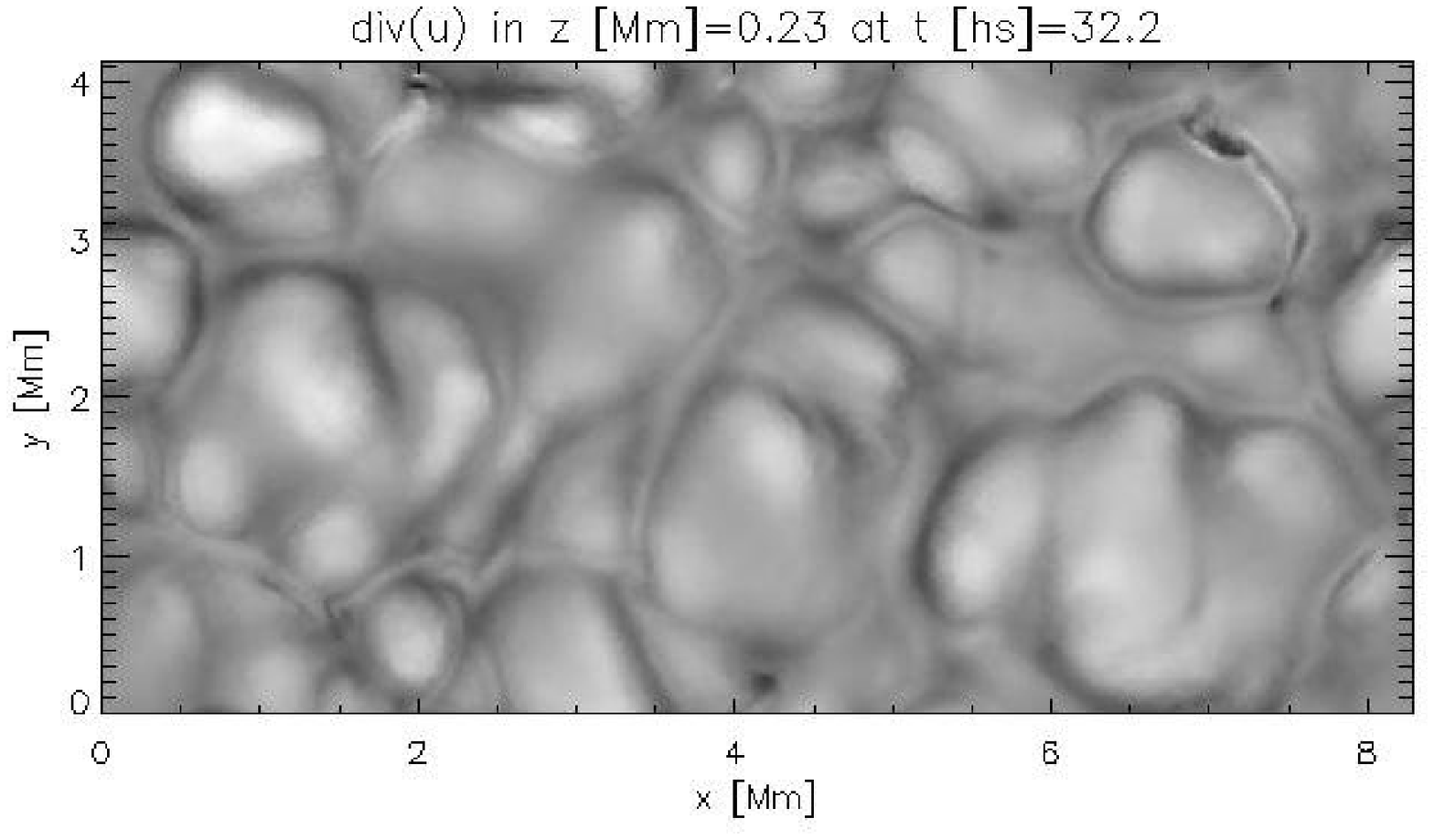}
\includegraphics[width=7.5cm]{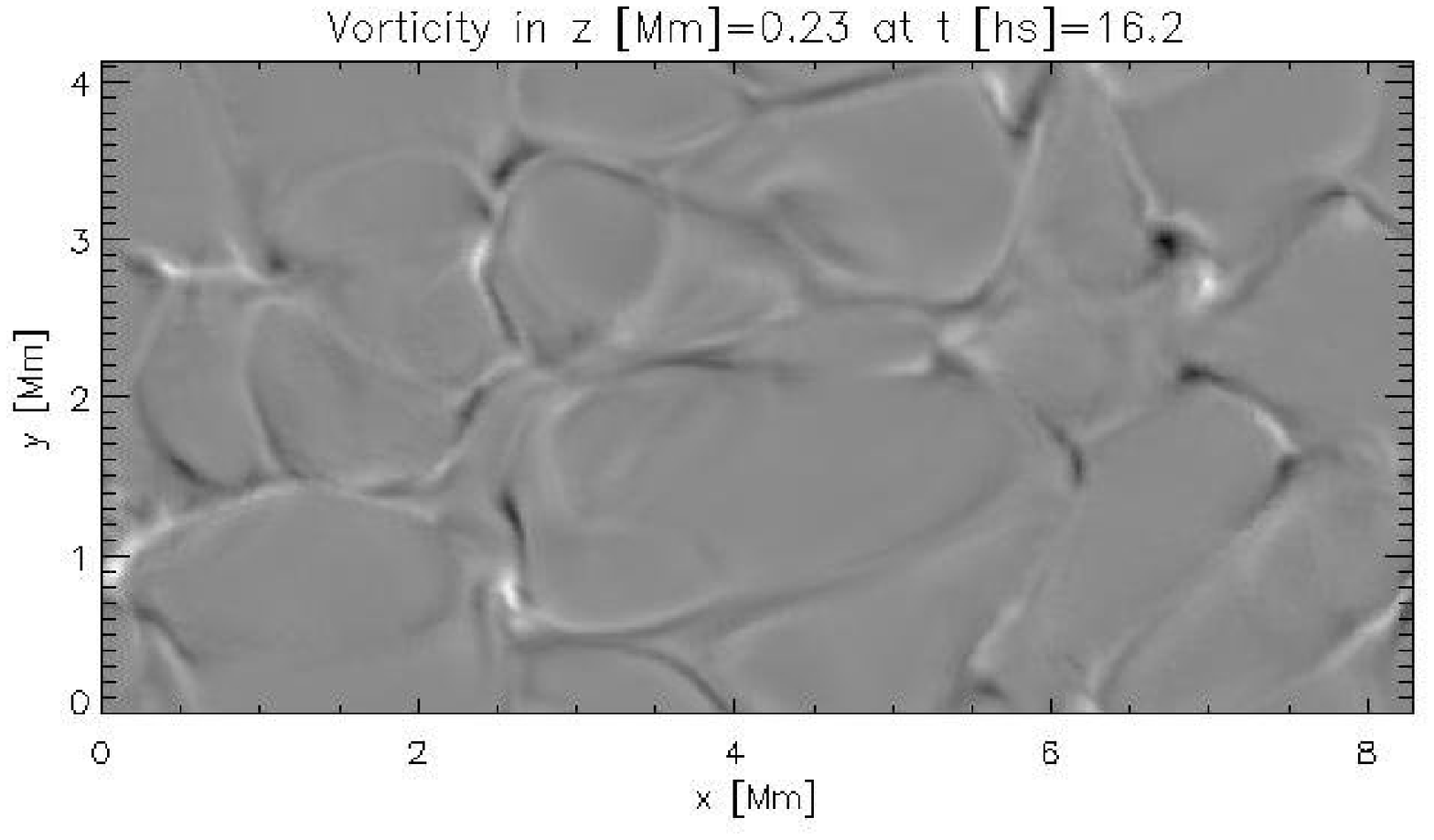}
\includegraphics[width=7.5cm]{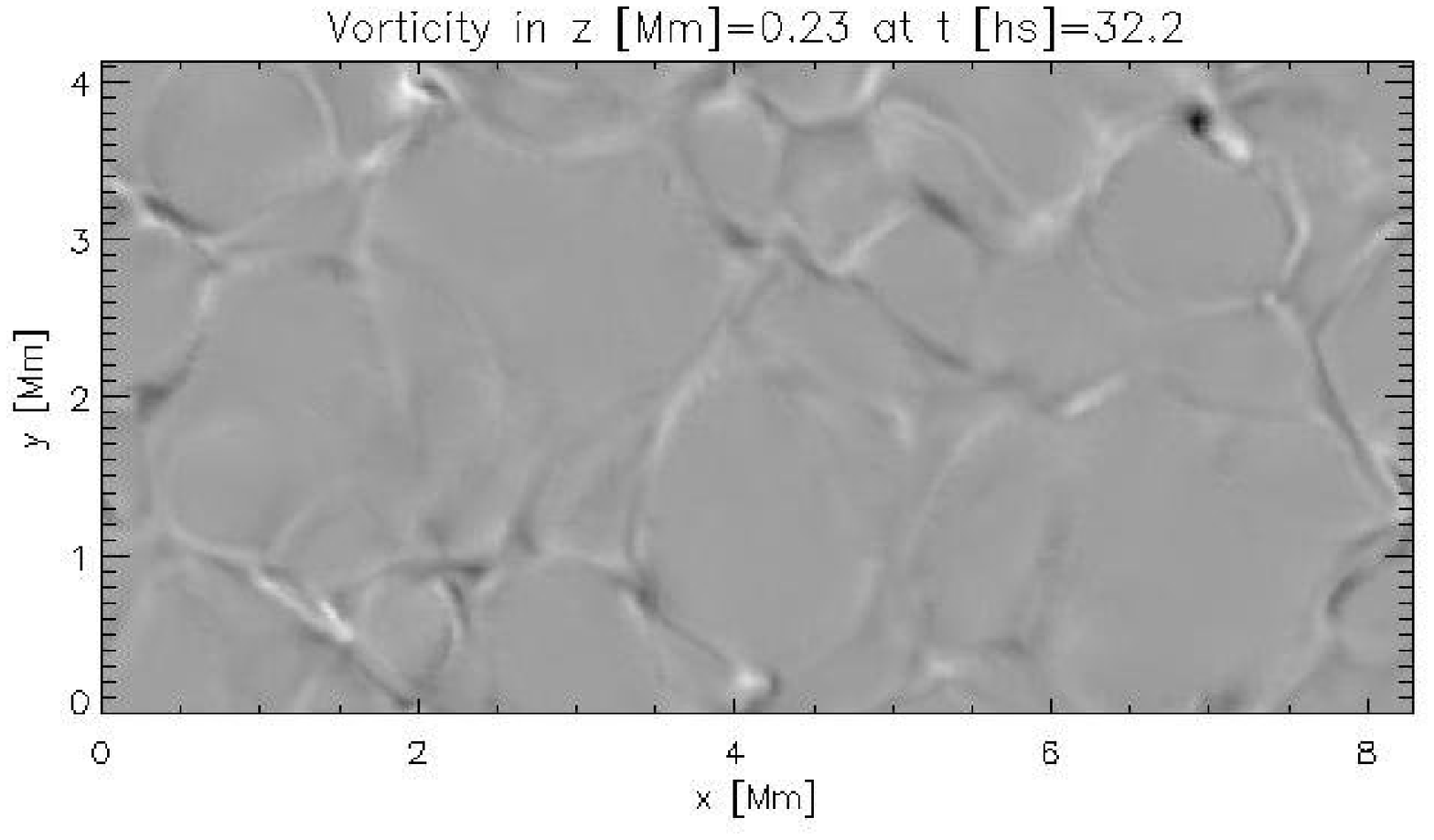}
  \end{center}
  \caption{\label{fig:revgran} Continuum intensity at 170~nm (top row),
  vertical velocity (second row), 
  divergence of the velocity (third row) and 
  the vertical vorticity (bottom row) at $z=234$~km 
  at time 1620~s (left column) and 3220~s (right column).}
\end{figure}

We define a horizontal layer at $z= 234$~km. At this height the granular 
pattern is still clearly visible, but it is reversed compared to the photosphere 
(figure~\ref{fig:revgran}). We see expansion of the cells beginning at time 920~s. 
The flux tube starts to go
through the reverse-granulation layer some 80~s after it crosses the photosphere,
{\it i.e.} at time 1380~s.
As in the photosphere, the reversed granules increase in size as the flux
tube approaches. The magnetic field lines of the tube are seen in figure~\ref{fig:3drev} 
to go from one side of the expanded cells to the other.

\begin{figure}
\begin{center}
	\includegraphics[width=7.5cm]{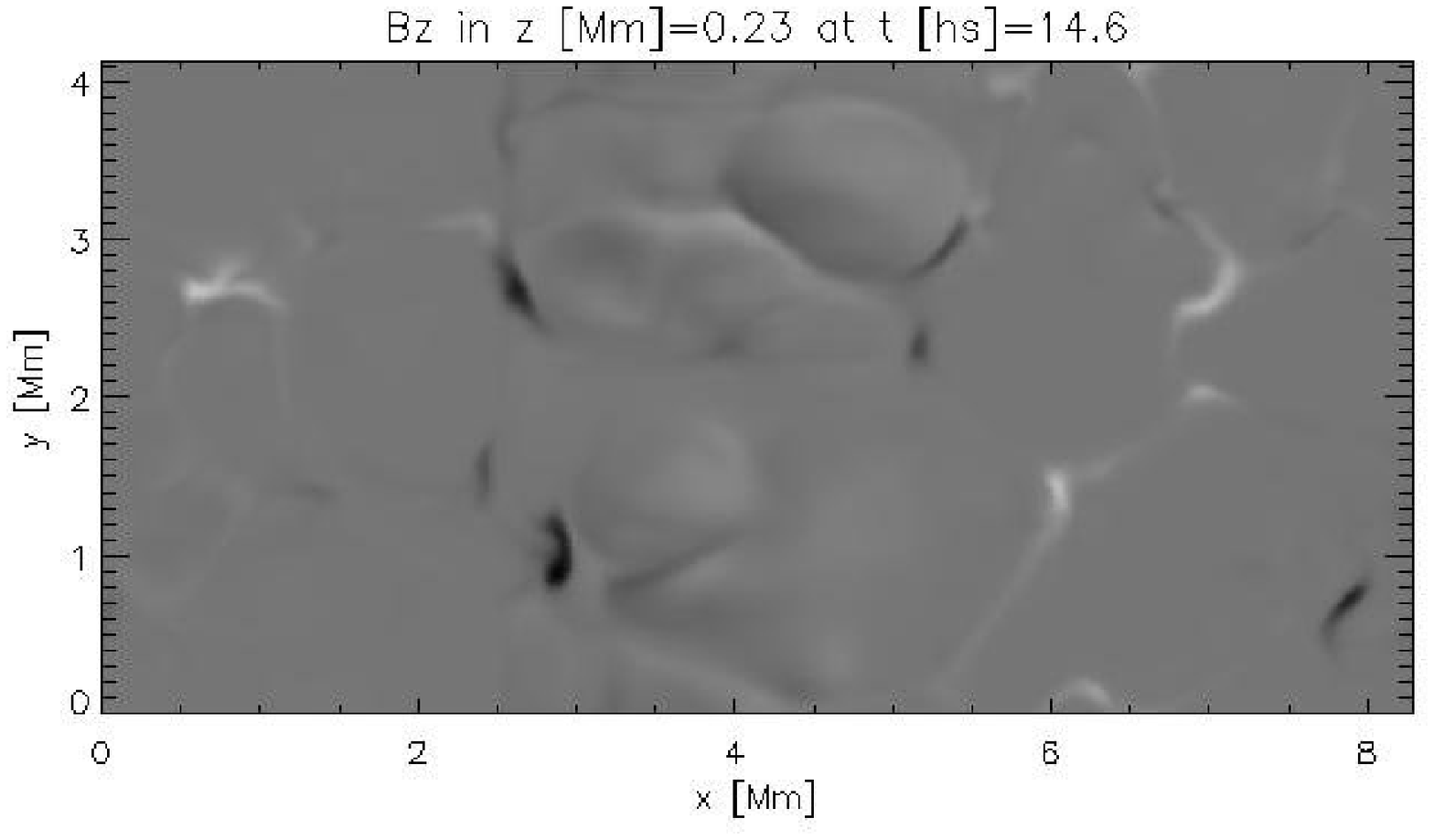}
	\includegraphics[width=7.5cm]{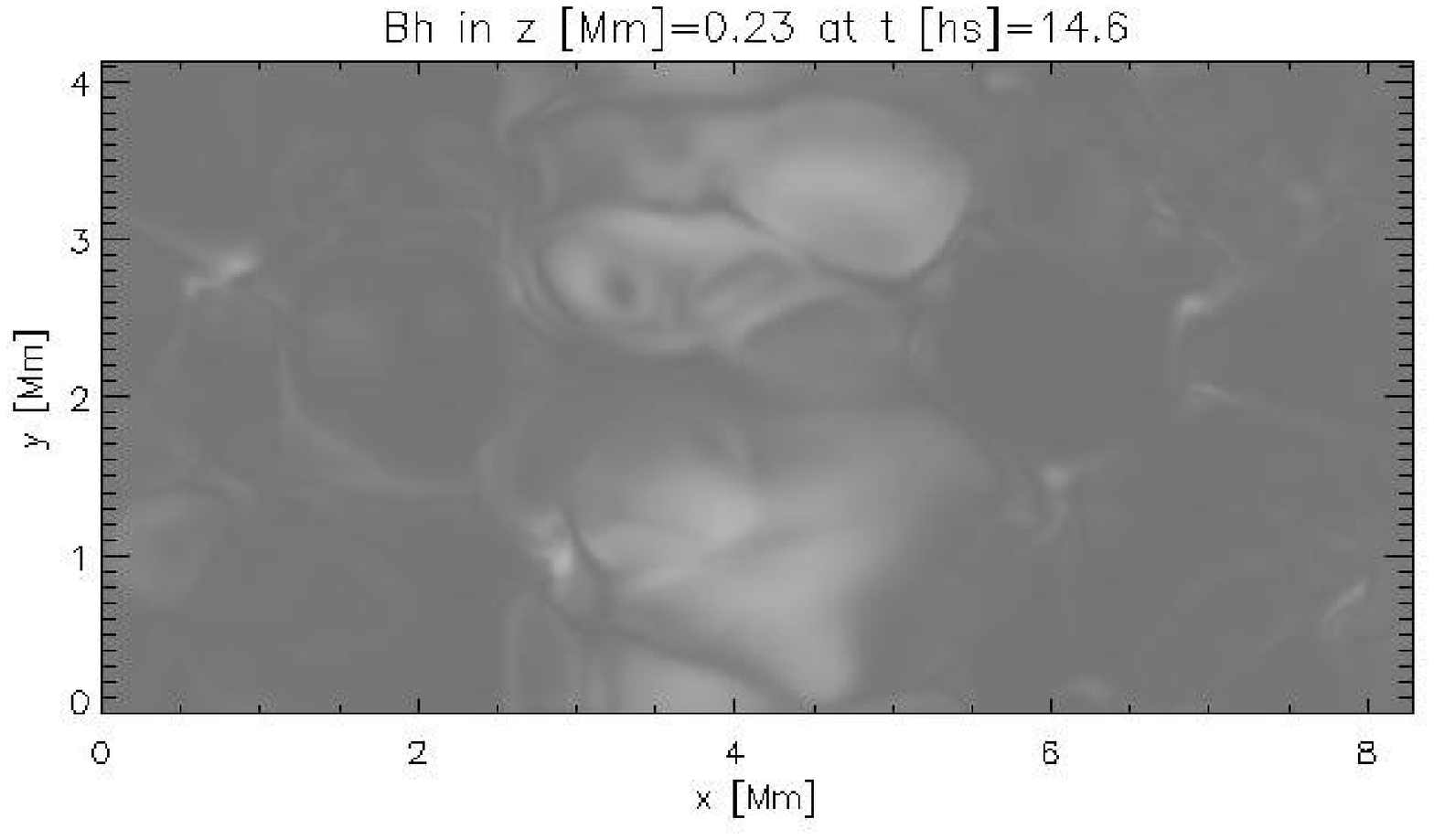}
	\includegraphics[width=7.5cm]{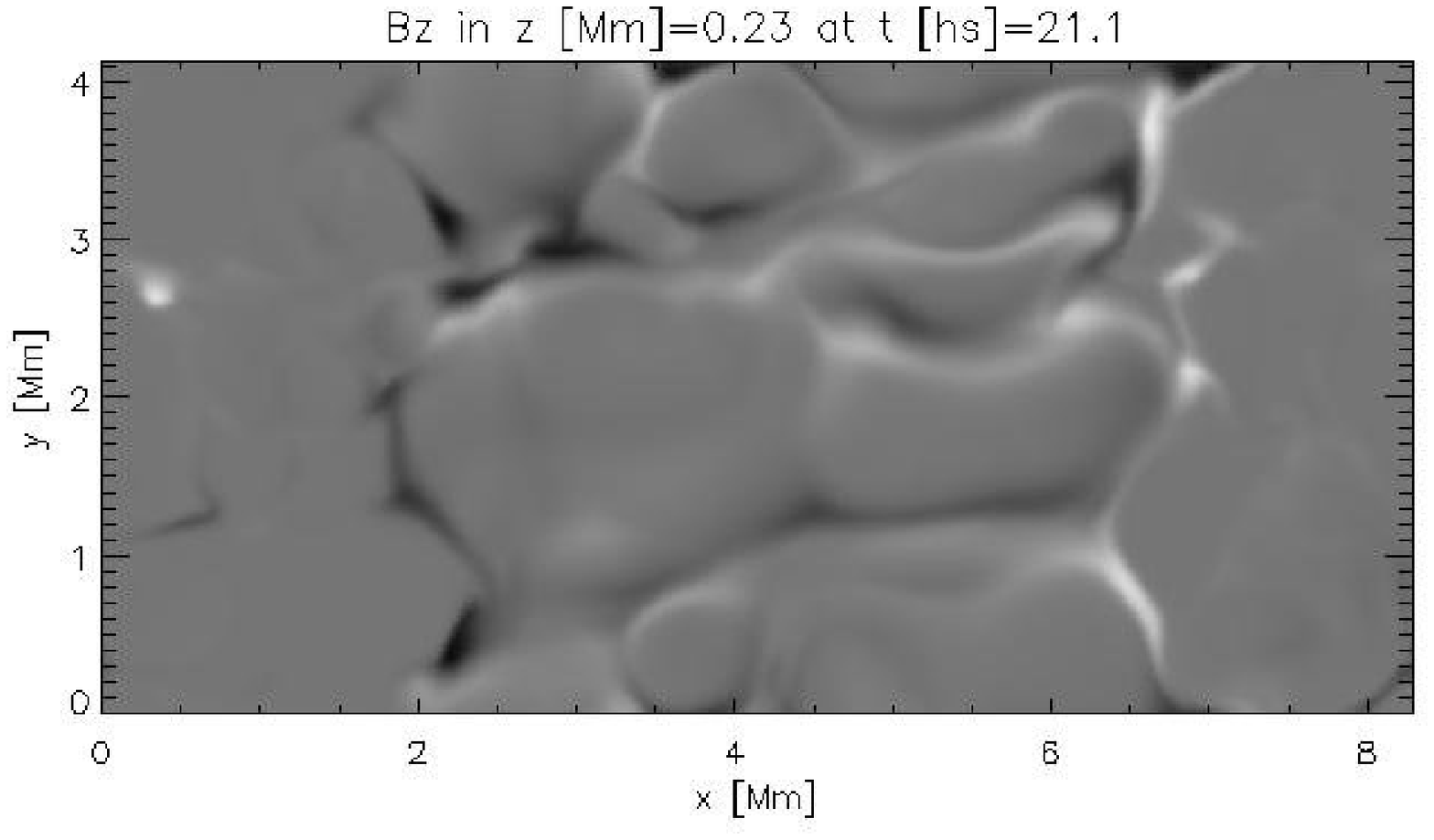}
	\includegraphics[width=7.5cm]{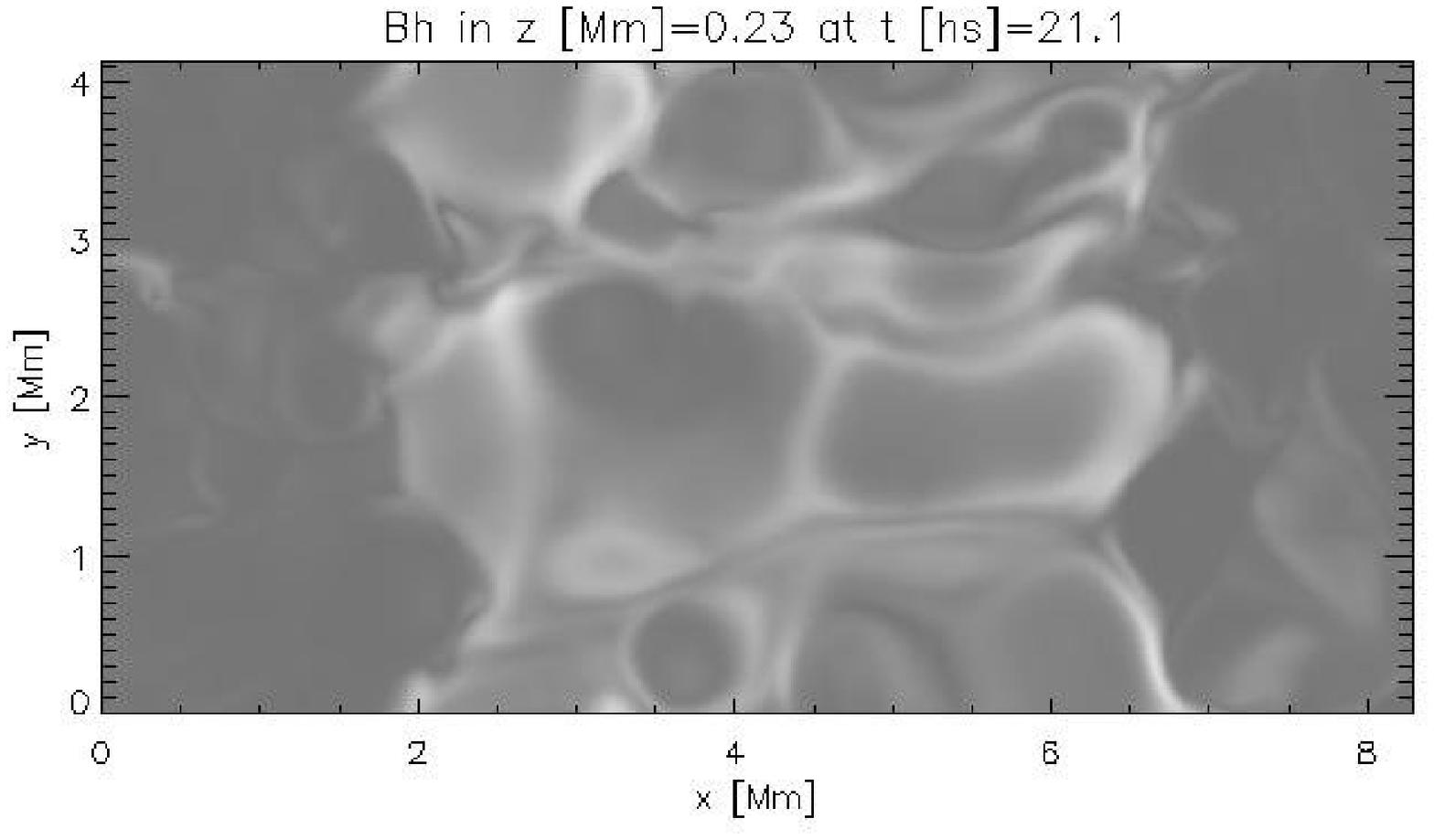}
	\includegraphics[width=7.5cm]{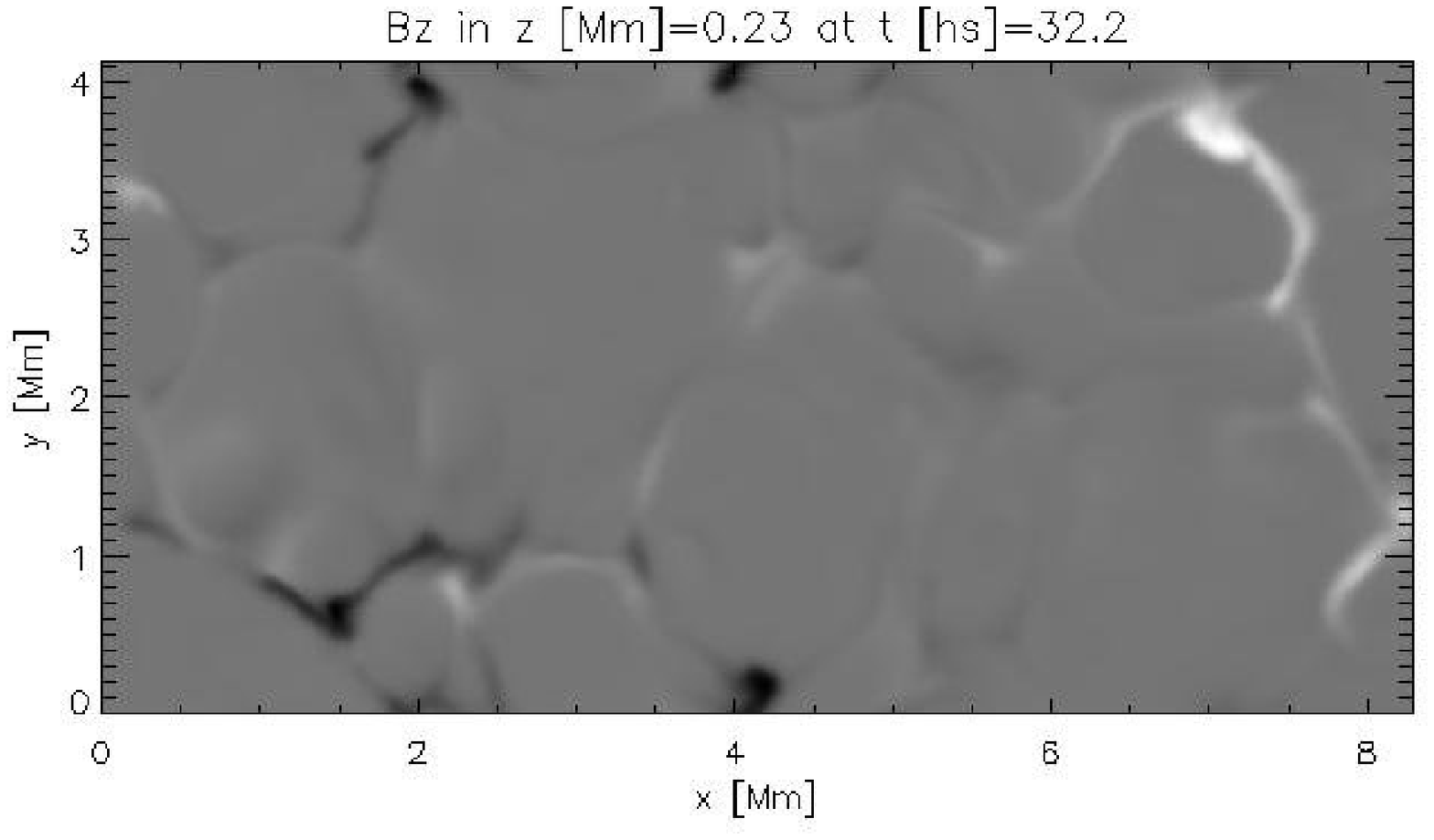}
	\includegraphics[width=7.5cm]{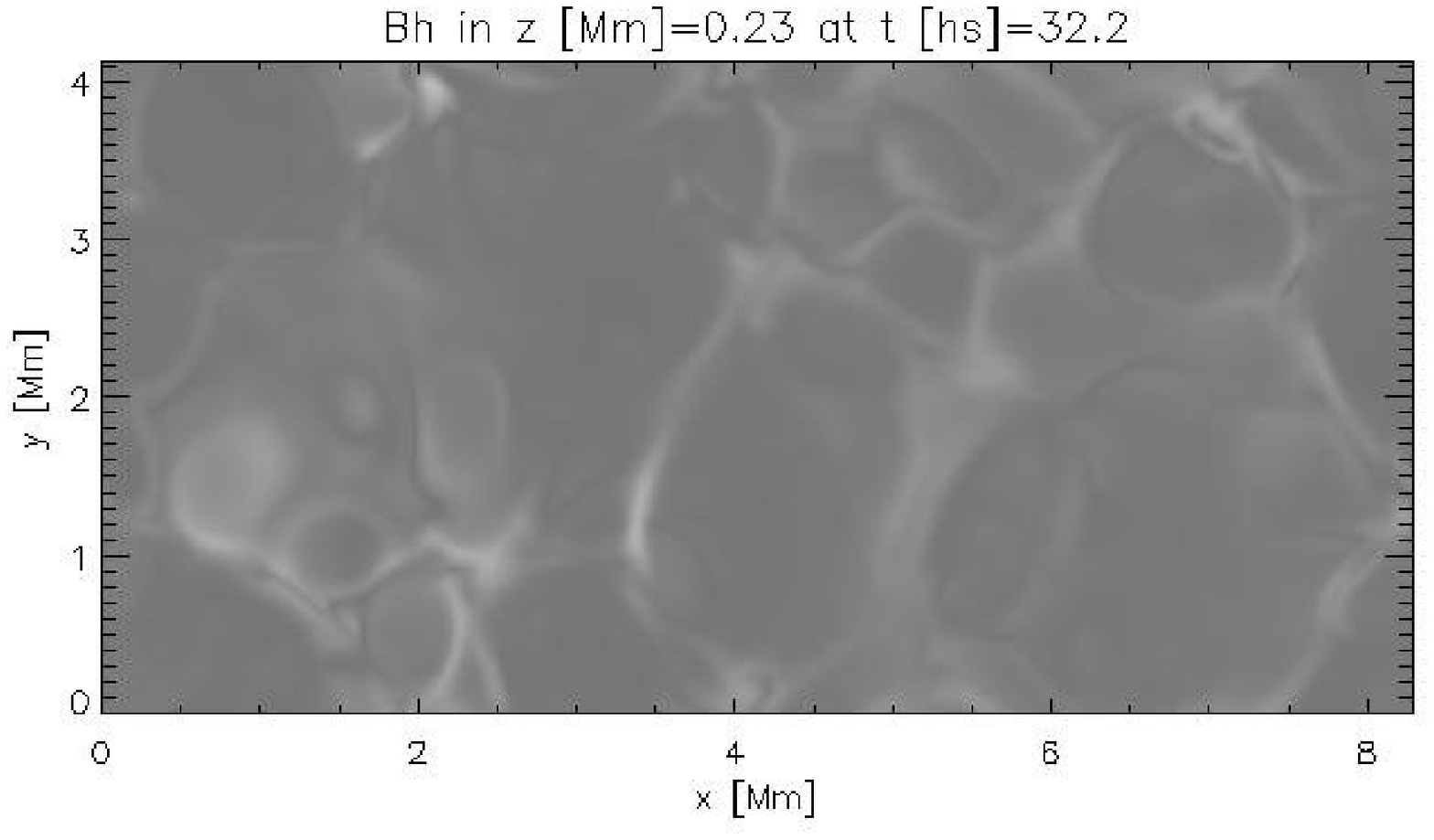}
 \end{center}
 \caption{\label{fig:fieldlwch} $B_z$ 
 at $z=234$~km (left panels) and $B_h$ (right panels).
 The grey-scale goes from -560.3~G to 659.5~G. 
 The field is shown at times 
 1460~s (top panels) when the tube start to cross the 
 reverse granulation layer, 
 at time 2110~s when the tube is crossing the reverse granulation (middle panels) and
  at time 3220~s when some rest of the flux returns to the reverse granulation (bottom panels)}
\end{figure}

Figure~\ref{fig:fieldlwch} shows the evolution of the vertical magnetic field $B_z$ in the
left column and the horizontal magnetic field in the right column, at three different
instants in time. The maximum magnetic field-strength, before the flux tube reaches the reverse
granulation, is of order 500~G. When the tube starts to cross the
height $234$~km the maximum rises to 600~G. 
At later times, but while parts of the tube are still to be found at this height,
the remaining magnetic field is confined to downflow regions and the maximum field strength 
rises to 650~G. After the tube has passed, some remaining confined flux rises to a  maximum
of up to 800~G. During the last 500~s of the simulation the maximum field strength oscillates
around 700~G. Again, as in the photosphere, the magnetic field is
initially horizontal and appears in the center of the granular cells 
(top panels). The magnetic field is then confined and moves with the fluid 
to the edges of the cells where it becomes vertical (see the second and third 
rows of the figure \ref{fig:fieldlwch}). 
At the cell edges 
a great number of bipoles form by $t=2100~s$.
However, at later times the bipoles are canceled
by magnetic diffusion; {\it i.e.} in the bottom panel on the left side of 
figure~\ref{fig:fieldlwch} we find generally that negative (downward) magnetic field 
is concentrated while we find positive (upward) flux on the right.

The top row of figure~\ref{fig:revgran} shows the continuum intensity at
170~nm as calculated from the simulation at times 1620~s (left) and
3220~s (right). The continuum opacity at this wavelength is dominated
by Fe bound-free opacity and the monochromatic optical depth unity is
around $250$~km height in the simulation. Note that real observations
around this wavelength (such as the TRACE 170~nm bandpass) are dominated
by line opacity, with the intensity thus coming from higher layers. The
intensity images shown here are, however, typical of a diagnostic formed
in the reverse granulation layer (like the inner wings of the resonance
lines from ionized calcium).
The figure also shows the vertical velocity in the second row with the same grey-scale
values, {\it i.e.} from 2.6~km s$^{-1}$ (white) to -4.1~km s$^{-1}$ (black), the divergence of 
the velocity in the third row and the vertical vorticity in the fourth row, all
these variables at height $234$~km. The synthetic 170~nm intensity shows reverse granulation
and it is clear that at $t=1620$~s the granular cells have expanded in regions co-spatial
with the flux tube. The same pattern is also found in plots of the temperature and the 
vertical velocity. Larger cells grow dimmer and colder due to the expansion as can be verified 
by considering the velocity divergence. While cool, dim regions are due expansion 
the hotter bright network points are seen to be related to contraction of the fluid at this height. 
At the end of simulation we find several bright points. This is seen clearest 
at time $t=3220$~s in the regions where we find concentrated vertical 
magnetic field (see  second column of the figure \ref{fig:revgran} at the position 
[$6.7,3.7$]~Mm). These bright points in intensity 
are related to large vertical velocities, but also to high temperatures compared 
to the surroundings, a large vertical vorticity and to 
compression of material (black color in the divergence of the velocity plots).  
The heating process for these bright points is clearly related to the high velocity 
convergence and high vorticity. 
We will find similar bright points at greater heights in the atmosphere, though 
with larger horizontal extent.

The temperature range is found to be [4.9,6.2]~kK before the tube reaches the $234$~km
layer,  while during and after the flux tube passes the temperature range is [4.6,6.1]~kK;
the lower limit decreases in time after 2400~s. The reason for this temperature drop is the
expansion as discussed above.

There are no
large changes in the upward maximum velocity with time before, during
or after the tube crosses the layer $234$~km height; it is found to be around 2.5~km s$^{-1}$. 
The maximum downward velocity does not change until after 
$t=1400$~s when it has a small increase for 200~s.
The maximum downward velocity then returns to normal before
increasing from 3~km s$^{-1}$ to 5~km s$^{-1}$ starting at $t=2200$~s.
These large downflow velocities are found to correspond to the small regions 
with a strongly confined magnetic field in the intergranular lanes. 
Examples of such a regions are found in the second row, right column, of figure \ref{fig:revgran} at position 
$[x,y]=[4.1,0.2]$~Mm and also at $[7.1,3.6]$~Mm. 
Towards the end of the simulation, at time 3100~s, the maximum downward velocity 
has returned to 3~km s$^{-1}$. 
The maximum horizontal velocity oscillates around 5.5~km s$^{-1}$.
However, it has a peak at 1480~s with a maximum velocity of 8~km s$^{-1}$ 
as the tube is crossing the layer $234$~km height. 

The expansion of the cells is observed at time $t=900$~s in the photosphere ($z=10$~km) and
roughly 25~s later in the reverse granulation layer $234$~km height. The separation between
these two layers is $224$~km. The propagation speed of the tube itself is much slower,
as table \ref{tab:events} shows, it spends 250~s in moving from the 
layer at $10$~km to the layer at $234$~km height. Another possibility for explaining 
the expansion of the cells at the layer $234$~km height is by the propagation of 
Alfv{\'e}n waves, but again, the time such waves need to propagate from 
the layer $10$~km to the layer $234$~km height is too long, because 
the mean Alfv{\'e}n speed is very small in that region ($<100$~m s$^{-1}$). 
On the other hand, the average sound speed between $10$~km to $234$~km height is 8.5~km s$^{-1}$, which 
gives a travel time close to the 25~s that fits with the time to move outward the 
horizontal expansion of the cells.
In other words we see the expansion of the granular cells some 400~s before the magnetic tube 
actually emerges through the photosphere, when the tube is situated 182~km below the photosphere. 
The speed of the sound below the photosphere is on average 10~km s$^{-1}$ which means the expansion
wave is produced roughly 420~s before the tube reaches the photosphere, {\it i.e} when the 
tube is roughly at -246~km. The average tube velocity from -246~km to 10~km is 0.5~km s$^{-1}$.
However, the tube decelerates when it approaches the photosphere as explained 
in the section \ref{sec:photosphere}.
The cells average horizontal expansion velocity is roughly 2.5~km s$^{-1}$ before the tube reaches
the photosphere.

The figure~\ref{fig:fluxretlow} shows the average magnetic flux in the same manner
as in figure~\ref{fig:fluxret}, but here the regions considered are from the 
layer $z=10$~km to the layer $z=234$~km (black color) and from the layer $z=234$~km 
to the layer $z=458$~km (red color). 
Each of the four displayed simulations conserve magnetic flux in the sense that 
when the magnetic flux tube crosses these regions no flux is lost or retained in the 
region, as opposed to what happens in the convection zone. The biggest 
difference between the two regions is seen to be in the slope of the flux
($d\langle\phi\rangle/dt$) as the tube moves through the region. The slope of  
the flux is much smaller than that found in figure~\ref{fig:fluxret} when 
the tube crosses those layers, {\it i.e.} more time is spent in crossing 
each region. The oscillation of the mean magnetic flux occurs more or 
less at the same time, when the tube crosses the corresponding region,
but the amplitude of the variation is greater in higher regions.

\subsection{End of the photosphere-beginning of the Chromosphere}
\label{upcrom}

The top of the convection zone, photosphere, and lower-chromosphere 
are the regions where acoustic waves are excited by convective motions. 
While the waves propagate upwards, they steepen into shocks, dissipate, 
and deposit their mechanical energy as heat in the chromosphere. 
In studying the quiet non-magnetic chromosphere
\citet{Wedemeyer:2004} corroborates in 3D models the finding by 
\citet{Carlsson+Stein1994} that the chromospheric temperature rise
derived from semi-empirical modeling of the time average intensity does not 
necessarily imply an increase in the average gas temperature but rather can 
be explained by the presence of substantial spatial and temporal temperature 
inhomogeneities, such as those caused by non-linear waves. 
We are interested in finding how the rise of a flux tube
affects chromospheric dynamics and energetics, {\it i.e.} if there are other 
energy sinks or sources that become active with the appearance of new magnetic flux,
or if the new flux modifies the propagation of acoustic waves into
the middle and high chromosphere.

In our simulations the magnetic flux tube enters the height $z=458$~km 
some 300~s after it has crossed the photosphere. The layer at $458$~km height suffers 
various perturbations in response to the appearance of the flux tube. 
Of the most important effects is the expansion 
and the cooling related to that expansion. Each of these processes are 
explained below.

\begin{figure}
\begin{center}
\includegraphics[width=7.5cm]{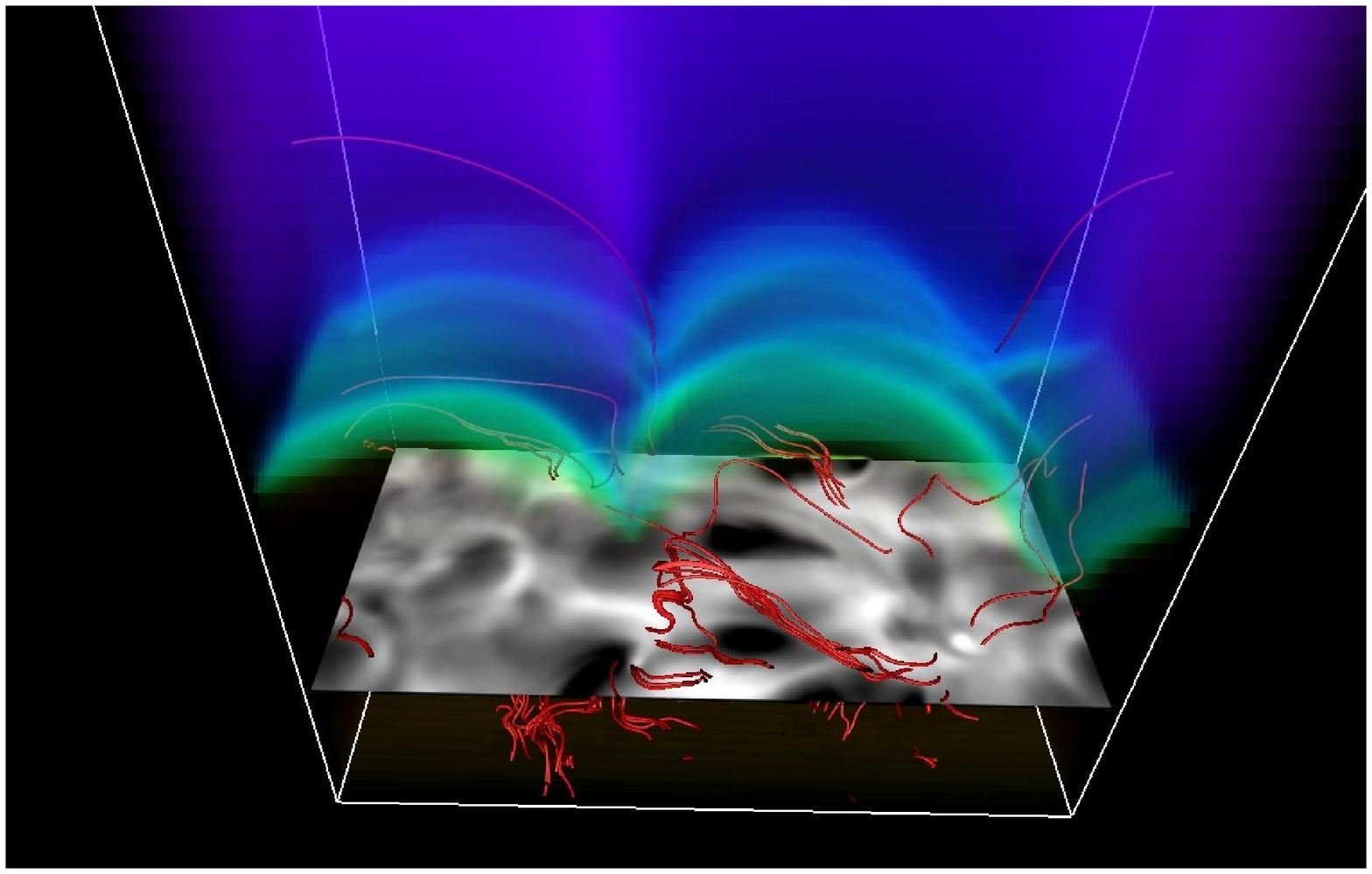}
\includegraphics[width=7.5cm]{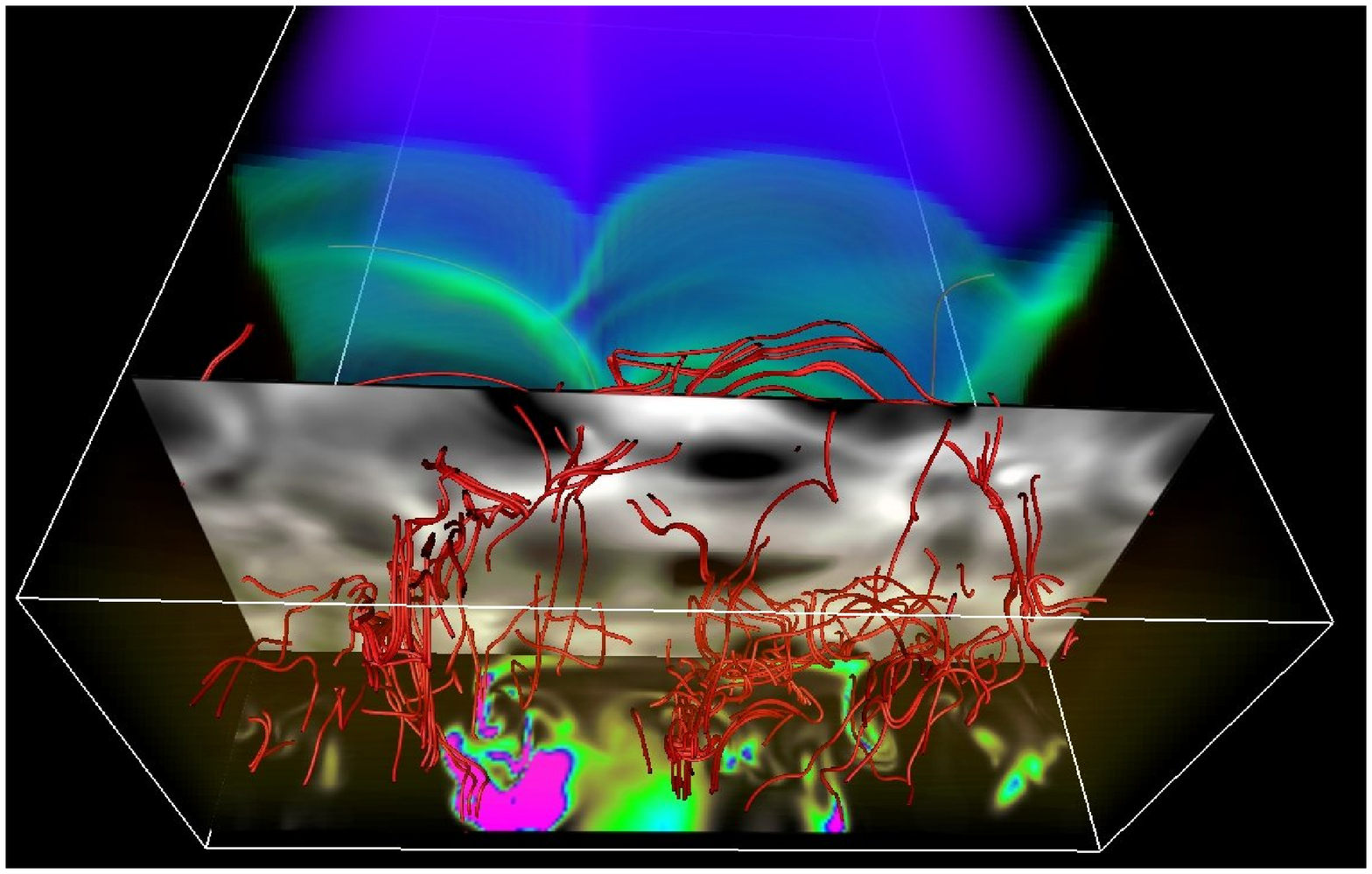}
  \end{center}
  \caption{\label{fig:3dcro} 3D image of the simulation at time $t=1980$~s.
  Temperature at $z=458$~km shown with
  grey-scale intensity. The rest of the color scale is the
  same as in figure~\ref{fig:3dpho}. 
  View from above (left panel) and from below (right panel).}
\end{figure}

Figure~\ref{fig:3dcro} shows two 3D images of the A4 simulation at
$t=1980$~s 
from above and from below a temperature slice made at height $z=458$~km
above the photosphere. The temperature structure at that height ($z=458$~km)
 has been changed by the presence of the tube and is much cooler in regions
where the tube is located. However, on the borders of the rising tube we find regions with 
higher temperatures. As in the previous 3D images, featuring lower heights,
we see how the structure of the magnetic flux tube has been perturbed as a result of 
rising through the atmosphere; comparison with these figures shows the structure of 
the tube is at this late time even more splintered and fragmented than at earlier times 
(see figures \ref{fig:3dpho} and \ref{fig:3drev}).

\begin{figure}
\begin{center}
	\includegraphics[width=7.5cm]{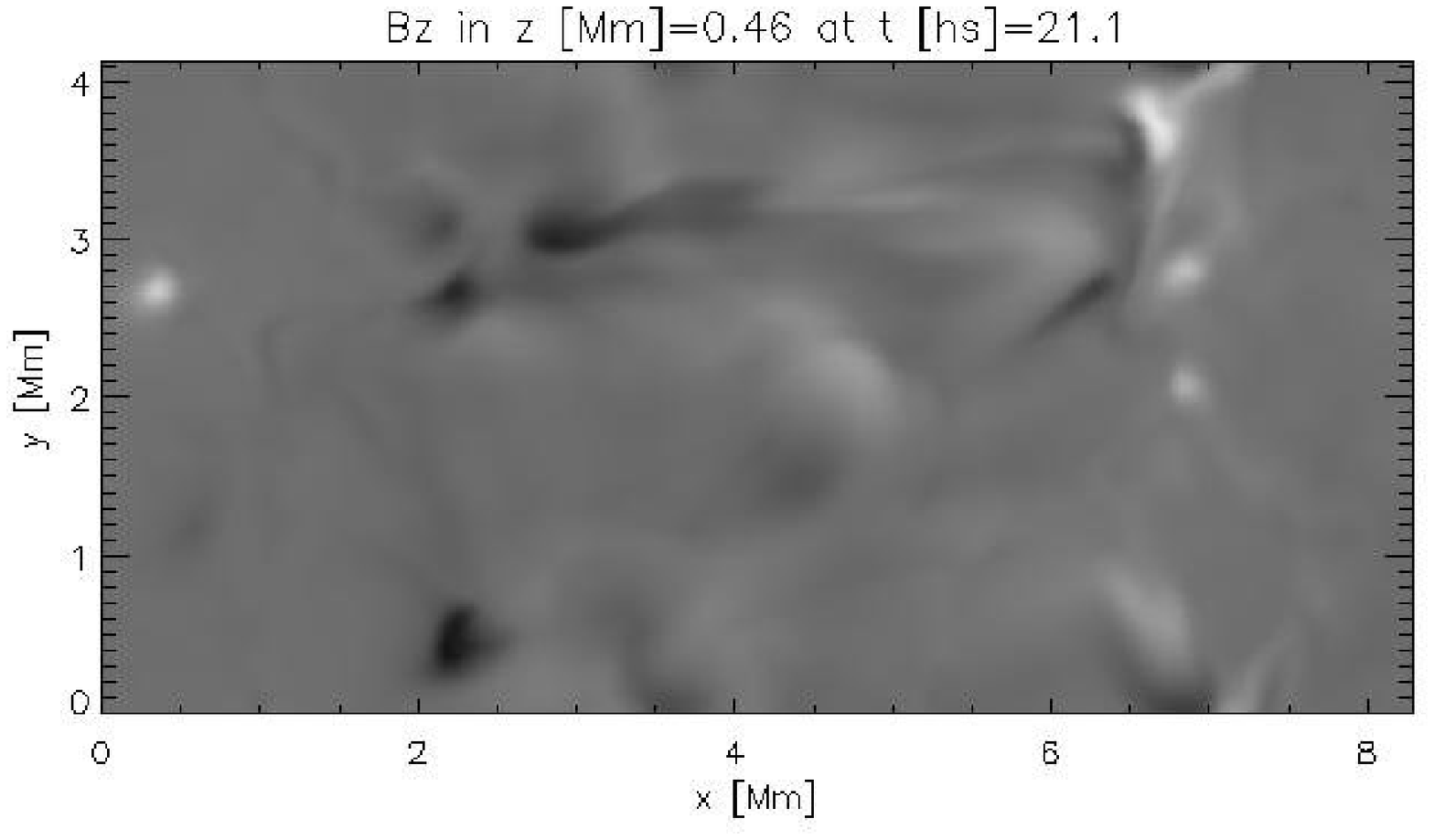}
	\includegraphics[width=7.5cm]{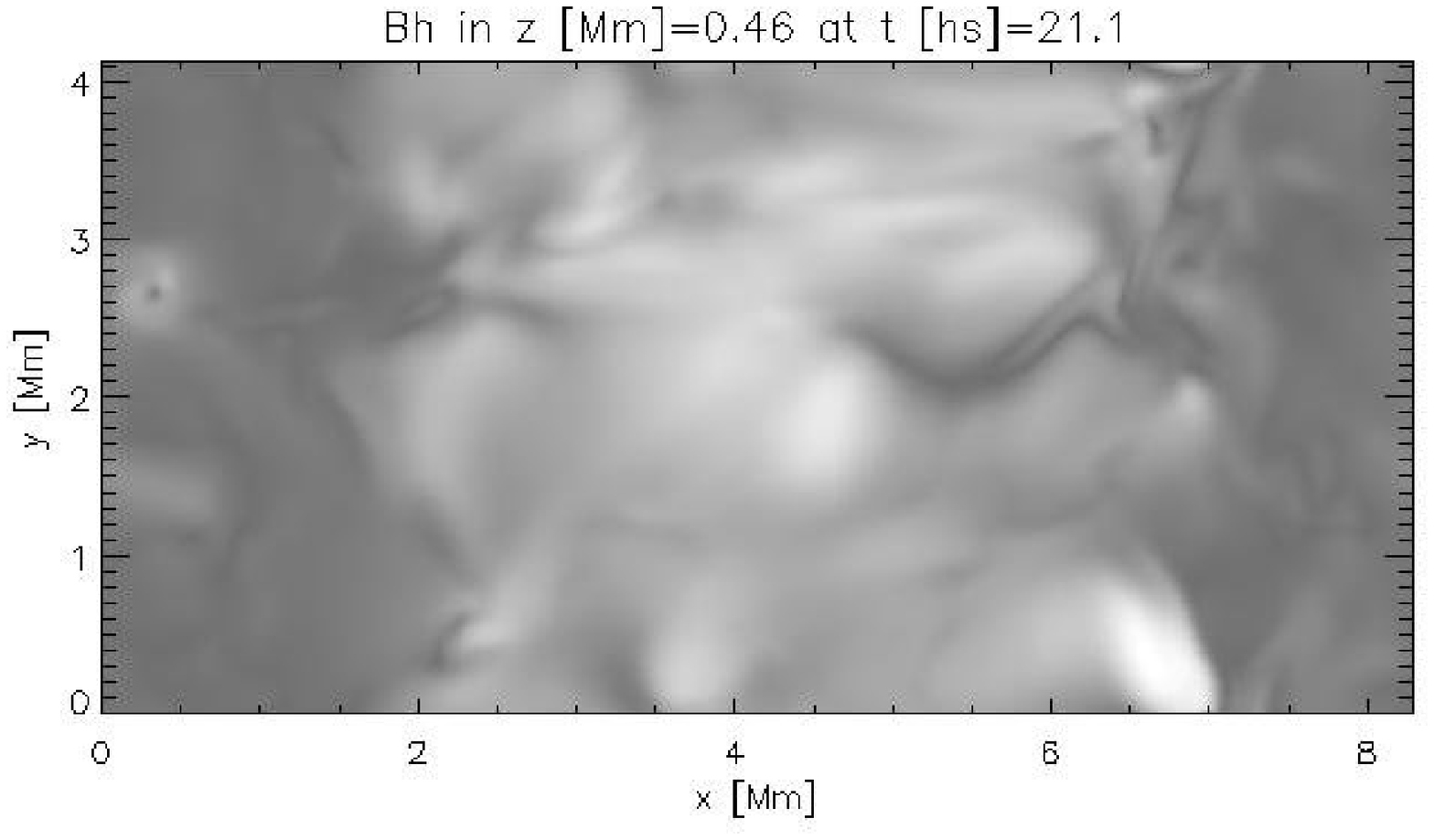}
	\includegraphics[width=7.5cm]{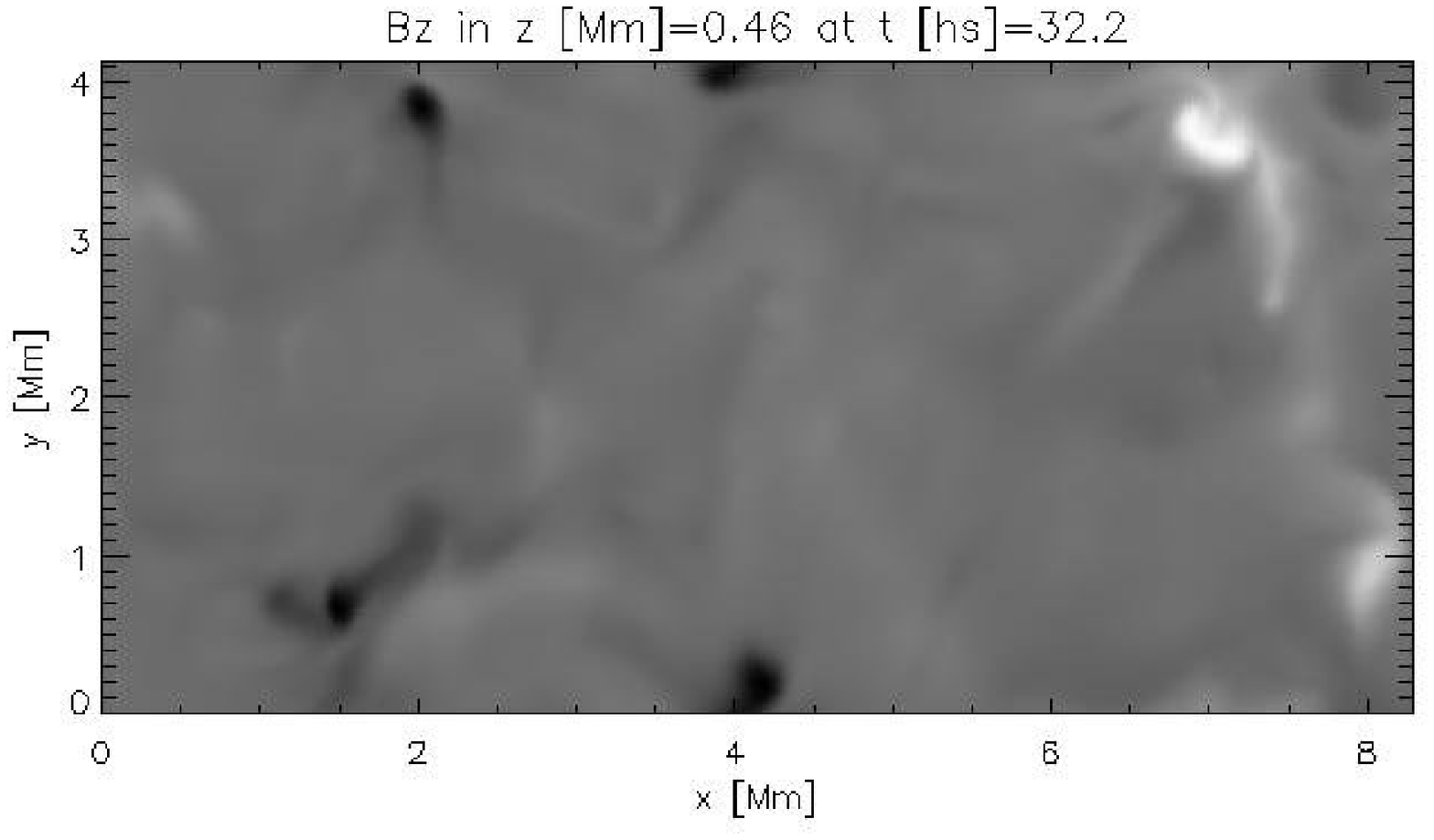}
	\includegraphics[width=7.5cm]{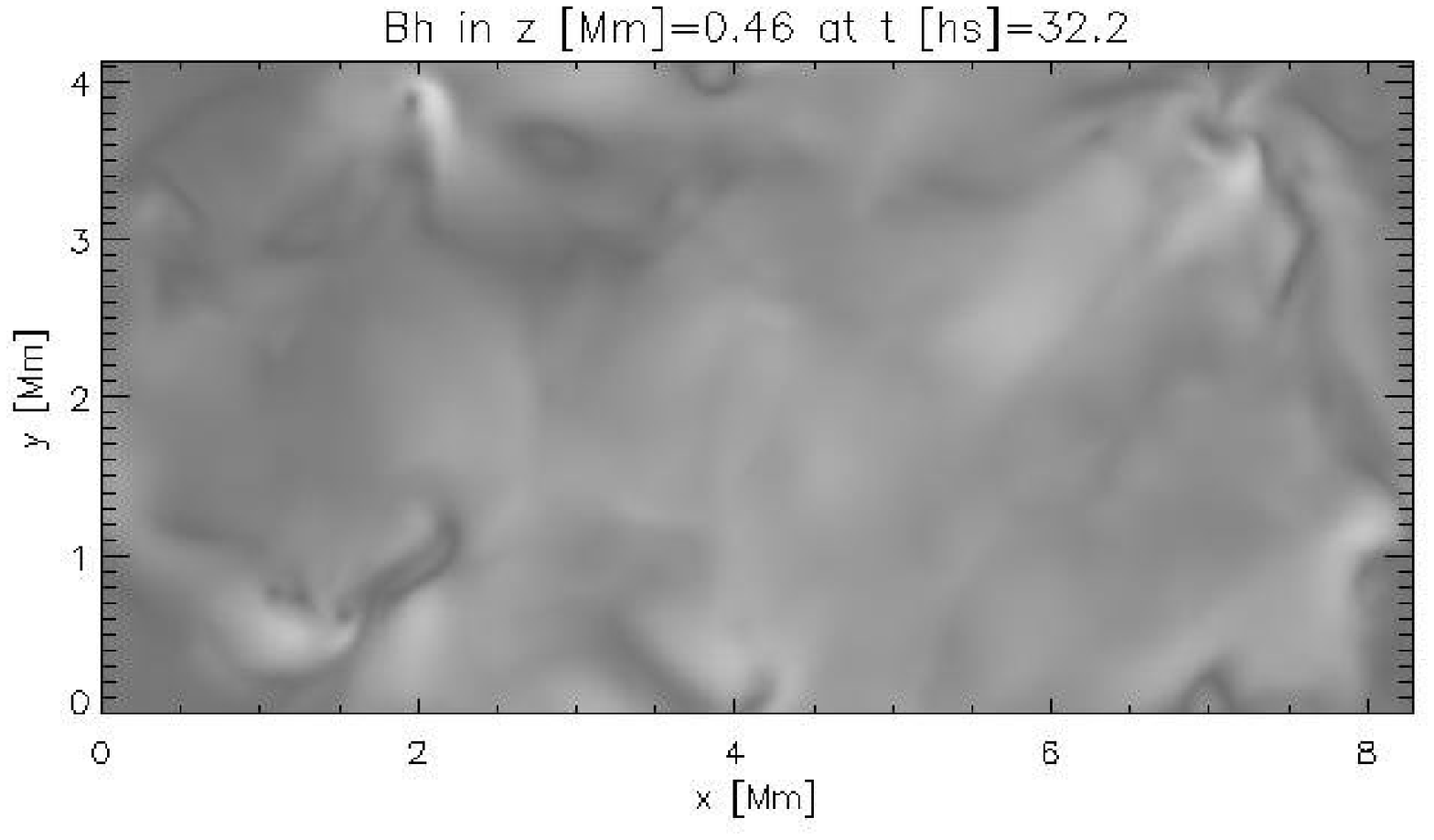}
 \end{center}
 \caption{\label{fig:fieldcr} $B_z$ at $z=458$~km (left panels)
 and $B_h$ (right panels) at times 
 2110~s when the tube is crossing the layer $458$~km height
 (top panels)
 and at time 3220~s (bottom panels). 
 The grey-scale range is [-217,284] G for all panels.}
\end{figure}

The vertical magnetic field is shown in the left column of figure~\ref{fig:fieldcr} 
and the horizontal magnetic field in the right column,
both for 2 different instants at height $z=458$~km. The maximum magnetic
field strength before the magnetic flux tube reaches this height is 170~G. 
During the flux emergence process the maximum field strength increases to 270~G, and
at later times, as the magnetic flux is compressed and confined to smaller regions, 
the field strength in some points increases to 350~G. 
At this height there is a cellular structure at any given time, but
the cells do not directly correspond to the photospheric granular 
cells, but rather to patterns in the velocity field that change from instant to 
instant. The vertical magnetic field is mostly confined to downflow regions and dominates
the field. However, as the flux tube crosses the layer at $458$~km height the horizontal 
magnetic field is strongest and fills the upflow cells. 
There are some regions or ``points'' with strong vertical magnetic field. For instance
the bottom left panel of the figure~\ref{fig:fieldcr} at position $[7,4]$~Mm.  
These
points have increasing maximum magnetic field-strength with time
after the flux tube has passed through the layer and 
they move around with the fluid.

\begin{figure}
\begin{center}
\includegraphics[width=7.5cm]{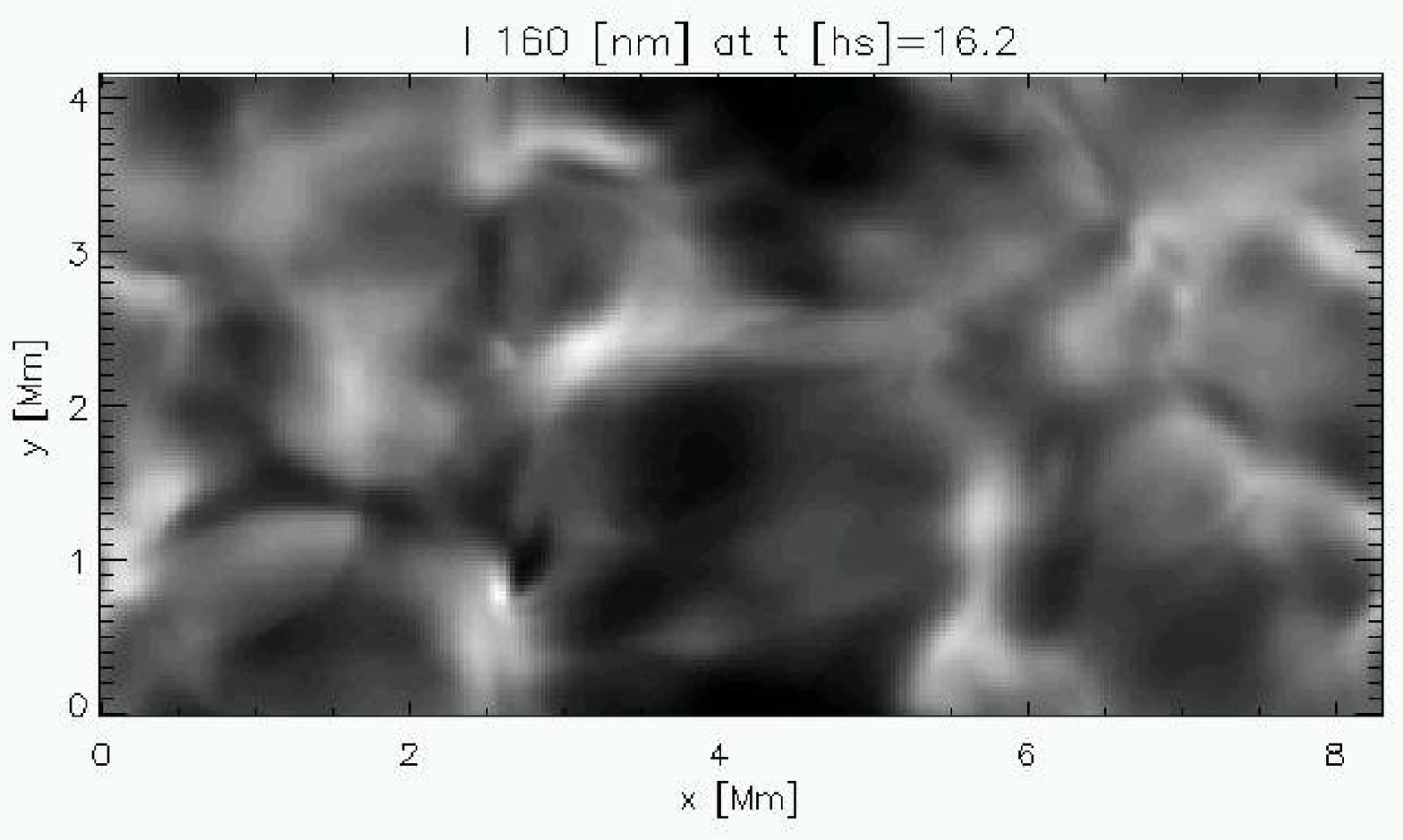}
\includegraphics[width=7.5cm]{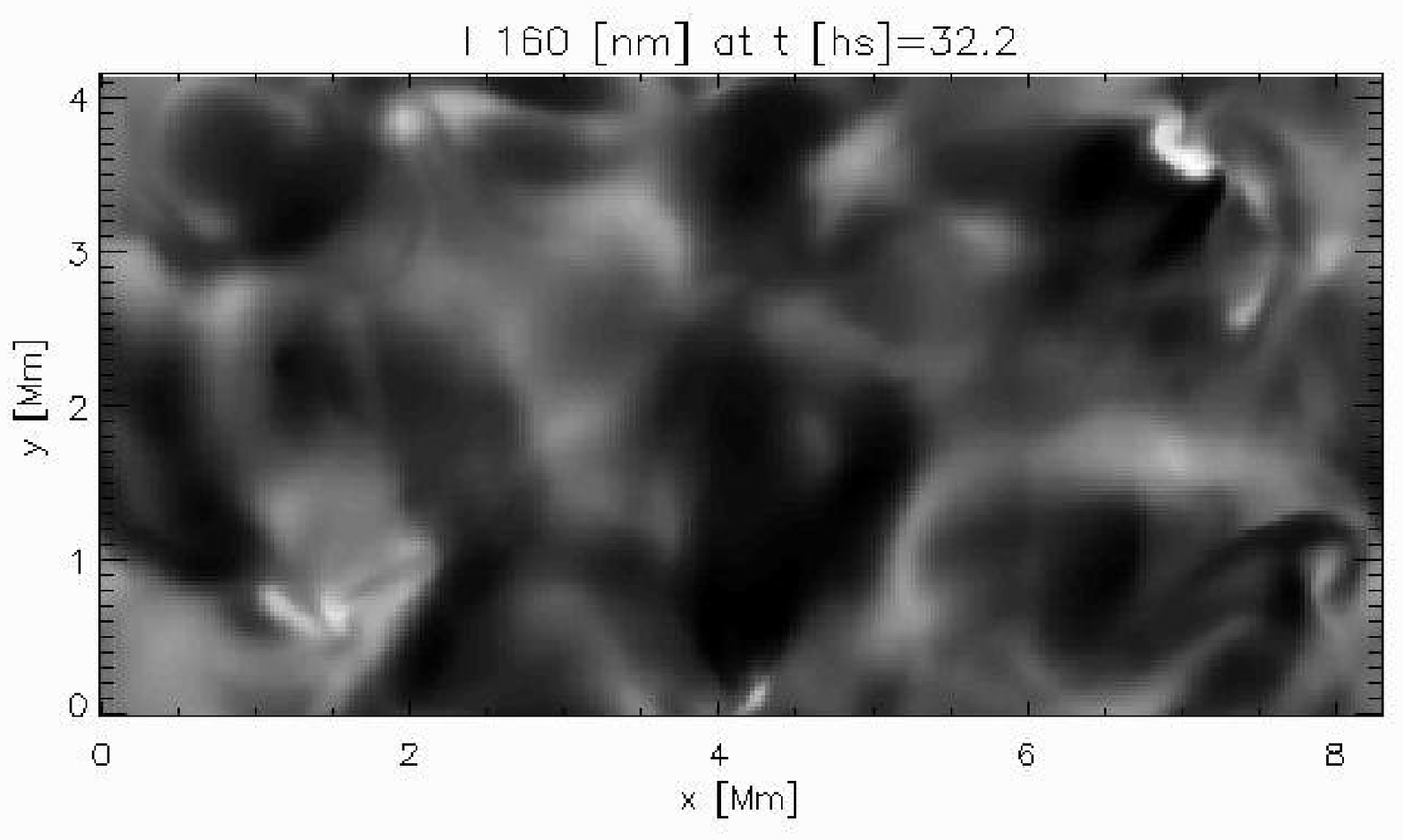}
\includegraphics[width=7.5cm]{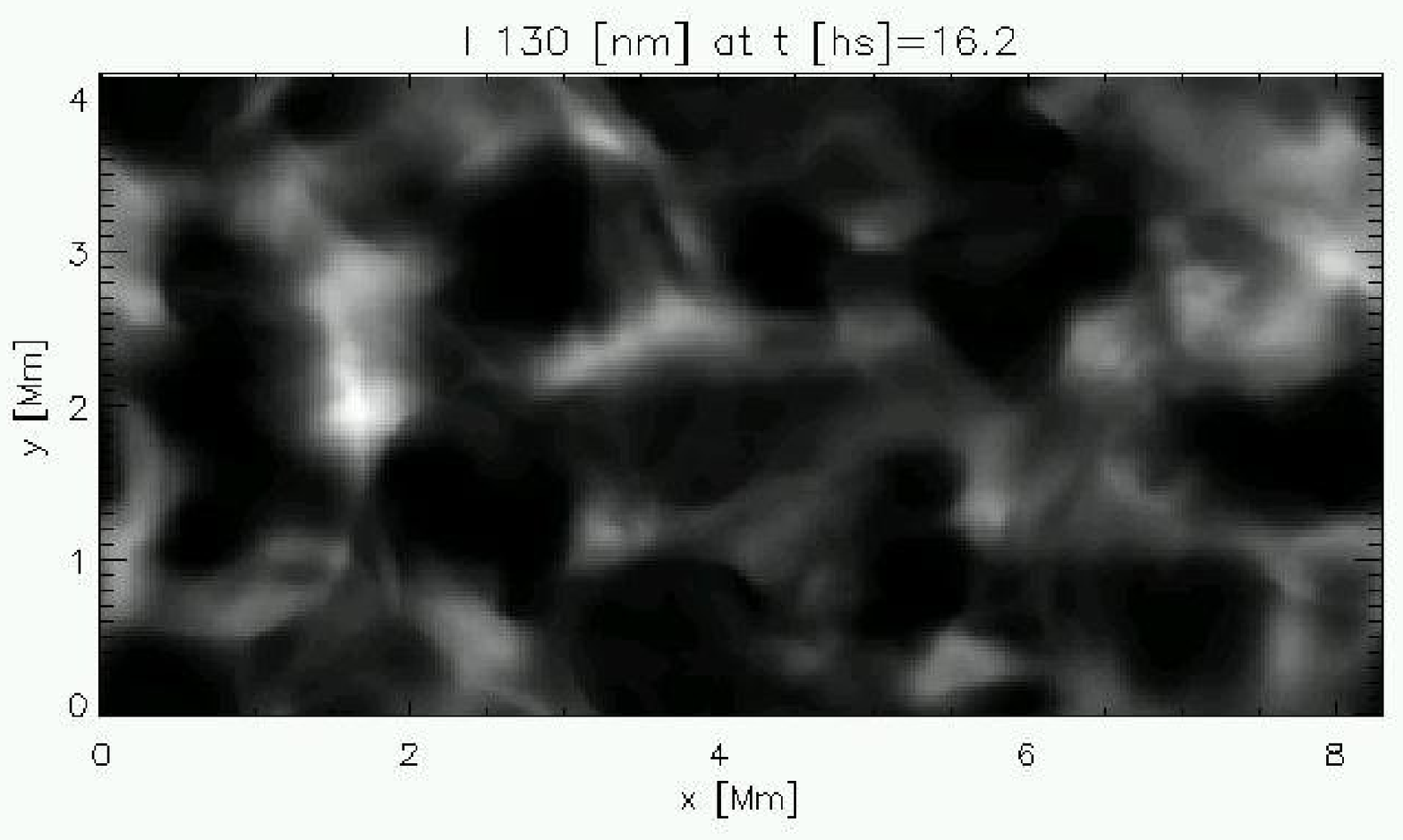}
\includegraphics[width=7.5cm]{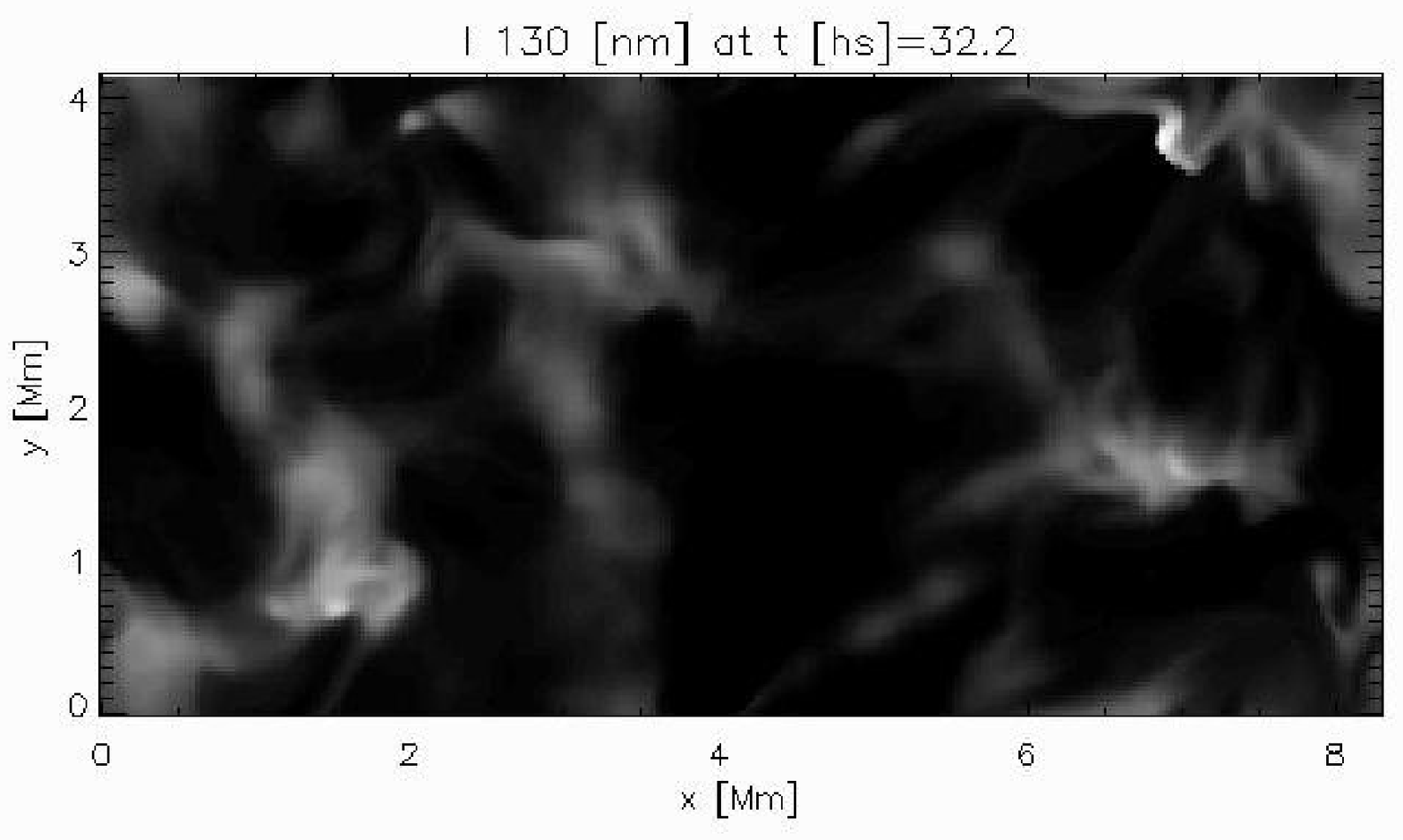}
  \end{center}
  \caption{\label{fig:tgcr} Continuum intensity to the power
  of 0.4 (equivalent to a gamma of 0.4 for an image) to better bring out
  the structure,
  at 160~nm (top row) and 130~nm (bottom row)
  at time 1620~s when the tube starts to
  enter the chromosphere (left column)
  and at 3220~s when the tube has crossed the height $z=458$~km
  (right column)}
\end{figure}

Figure~\ref{fig:tgcr} shows the calculated continuum intensity as described 
in the figure at
160~nm and 130~nm for two snapshots of the simulation. These 
images show what could
be observed in the TRACE 160~nm filter and in a SUMER raster 
at 130~nm. At both wavelengths the continuum opacity is dominated
by bound-free opacity from silicon: at 160~nm from the first
excited level and at 130~nm from the ground state.
The monochromatic optical depth unity is
on average around $480$~km and $790$~km height in the simulation at 160~nm and 130~nm respectively 
(in the VALC model atmosphere the corresponding heights are $456$~km and $735$~km, respectively).
The intensity at $t=1620$~s shows expansion
of the plasma with subsequent dimming, somewhat more clearly at 
160~nm. At 3220~s both the continua show a bright point
at position $[7,3.8]$~Mm. It is also possible that the tube expansion
could be affected by the finite horizontal size of the computational domain;
figure~\ref{fig:fieldcr} shows that the horizontal magnetic field almost reaches 
the boundary layers of the box. On the other hand the (limited number of) models we have run
with a larger box, $16\times 8\times 16$~Mm$^3$ do not show a fundamentally different 
behavior, so we believe the effects of limited simulation size are minor.

\begin{figure}
\begin{center}
\includegraphics[width=7.5cm]{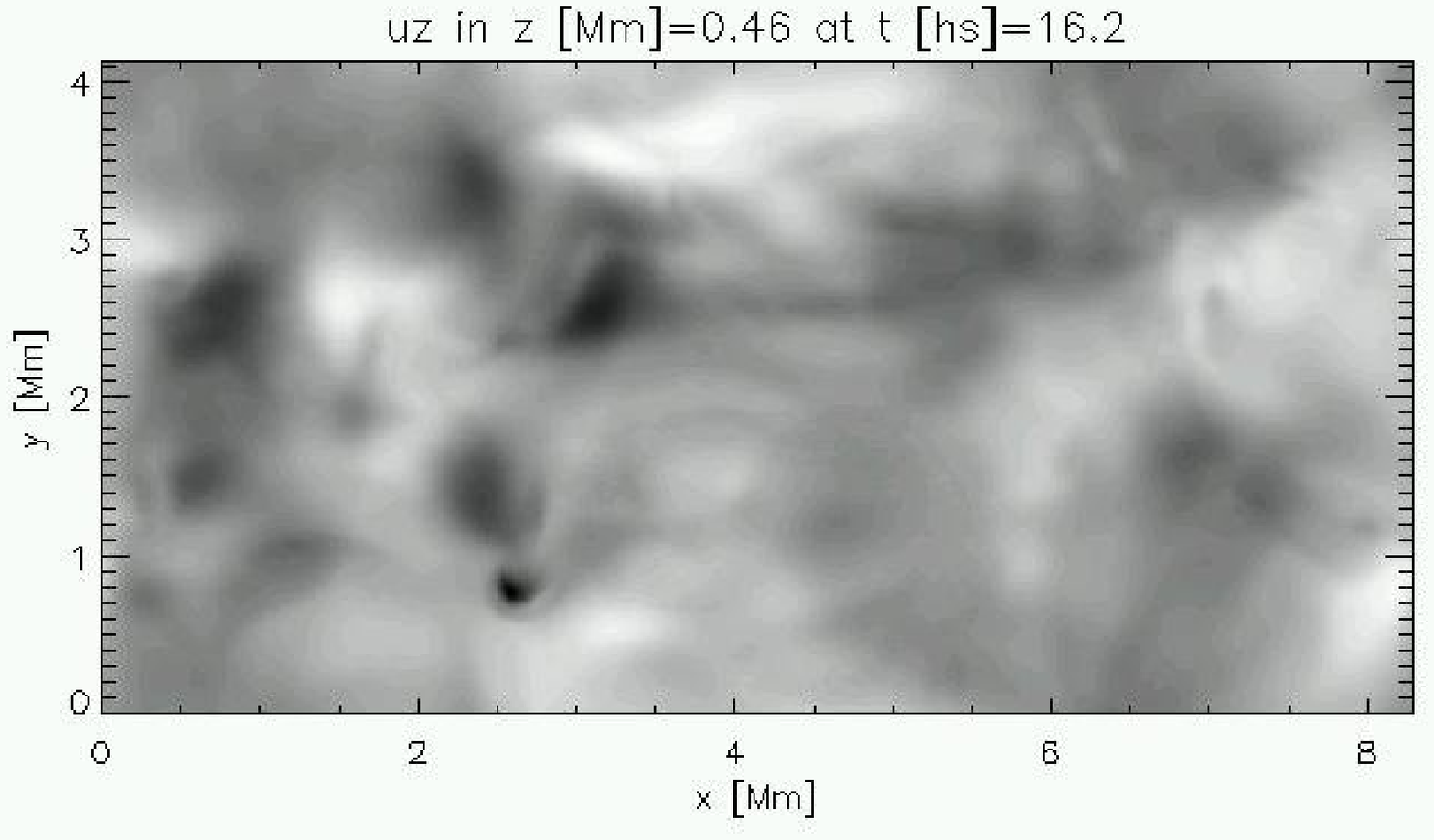}
\includegraphics[width=7.5cm]{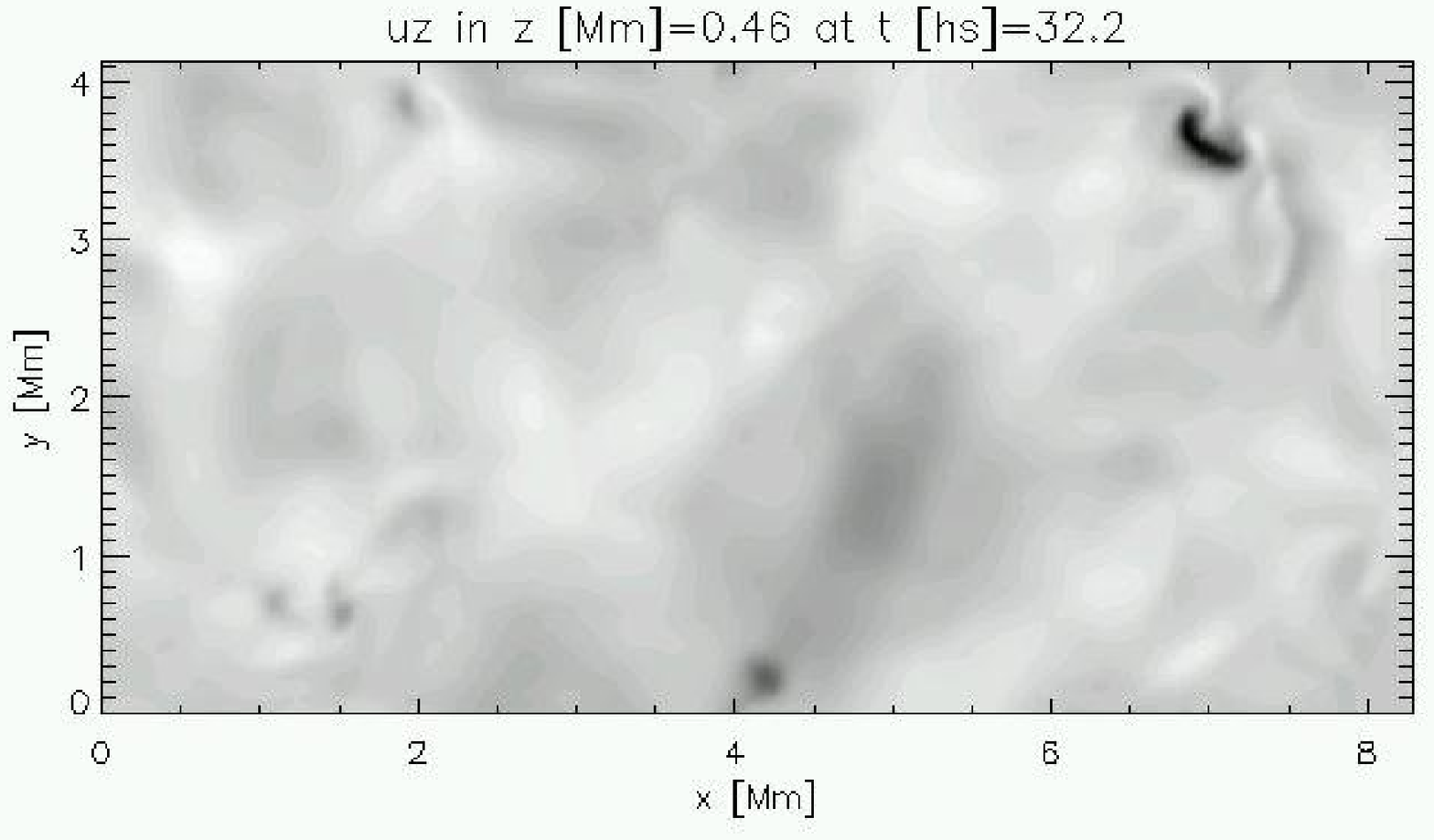}
\includegraphics[width=7.5cm]{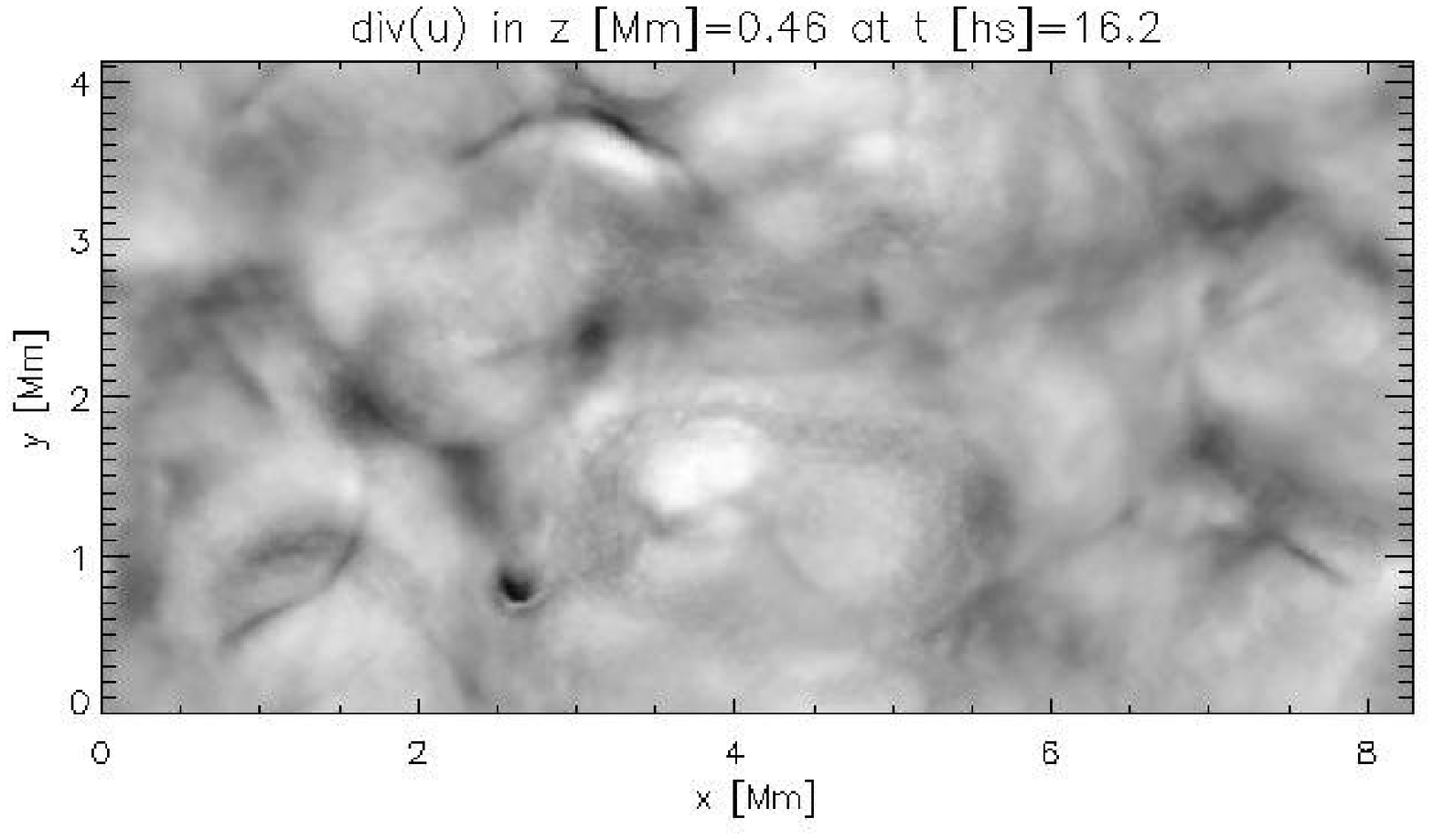}
\includegraphics[width=7.5cm]{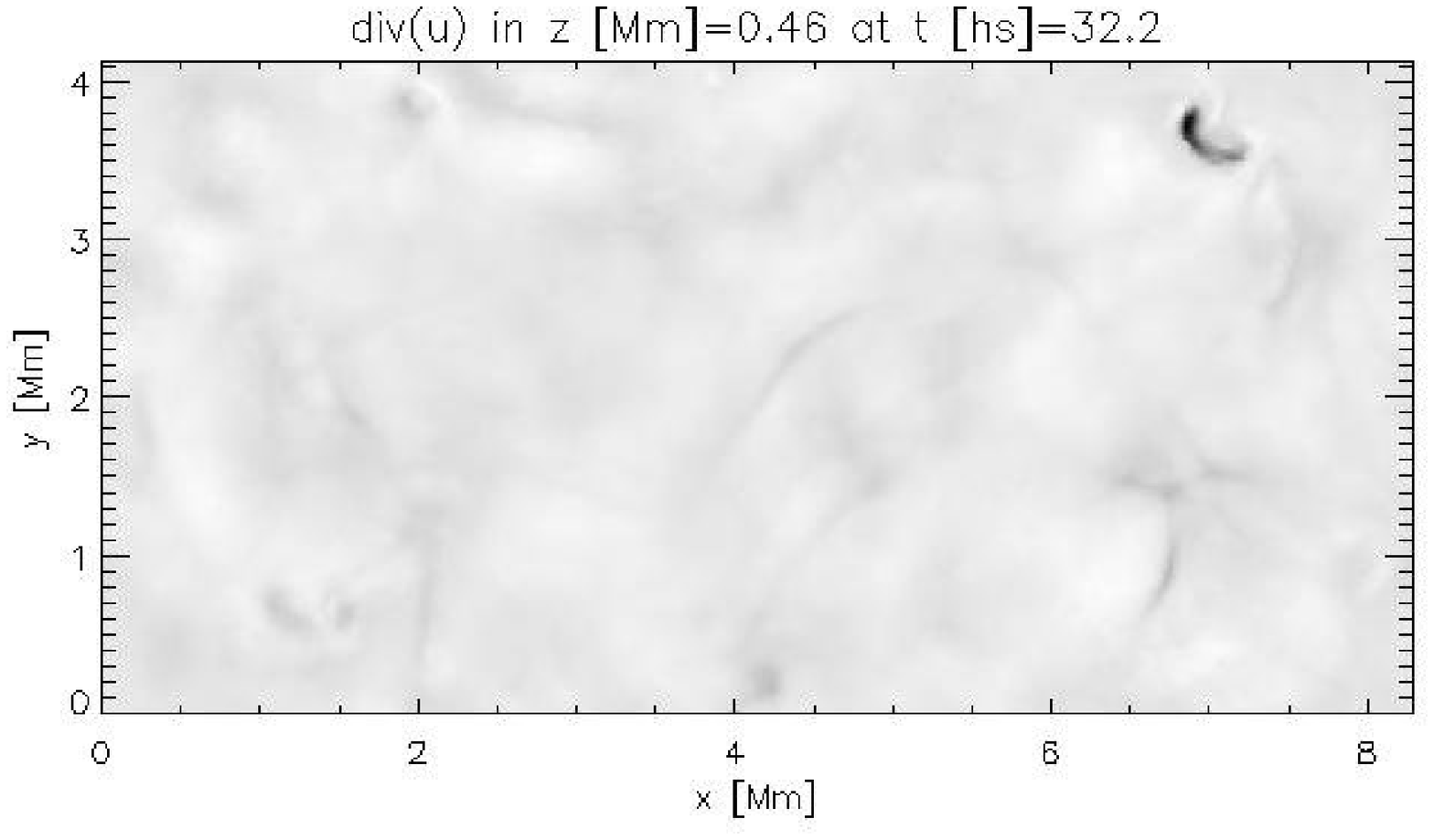}
\includegraphics[width=7.5cm]{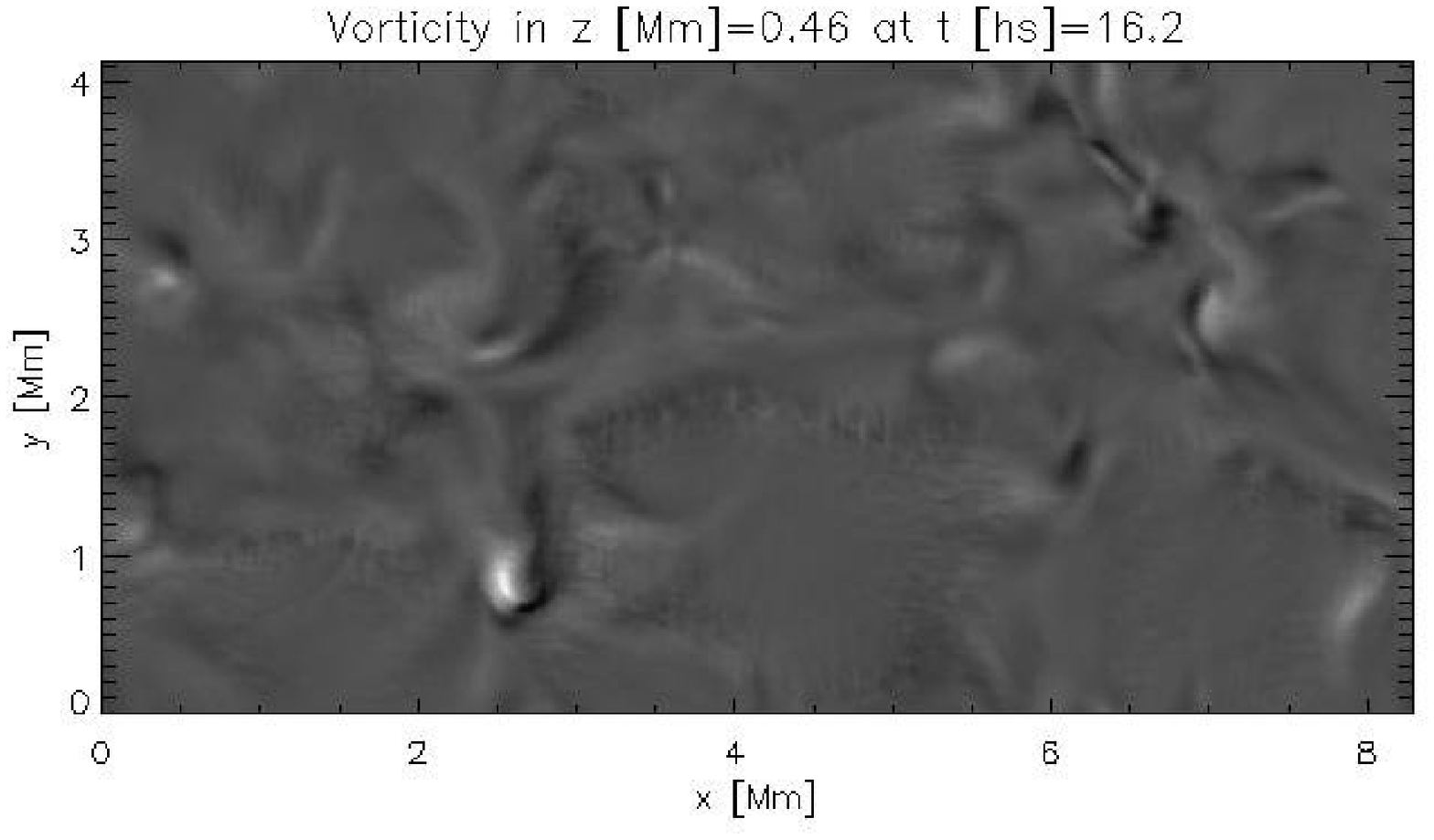}
\includegraphics[width=7.5cm]{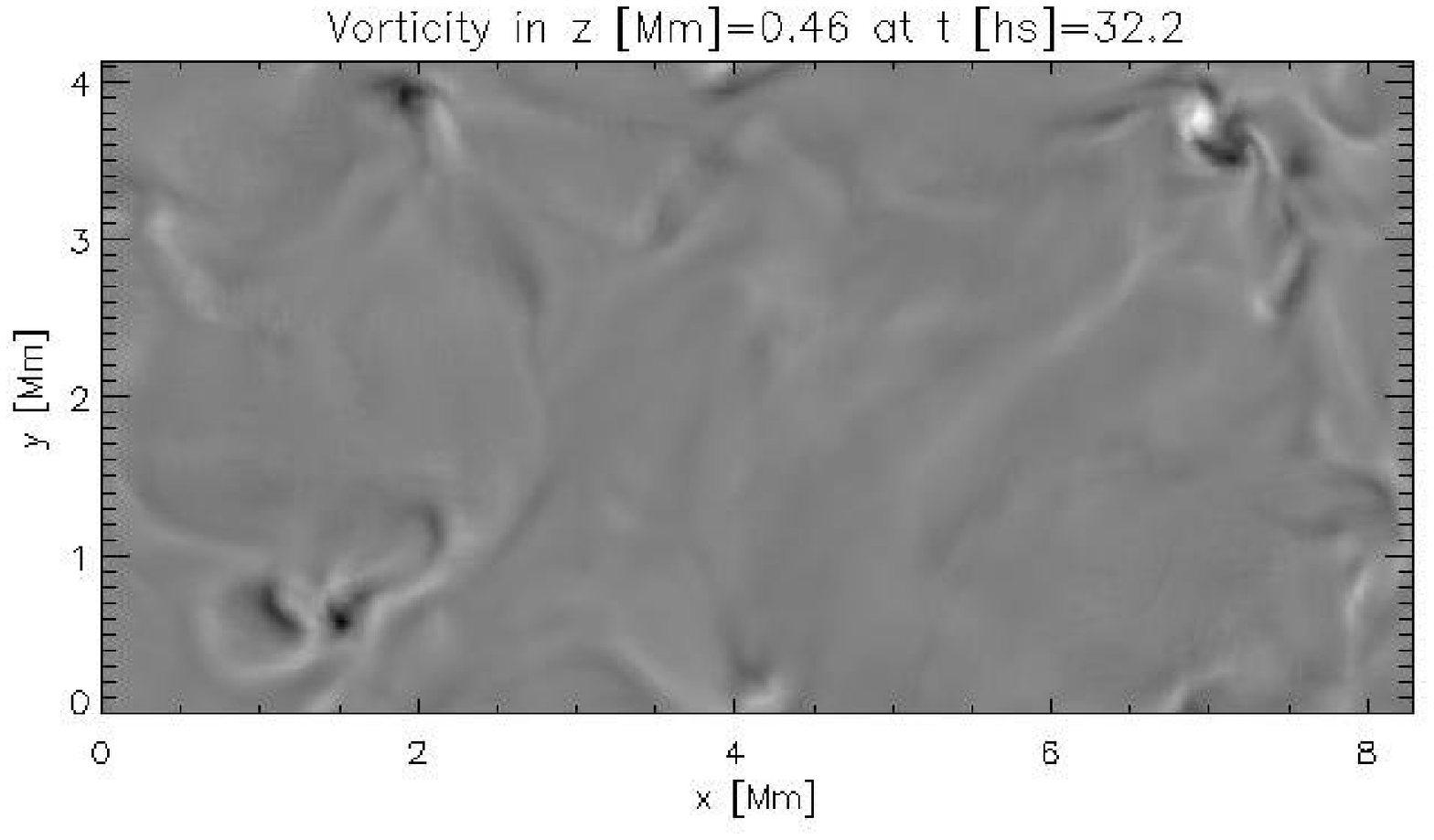}
  \end{center}
  \caption{\label{fig:divucr} 
  Vertical velocity (top row), 
  divergence of the velocity (middle row) and 
  the vertical vorticity (bottom row) at $z=458$~km 
  at time 1620~s, when the flux tube is close to the 
  layer at height $z=458$~km (left column), 
  and at time 3220~s, when the tube has crossed the layer $458$~km height
  (right column).}
\end{figure}

The vertical velocity is shown in figure~\ref{fig:divucr} in the first
row of panels using the same grey-scale intensity, from 1.5~km s$^{-1}$ (white) 
to -5.88~km s$^{-1}$ (black); the divergence of the velocity in the second row
and the vertical vorticity in the last row of panels, all at time 
1620~s in the first column and at time 3220~s in the second column. 
We find similar structure to those found in the synthetic observations 
(figure~\ref{fig:tgcr}) before the tube crosses the layer at $458$~km height. 
The structure of the cells is strongly modified as the tube goes
through that layer. We find that the cell expansion is related
to the divergence of the velocity, also at this height, but that the 
correspondence between temperature/continuum emission and velocity 
divergence is not as clear as in the lower layers. There are 
large vertical velocities (first row of the figure~\ref{fig:divucr}), both up or down 
(see the left-panels of the figure~\ref{fig:divucr}  at position [$2.1,0.8$] 
and right panels  at position [$7,3.7$]) which
correspond to the bright points in the synthetic observations and also fit 
with the large concentration of magnetic flux and vertical vorticity. 

The maximum upward velocity oscillates around
1.5~km s$^{-1}$ with no important changes in time except at $t=2420$~s when the
peak rises to 4~km s$^{-1}$. This peak is due to the rapid movement of the tube as
it rises. The maximum downward velocity oscillates around 2~km s$^{-1}$. Later,
the maximum downward velocity increases slowly with time from $t=1800$~s, up to
5~km s$^{-1}$ towards the end of the simulation at $t=3220$~s. The downflowing plasma
is located in the same region as the bright points with high magnetic field strength and 
vorticity are found. The maximum horizontal
velocity varies around 3.5~km s$^{-1}$ in the period from 0~s to 1400~s.
After that time it 
increases with time and rises to 7~km s$^{-1}$ at time 1940~s due to the
expansion process. Afterwards the maximal upward velocity returns to 
roughly 5.5~km s$^{-1}$, though with big variations.

The temperature range before the tube crosses the layer $z=458$~km is 
[4,6]~kK from 0~s to 1300~s. As the tube approaches the 
lower limit decreases to 3~kK due to expansion. When the tube arrives,
the temperature range changes to  [3.2,6.2]~kK at time 1900~s.  The lower limit
increases to 3.9kK after 2500~s. 

After the magnetic flux tube has crossed the layer at $458$~km height the structure of the
plasma no longer appears granular. The magnetic field, shown in the figure~\ref{fig:fieldcr},
is almost horizontal and only in the boundary layer of the tube can one see any 
vertical component. 
With greater initial twist the tube 
remains more horizontal in the chromosphere compared to the cases with less initial 
twist, {\it i.e.} the ratio of the horizontal magnetic field
component to the vertical magnetic field component, in regions where the tube is located, 
increases with increasing twist. 

Figure~\ref{fig:divucr} also shows the expansion of the fluid as the tube 
crosses the layer at $458$~km height. The center of these cells are colder than 
the edges as the tube expands. 
However, as the tube crosses the layer at $458$~km height, 
the small cells seen in the initial model disappear and cold larger zones form near 
the center of the tube. This is presumably due to expansion 
(second row of the figure~\ref{fig:divucr}), but the velocity divergence
does not fit as well with the temperature or intensity maps as found in the lower 
layers of the height $458$~km. This suggests that at this height also other processes 
could be active in cooling the chromosphere. We will 
go into more detail on these alternate processes in the next section,
as they become steadily more important with increasing height into the chromosphere.

At the boundary layer of the magnetic flux tube the temperature
rises. This heating could be due to a wave, compression of the plasma, or reconnection
where the magnetic field is almost vertical.
Once 
the tube has disappeared from the layer at $458$~km, at time $t\approx 2300$~s, the temperature 
displays an irregular structure with some high temperature points. These points are coincident
with a high vertical vorticity and with strong vertical magnetic field 
(e.g. at position [7,3.8] Mm at time 3220~s, see figures~\ref{fig:fieldcr}-\ref{fig:divucr}). 
Note that the temperature structure has not returned to its 
pre emergence state. This could be due to the fact that the magnetic structure of the atmosphere
has changed: before the tube appeared the energy density of the
magnetic field was low and $\beta > 1$, but around $t=2200$~s we find several regions 
where $\beta \approx 0.5$ and the field dominates the chromospheric structure.

Bright points are seen to occur in the synthetic observations (figure~\ref{fig:tgcr}). 
In the layer at height $458$~km the bright points have greater contrast than found in the regions
below and also much greater temperature than in the surrounding plasma (The bright point in figure
\ref{fig:tgcr} right-top panel at the position $[7,4]$~Mm has a temperature at this height of
6500~K compared with the surrounding temperature of 5100~K. At the same position at height
$234$~km (see figure \ref{fig:revgran}) the temperature is 5700~K compared with 5200~k in the surroundings)
As explained earlier for the lower layers, bright points are related to strong magnetic field, 
to vertical velocity, and vertical vorticity, and to a larger compression 

than in the surroundings. The heating seems to come primarily from the 
convergence of the velocity field as we find the most important heating term is the advection term; 
it is an order of magnitude greater than the joule and viscous heating terms.

\begin{figure}
\includegraphics[width=7.5cm]{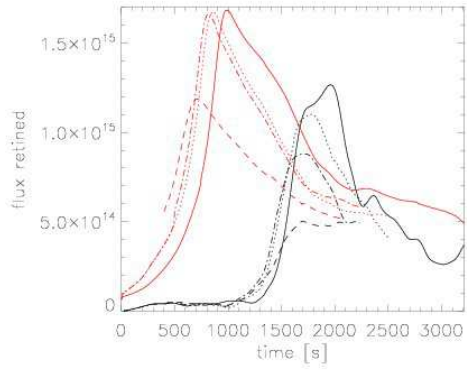}	
  \caption{\label{fig:fluxret} Mean magnetic flux per unit area
  perpendicular to the tube as function of time. Region confined between the
  $z=10$~km and $z=254$~km in black color
  and below the photosphere in red color for the runs A1 (dash-dot line),
  A2 (dashed line), A3 (dotted line) and A4 (solid line).}
\end{figure}

\begin{figure}
\begin{center}
\includegraphics[width=7.5cm]{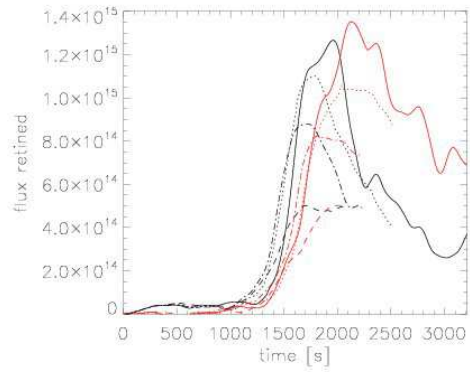}	
  \end{center}
  \caption{\label{fig:fluxretlow} Variation in time of the mean magnetic 
  flux by area  as defined in equation \ref{eq:phi2} from the photosphere ($z=10$~km) to the
  height $z=234$~km with black color and from the height  
  $z=234$~km to the height $z=458$~km with red
  color for the runs A1 (dash-dot line), A2 (dashed line), A3 (dotted
  line) and A4 (solid line).}
\end{figure}

\begin{figure}
\includegraphics[width=7.5cm]{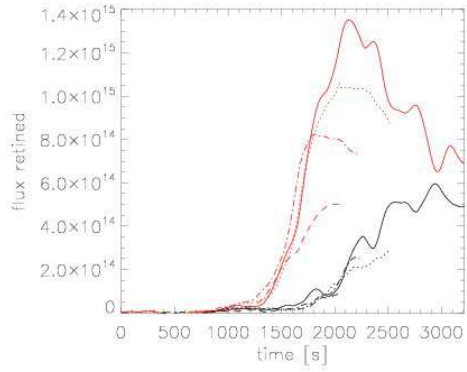}	
  \caption{\label{fig:fluxretup} Mean magnetic flux per unit area
  perpendicular to the tube as function of time. Region confined between
  $z=234$~km to $z=458$~km in red color and
  from $z=458$~km to $z=901$~km in black color for the
  runs A1 (dash-dot line), A2 (dashed line), A3 (dotted line) and A4
  (solid line).}
\end{figure}

Figure~\ref{fig:fluxretup} shows the total flux for the regions from 
$z=234$~km to $z=458$~km in red, and from  
$z=458$~km to $z=905$~km, in black. 
The  biggest difference between the figures \ref{fig:fluxret}, 
\ref{fig:fluxretlow} and \ref{fig:fluxretup} is the slope of the flux with time 
($d\langle\phi\rangle/dt$). The total flux entering the chromosphere is much smaller than 
that reported in figures~\ref{fig:fluxret} and \ref{fig:fluxretlow} because the tube spends
more time in crossing the upper regions. This is presumably due to the horizontal 
expansion of the magnetic flux tube. The amplitude of the slope of the curves in the figure 
\ref{fig:fluxretup} increase with magnetic field strength. After the initial
large rise, when the tube crosses the corresponding section, the curve shows an oscillation (see
figure \ref{fig:fluxretup}) that is quite similar to that seen in the layers 
below (see figures \ref{fig:fluxret} and \ref{fig:fluxretlow}). However, by comparing 
each curve it is possible to see that there is a small delay between the different lines 
in the figures \ref{fig:fluxret}, \ref{fig:fluxretlow} and \ref{fig:fluxretup}. 
This retardation of the oscillation between the different layers is what we would 
expect from a wave that propagates at the sound speed between each layer. 
The variation of the mean magnetic flux strength increases with increasing height, 
twist, and/or the magnetic field strength.
The oscillations seem related to the oscillations in the downward maximum
velocity with time and also to oscillations in the ratio between the vertical and horizontal 
magnetic field, as mentioned in the previous sections.

Finally, we find that the slope of the flux with time, $d\langle\phi\rangle/dt$, at the 
time the tube crosses into the chromosphere, 
is larger with greater values of the twist parameter and with larger magnetic 
field strength. This effect is clearer in the chromosphere than it is in the photosphere.   
Previous 2D flux emergence simulations by \citet{Shibata:1989lr} and \citet{magara2001} have 
also identified magnetic buoyancy 
instabilities as important in the rise of magnetic field to the chromosphere and coronal layers.

\subsection{Chromosphere, Transition Region and Corona}

\begin{figure}
\begin{center}
	\includegraphics[width=7.5cm]{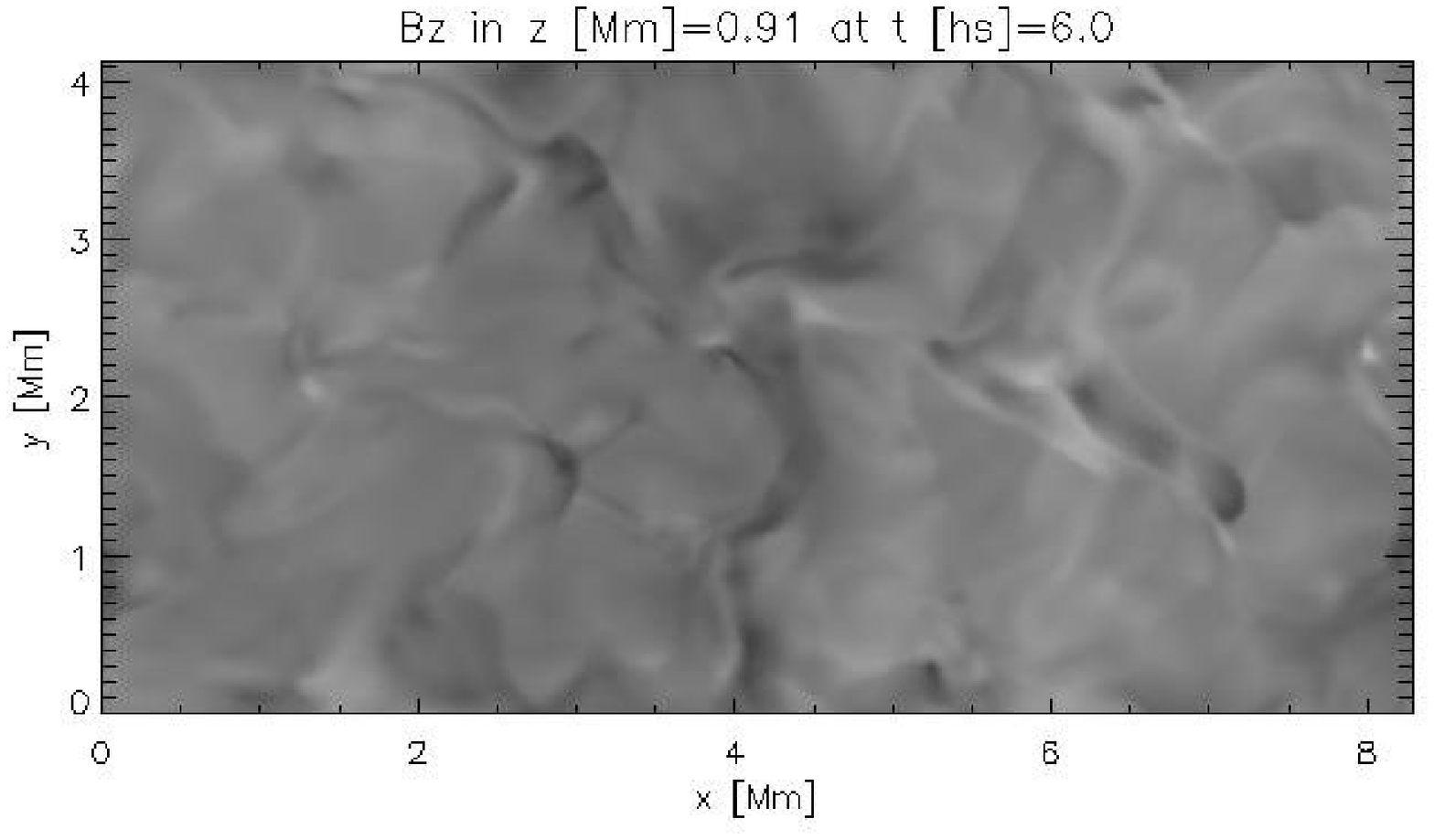}
	\includegraphics[width=7.5cm]{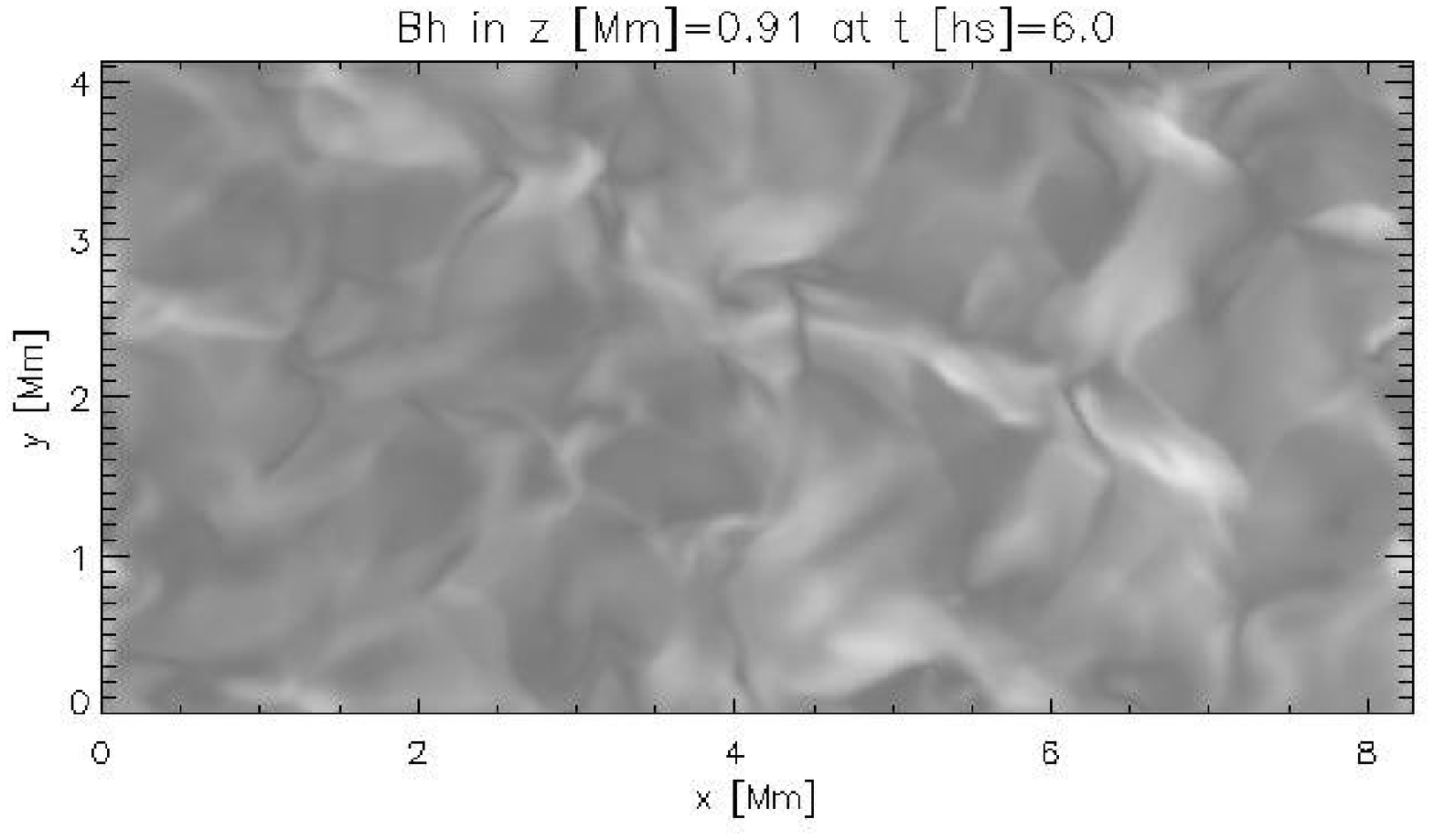}
	\includegraphics[width=7.5cm]{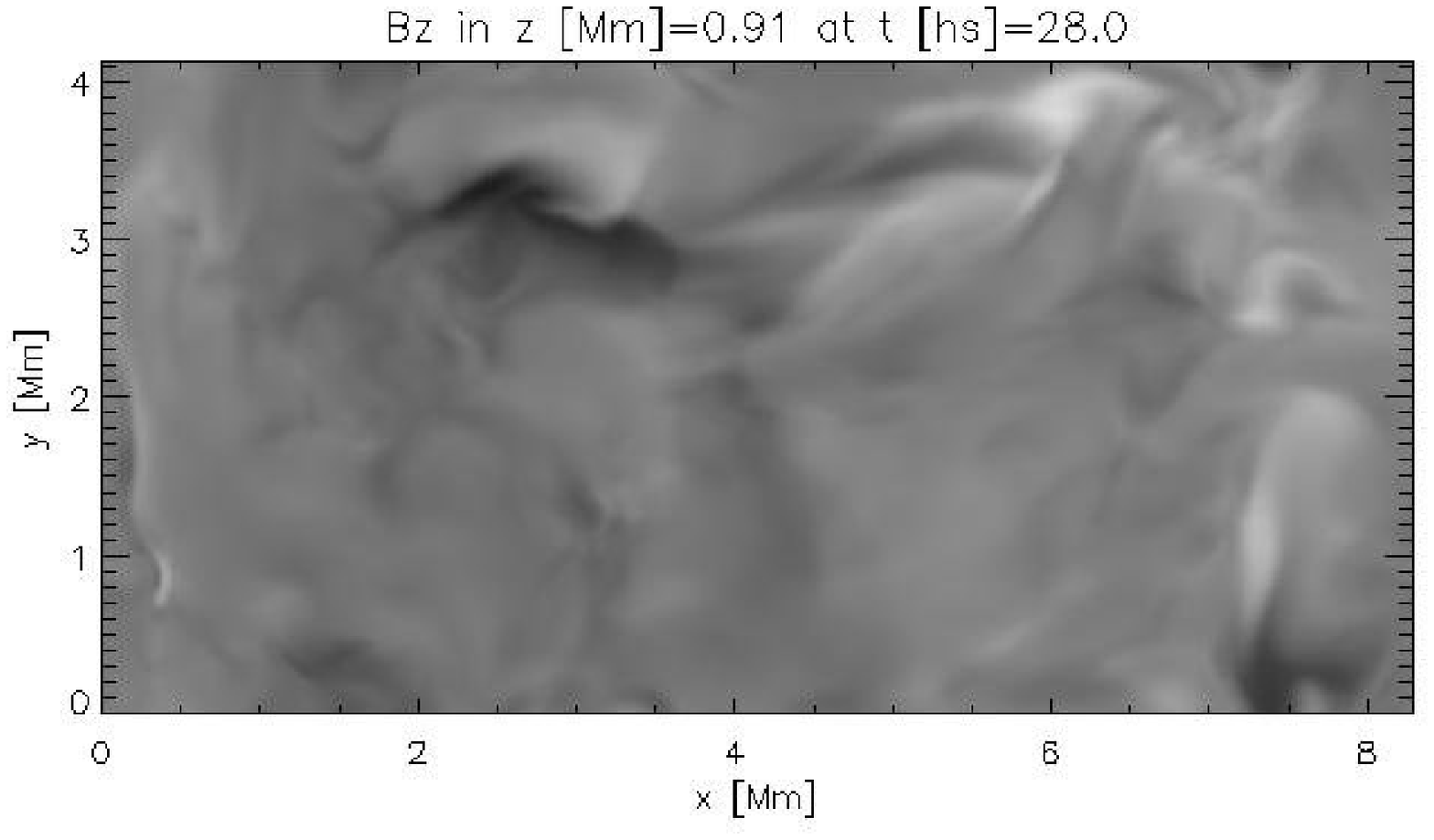}
	\includegraphics[width=7.5cm]{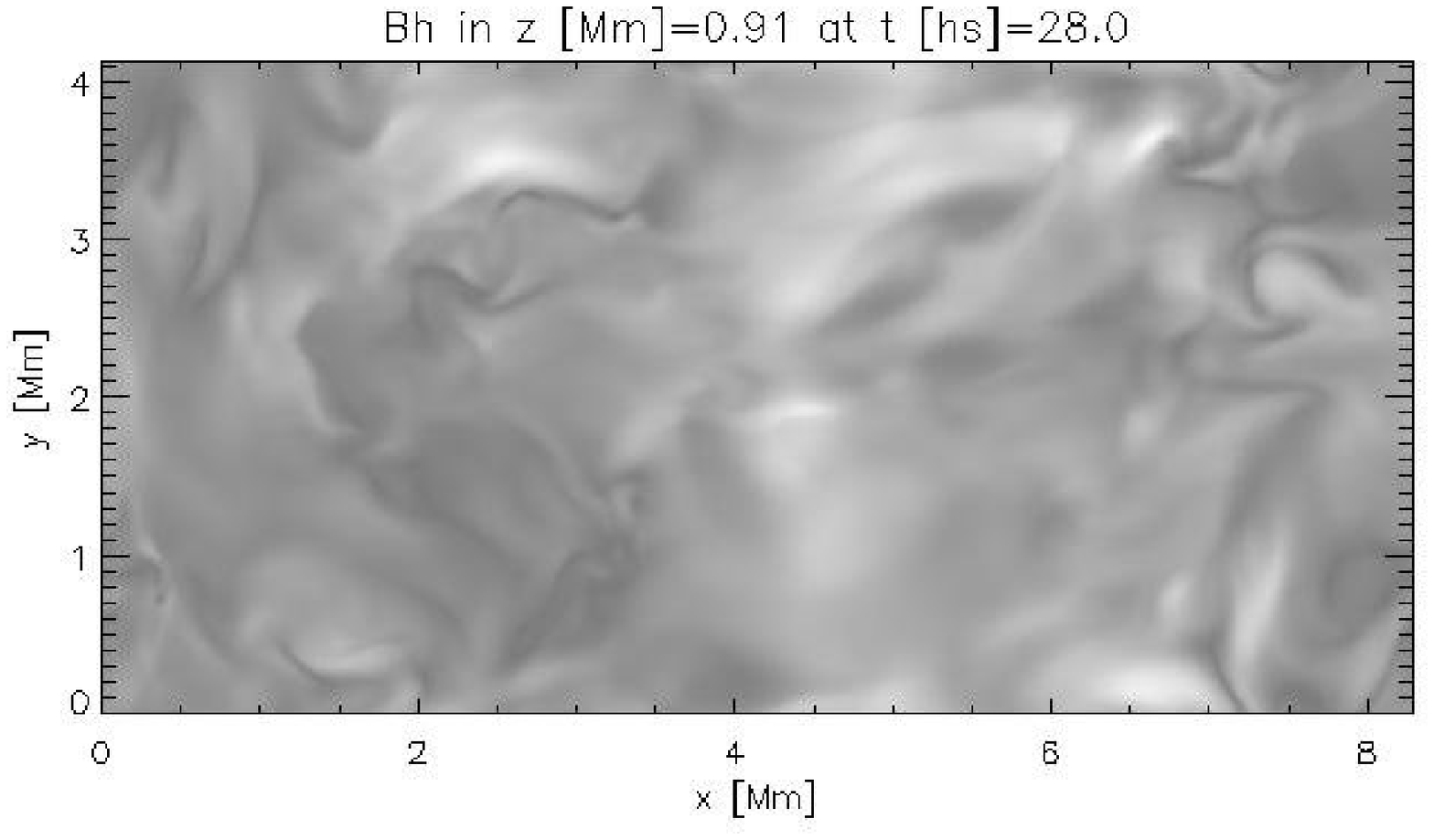}
	\includegraphics[width=7.5cm]{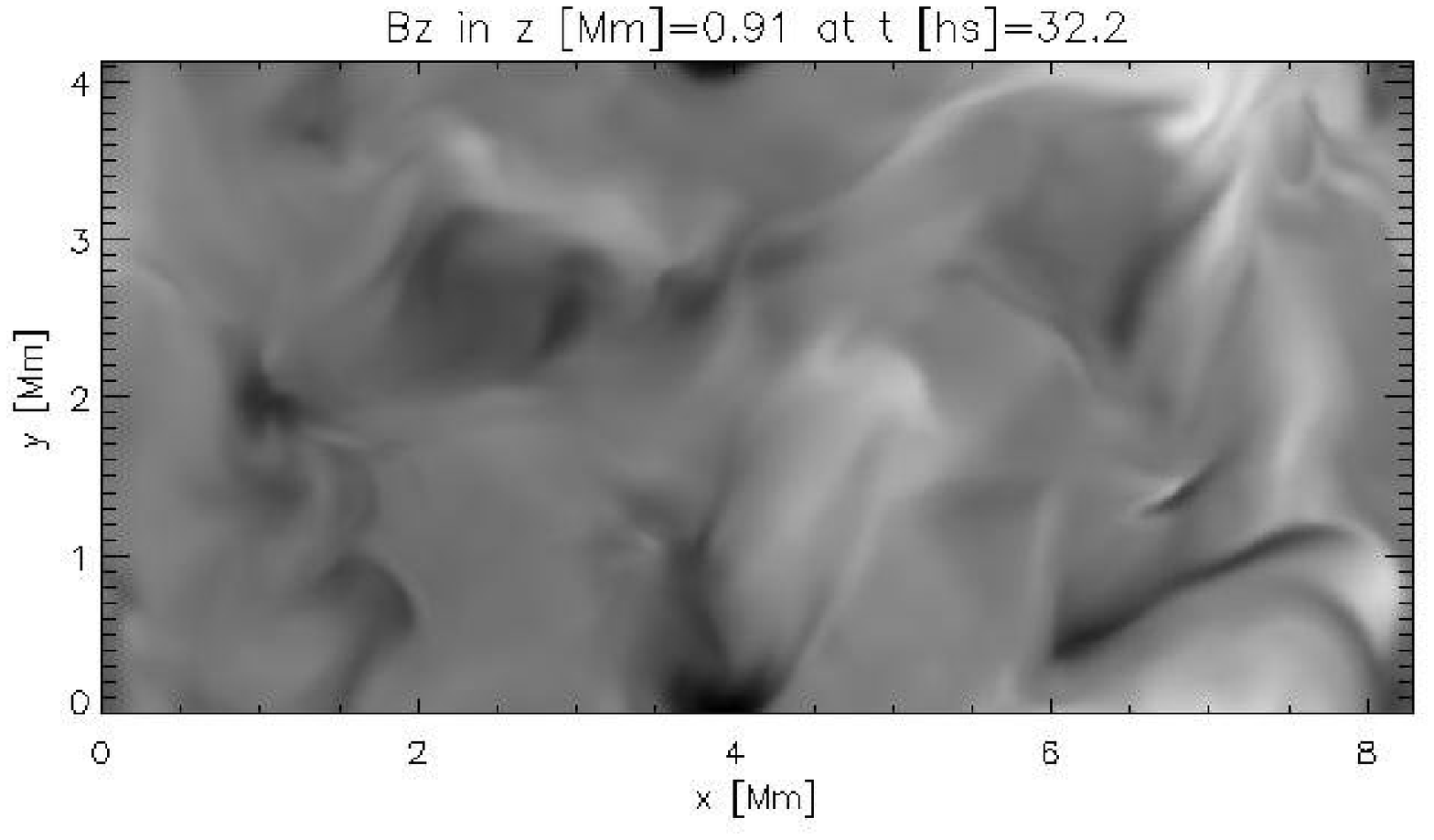}
	\includegraphics[width=7.5cm]{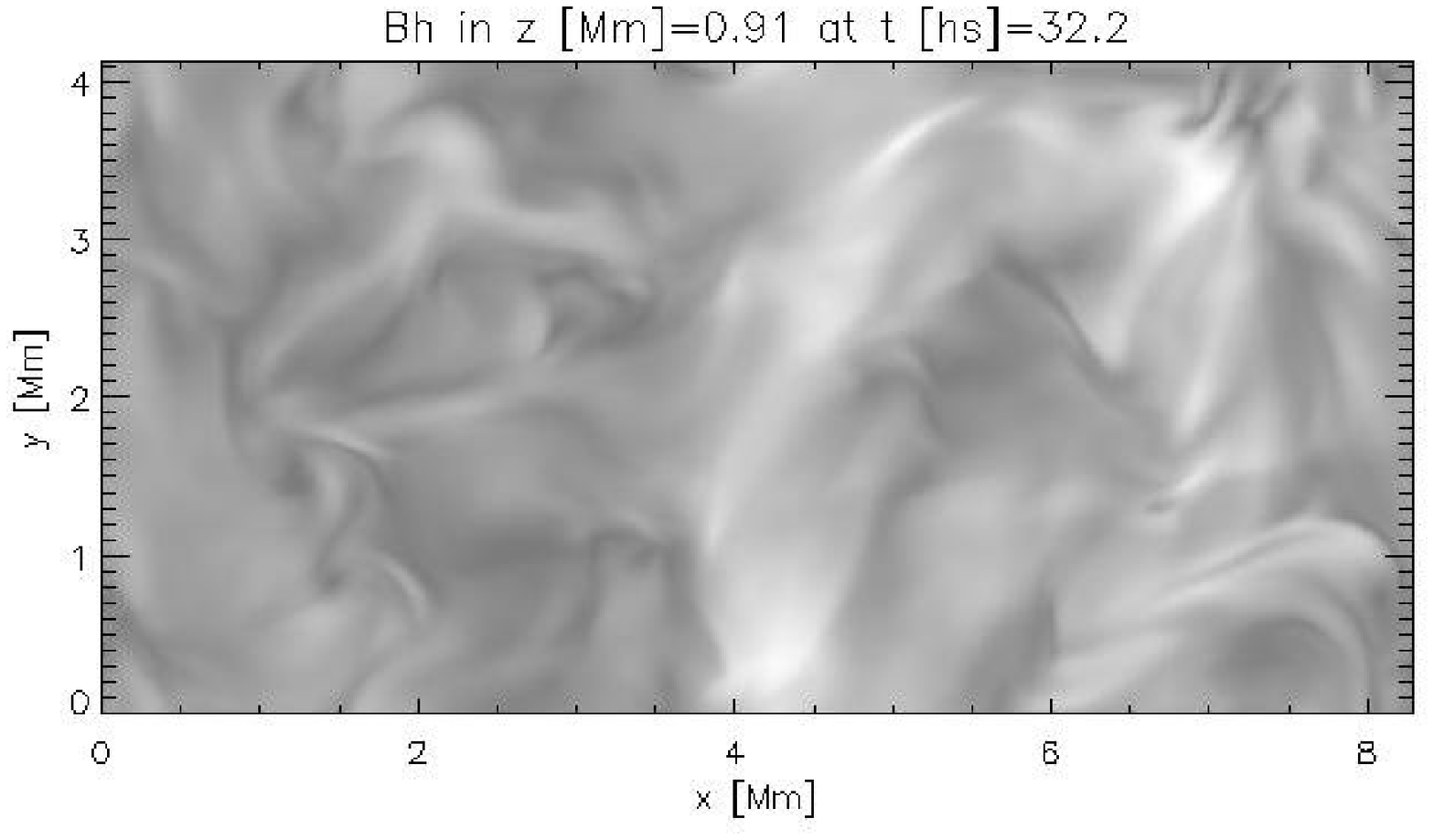}
 \end{center}
 \caption{\label{fig:fieldupcr} $B_z$ at $z=906$~km 
  (left panels), and $B_h$ (right panels) at times 600~s, before the flux
 tube starts to cross the chromosphere (top panels),
 at time 2800~s when the tube is crossing the photosphere (middle panels) 
 and at time 3220~s (bottom panels). 
 The grey-scale goes from -60.46~G to 65.91~G}
\end{figure}

Let us finally discuss the effects of the magnetic flux tube as it reaches the 
upper chromosphere, transition region and lower corona. The variation of the
magnetic field in the chromosphere, at $z=906$~km, is shown in 
figure~\ref{fig:fieldupcr} with the vertical field in the left column and the
horizontal field in the right column, both at three instants; times 600~s, 2800~s and
3220~s. The first signs of the magnetic flux tube reaching the chromosphere 
are evident at time $t=2100$~s. However, the tube does not appear uniformly 
in the region from $x=2$~Mm to $x=6$~Mm as it does in the levels below. Rather
patches of magnetic flux rise into different places randomly along the tube 
region at different times. The maximum absolute magnetic field, before the tube reaches 
the layer is 43~G. This value increases after the tube comes into the chromosphere
to 80~G. During the flux emergence into this level and after the tube has passed 
the average of the maximum field is of order of 65~G. 
At this height the magnetic field does not have any singular isolated points with strong 
amplitude. The field becomes generally more horizontal once the tube crosses this 
layer and has a smoother structure than in the layers below.

\begin{figure}
\begin{center}
\includegraphics[width=7.5cm]{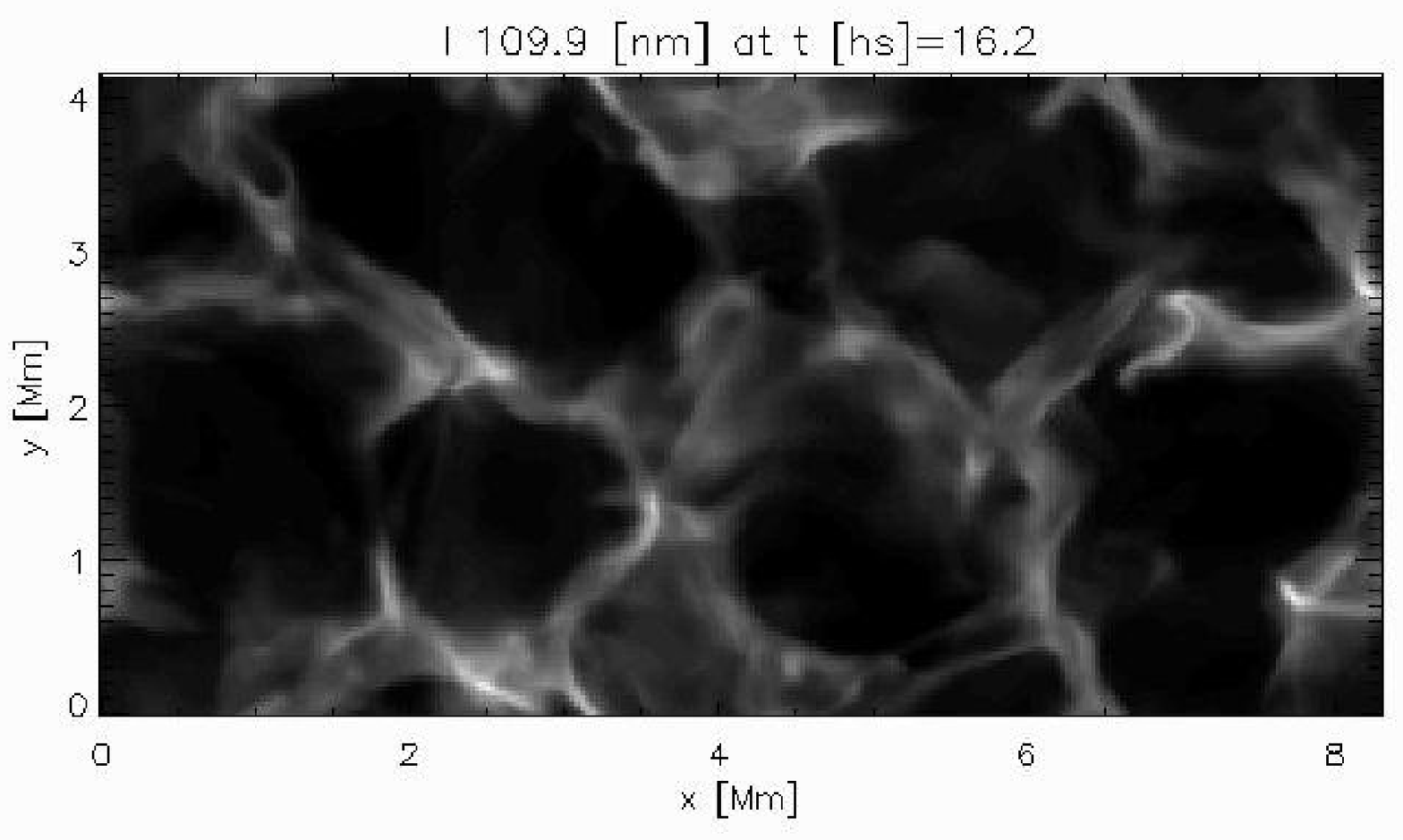}
\includegraphics[width=7.5cm]{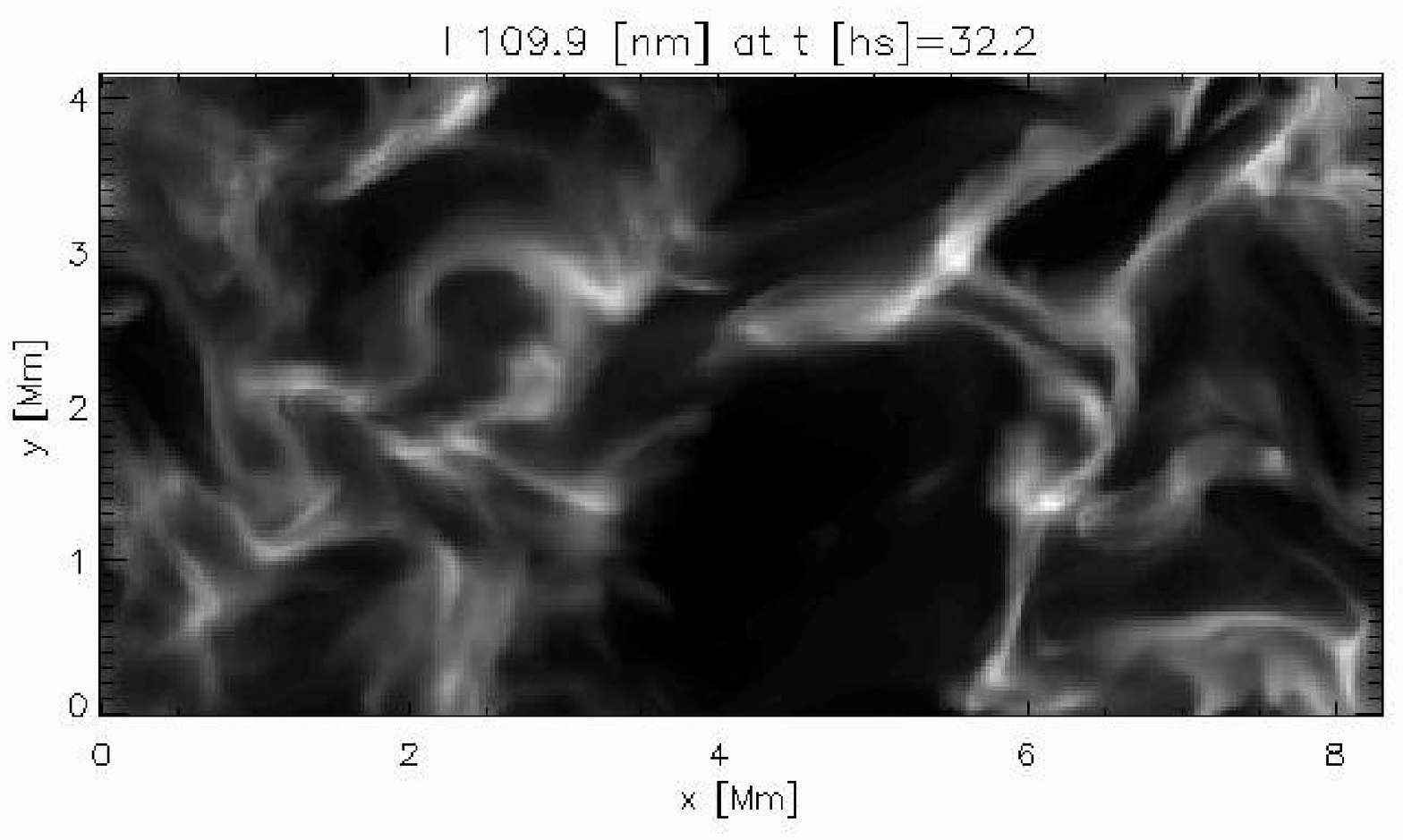}
\includegraphics[width=7.5cm]{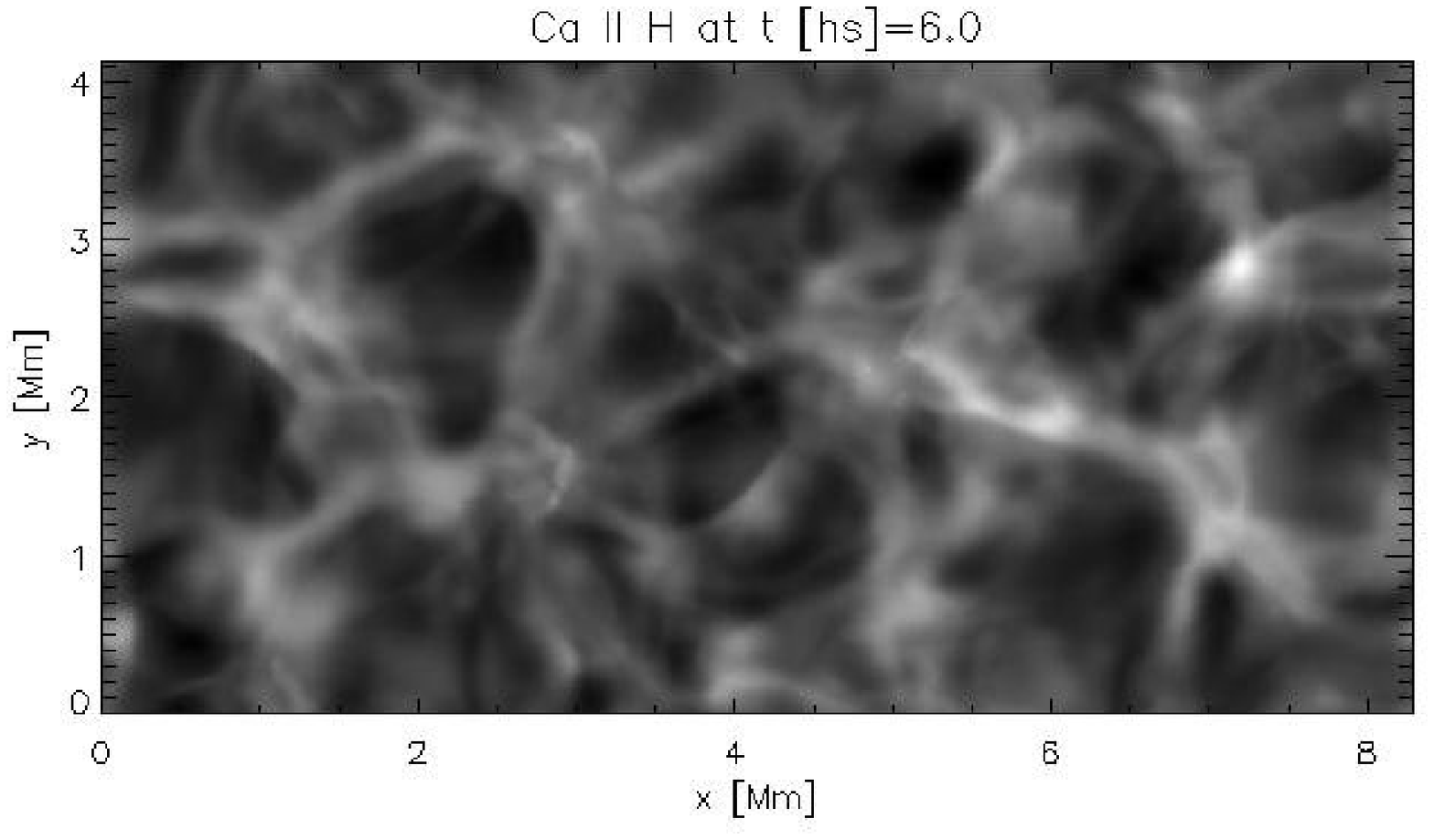}
\includegraphics[width=7.5cm]{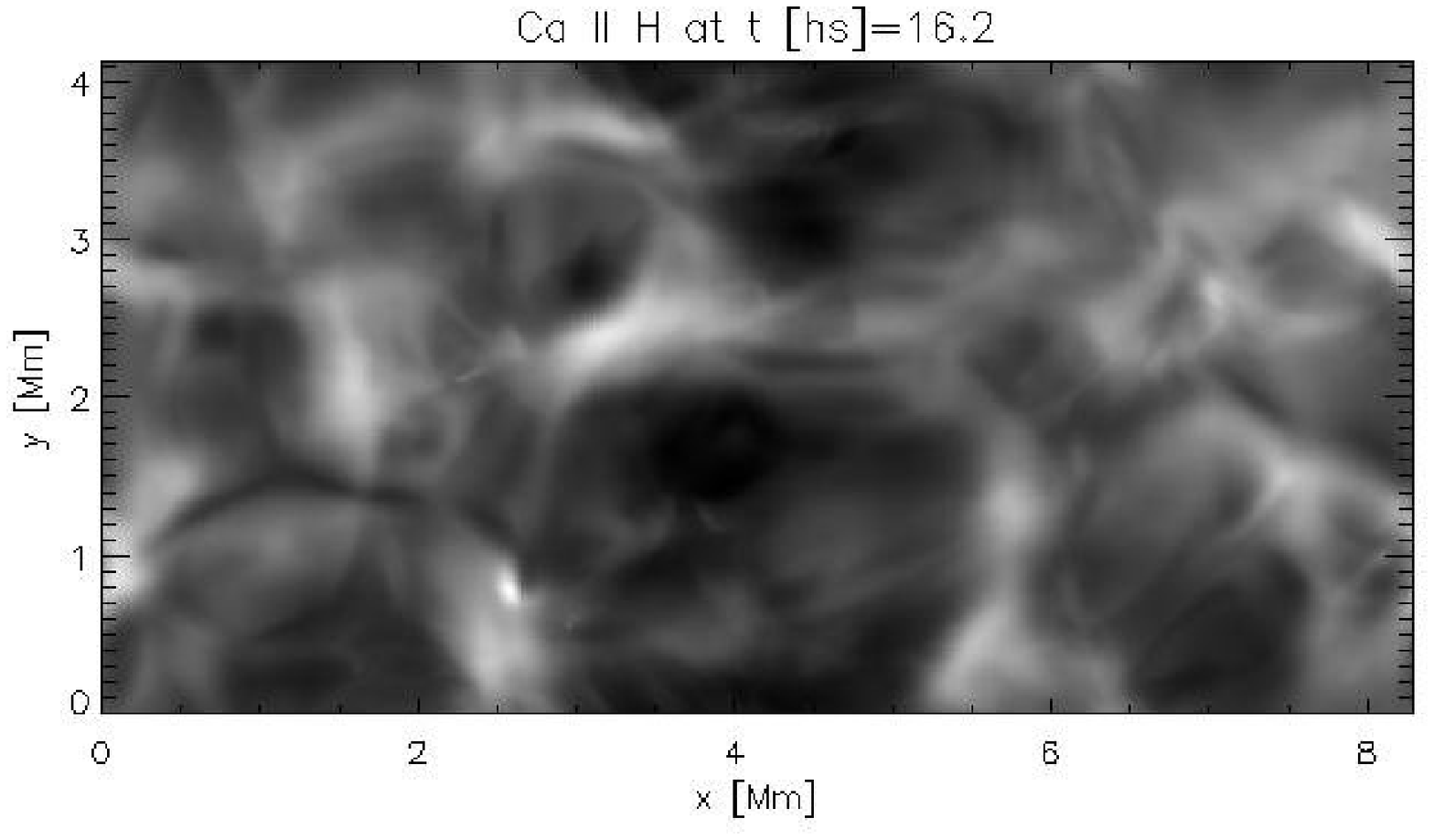}
\includegraphics[width=7.5cm]{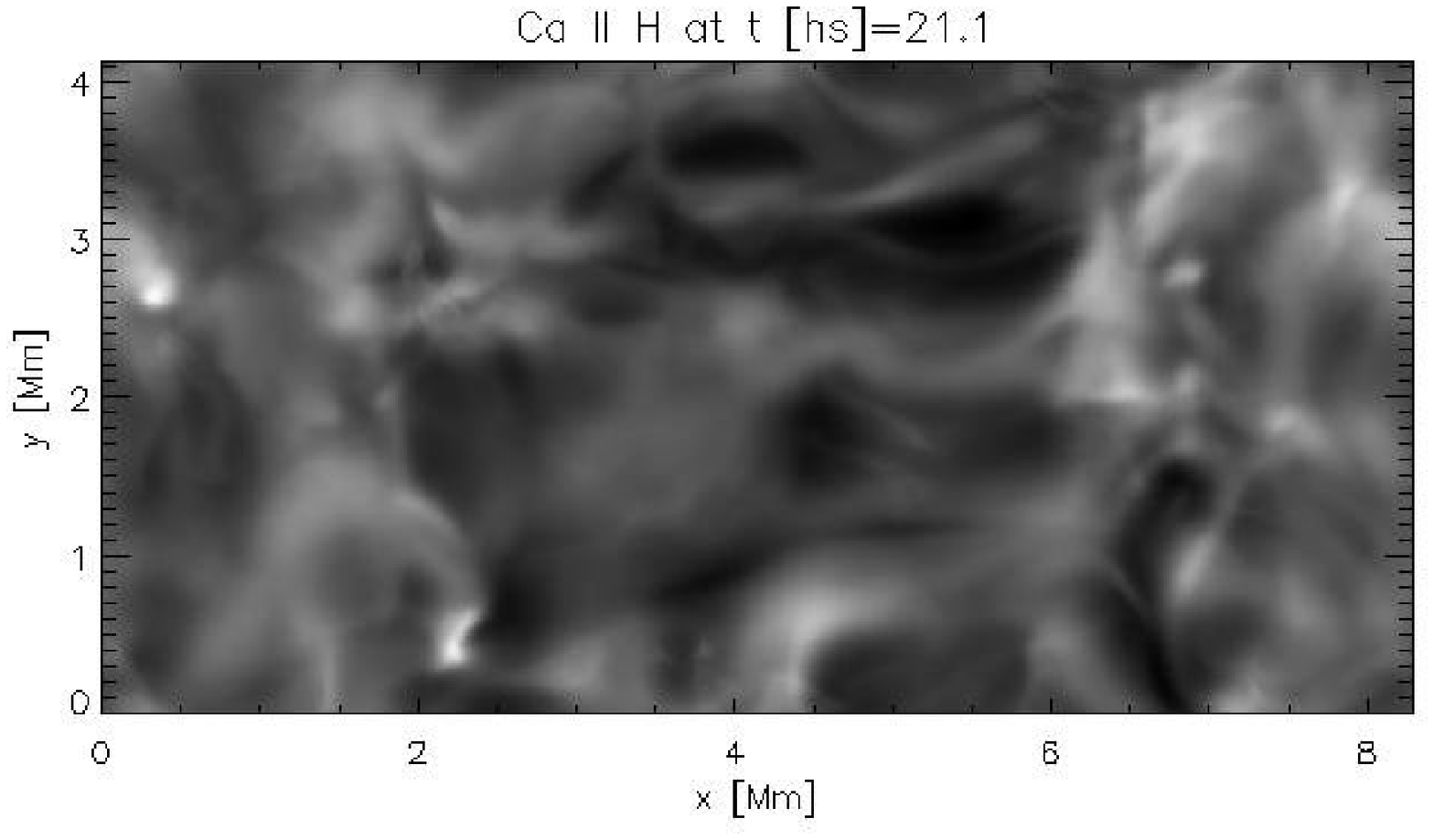}
\includegraphics[width=7.5cm]{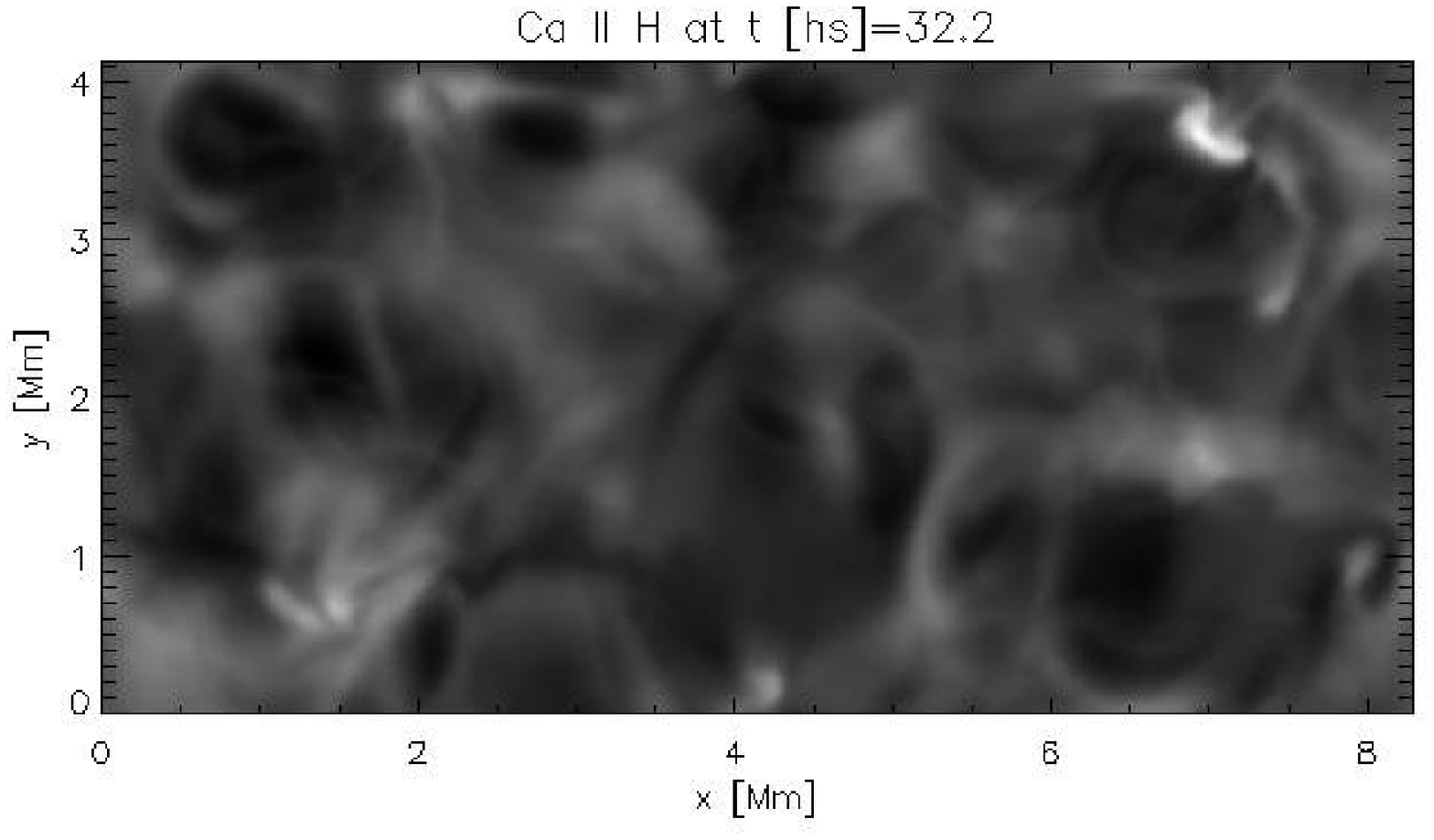}
  \end{center}
  \caption{\label{fig:tgupcr} Continuum intensity of the C~{\sc i} to the 
  power of 0.4 (equivalent to a gamma of 0.4 for an image) to better
  bring out the structure at 109.9~nm
  at time 1620~s (top left) and
  at time 3220~s (top right). The synthetic Ca II H line intensity 
 at time 600~s, before the flux tube goes through 
  the chromosphere, 
  at time 1620~s when the tube starts to cross, 
  at time 2110~s as the tube is crossing,
  and at 3220~s when the tube has crossed the chromosphere, 
  from the left to the right and top to bottom respectively.}
\end{figure}

In figure \ref{fig:tgupcr} we show the intensity of the C~{\sc i}  continuum 
at  109.9~nm (as could be observed with SUMER). We have
calculated intensities at $t=1620$~s before the emerging magnetic flux
has entered the chromosphere  and at $t=3220$~s, some 15 minutes after the
emerging flux first appears. The mean monochromatic optical depth unity is
around $1100$~km height in the simulation of the C~{\sc i} continuum at 109.9~nm 
(in the VALC atmosphere model it is at $z\approx975$~km).
We also show synthetic Hinode observations of the 
Ca~{\sc ii} H-line at times 600, 1620, 2110 and 3220~s. Note that while the carbon continuum
is formed mainly in the chromosphere
the Ca~{\sc ii} as observed with the 0.22~nm wide Hinode filter has contributions from 
the entire atmosphere, stretching from the upper photosphere to the chromosphere
\citep{Carlsson+etal2007}.
The structure of the intensity formed at this height shows large changes before and 
after the magnetic flux tube reaches the chromosphere. 
The temperatures range in the chromosphere at height $906$~km before the tube reaches this level
is [2.2,7.2]~kK, while after the tube has reached this height
the range is [2.0,7.0]~kK.
However, note especially the large
dim (and cold) zones that become evident after the tube crosses into the chromosphere
at time 3000~s. These dim, cold, regions are much larger than in the lower layers (see
section \ref{sec:revgran} and \ref{upcrom}). 

It is worth remarking that we do not find the concentrated bright points that 
appear in the lower lying layers. This is presumably because the magnetic field 
is mainly horizontal and we do not observe the confinement of the magnetic field that
occurs in the regions below due to plasma motions. On the other hand,
the chromosphere and corona suffer large changes in size, structure
and temperature after the tube enters this zone. 
The cooling in the layers between $234$~km and $458$~km height is due to radiative losses and
superadiabatic expansion, but in the chromosphere the big cold 
regions do not fit as well with the local plasma expansion as in the lower layers. 

\begin{figure}
\begin{center}
\includegraphics[width=7.5cm]{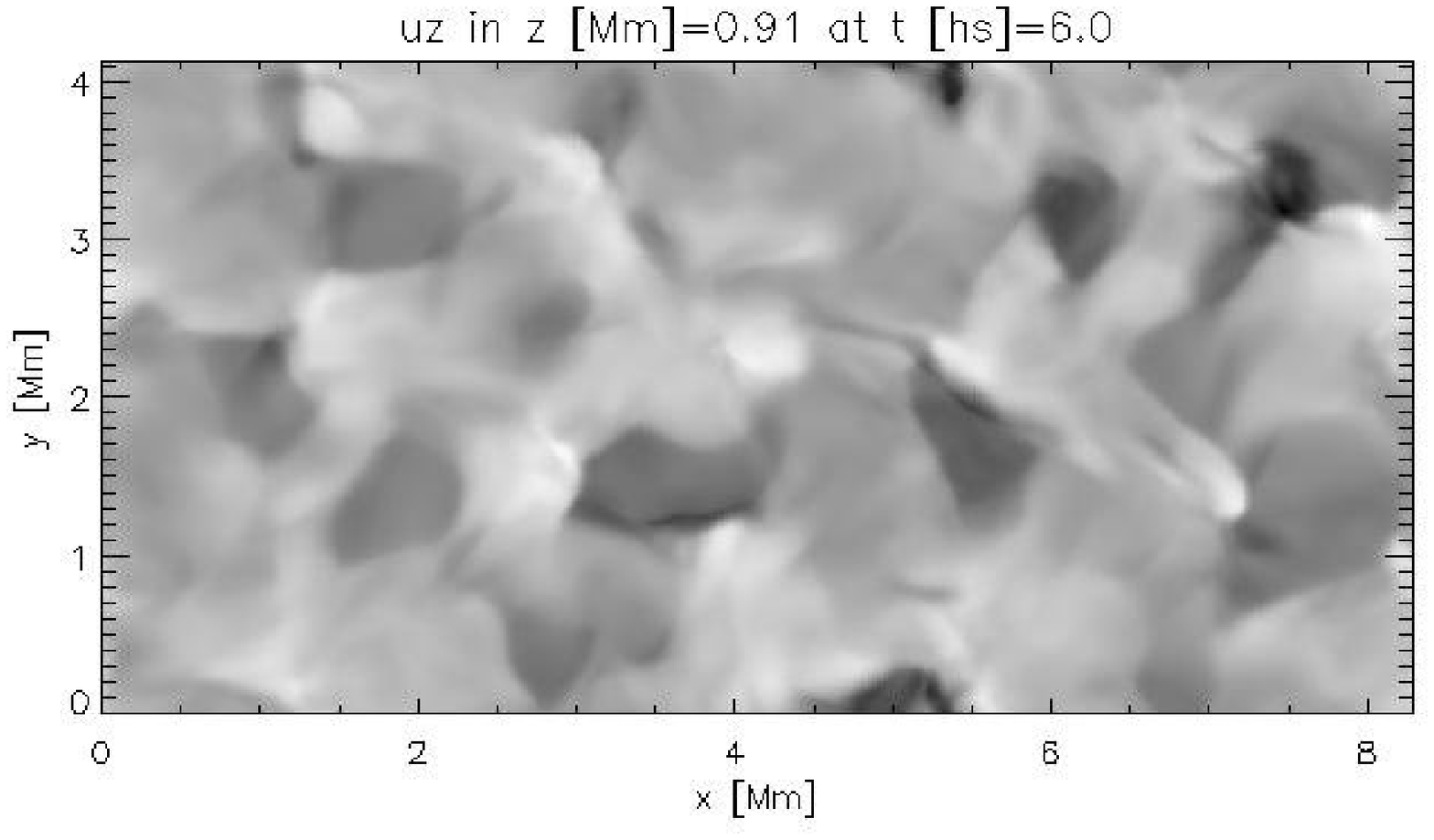}
\includegraphics[width=7.5cm]{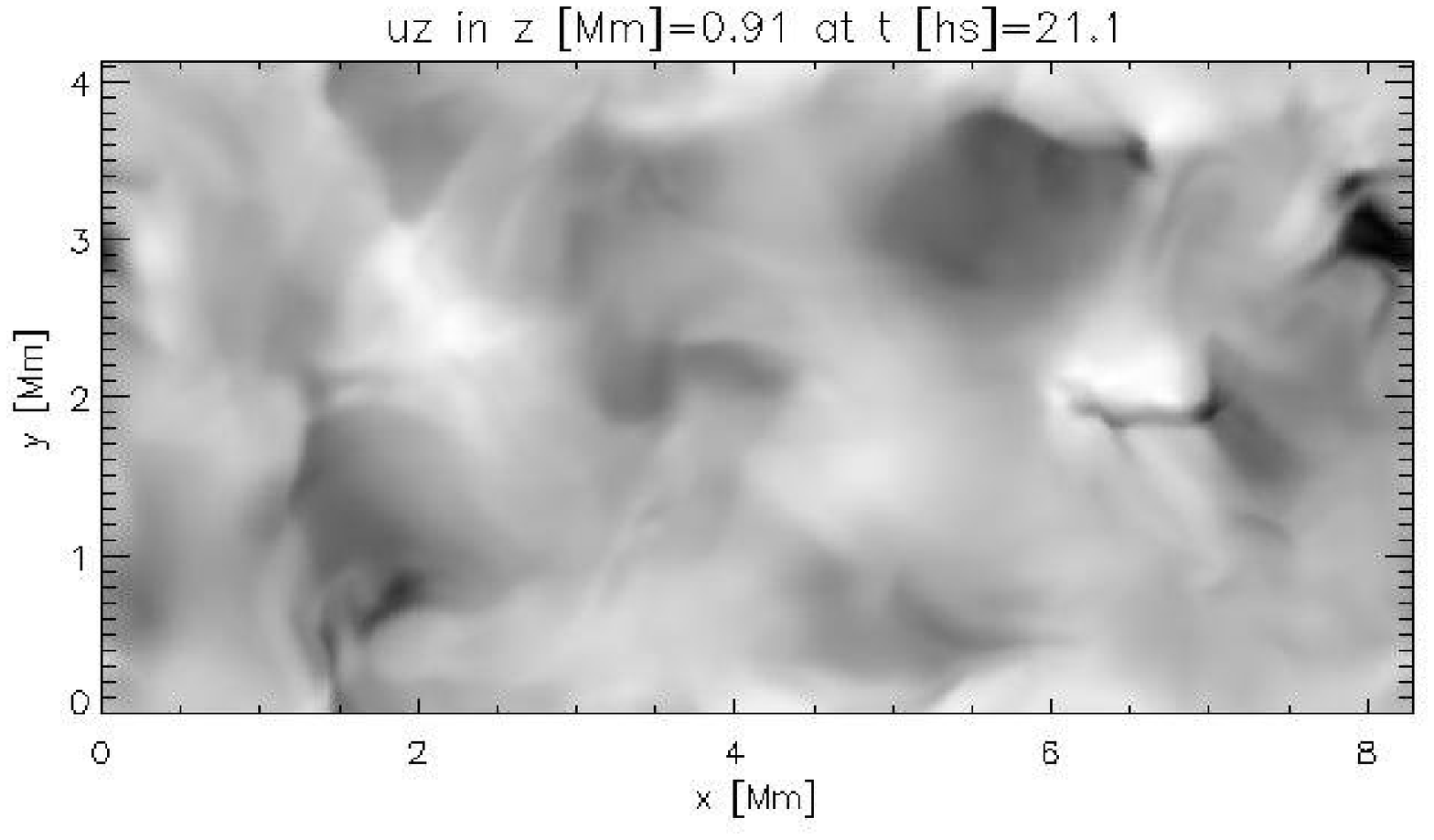}
\includegraphics[width=7.5cm]{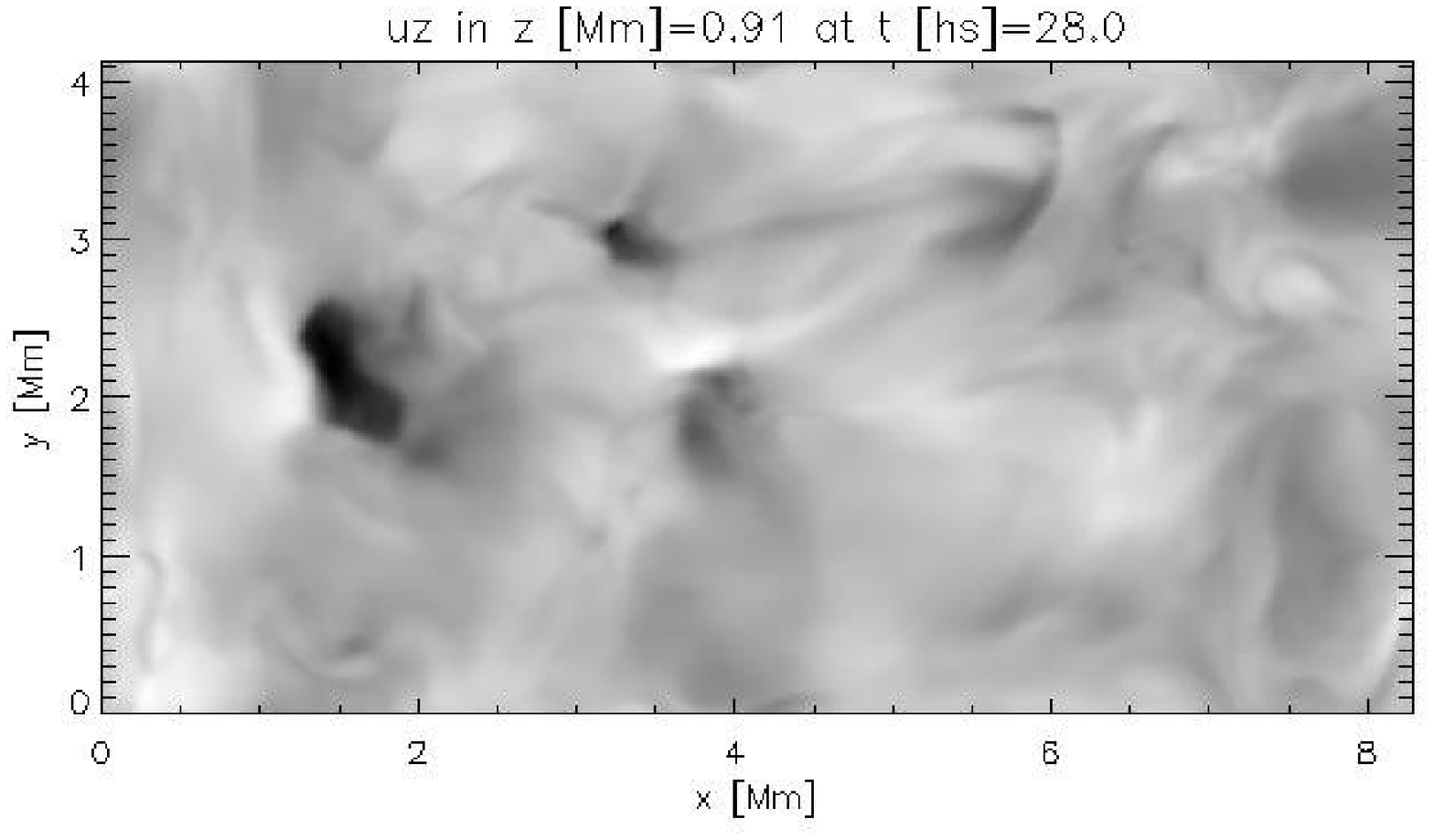}
\includegraphics[width=7.5cm]{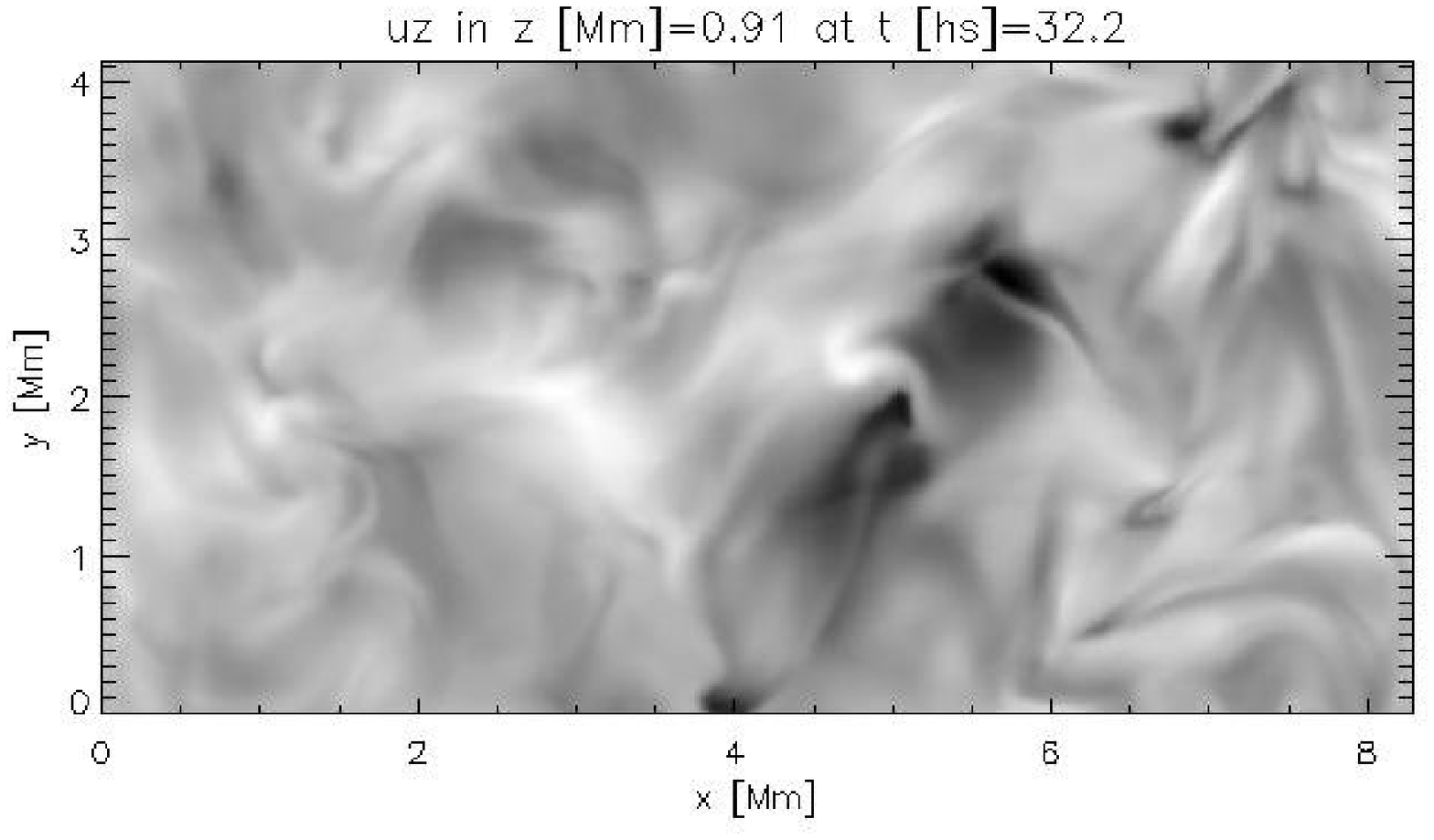}
  \end{center}
  \caption{\label{fig:uzupcr} Vertical velocity at $z=906$~km 
  at times 600~s, before the flux tube goes through the chromosphere (top left), 
  at time 2110~s, when
  the flux tube is closed to the chromosphere (top right), 
  at time 2800~s, when
  the big cells are cooling through expansion (bottom left),
  and at time 3220~s, after the tube has passed through the
  chromosphere (bottom right).}
\end{figure}

The vertical velocity in the chromosphere at height $906$~km is shown in figure~\ref{fig:uzupcr}
at four different instants; 600, 2110, 2800 and 3220~s. The maximum downward velocity 
is around 10~km s$^{-1}$ and the maximum upward velocity is around 7~km s$^{-1}$ throughout the whole 
simulation.
It is difficult to see any effects due the emergence of flux in the maximum 
horizontal velocity as these have large changes with time. The average maximum horizontal 
velocity is 13~km s$^{-1}$ and varies between 7~km s$^{-1}$ and 20~km s$^{-1}$. On the other hand the structure 
of the velocities change radically as is evident by comparing the pre- and post- emergent 
panels of figure~\ref{fig:uzupcr}. 
Shock waves dominate the upper chromosphere before
the tube crosses the chromosphere ($t=600$~s). However, during and after the tube
crosses the chromosphere the velocity is much smoother in the regions where the field
strength increases due to the rise of the flux tube into to upper chromosphere.

\begin{figure}
\begin{center}
\includegraphics[width=7.5cm]{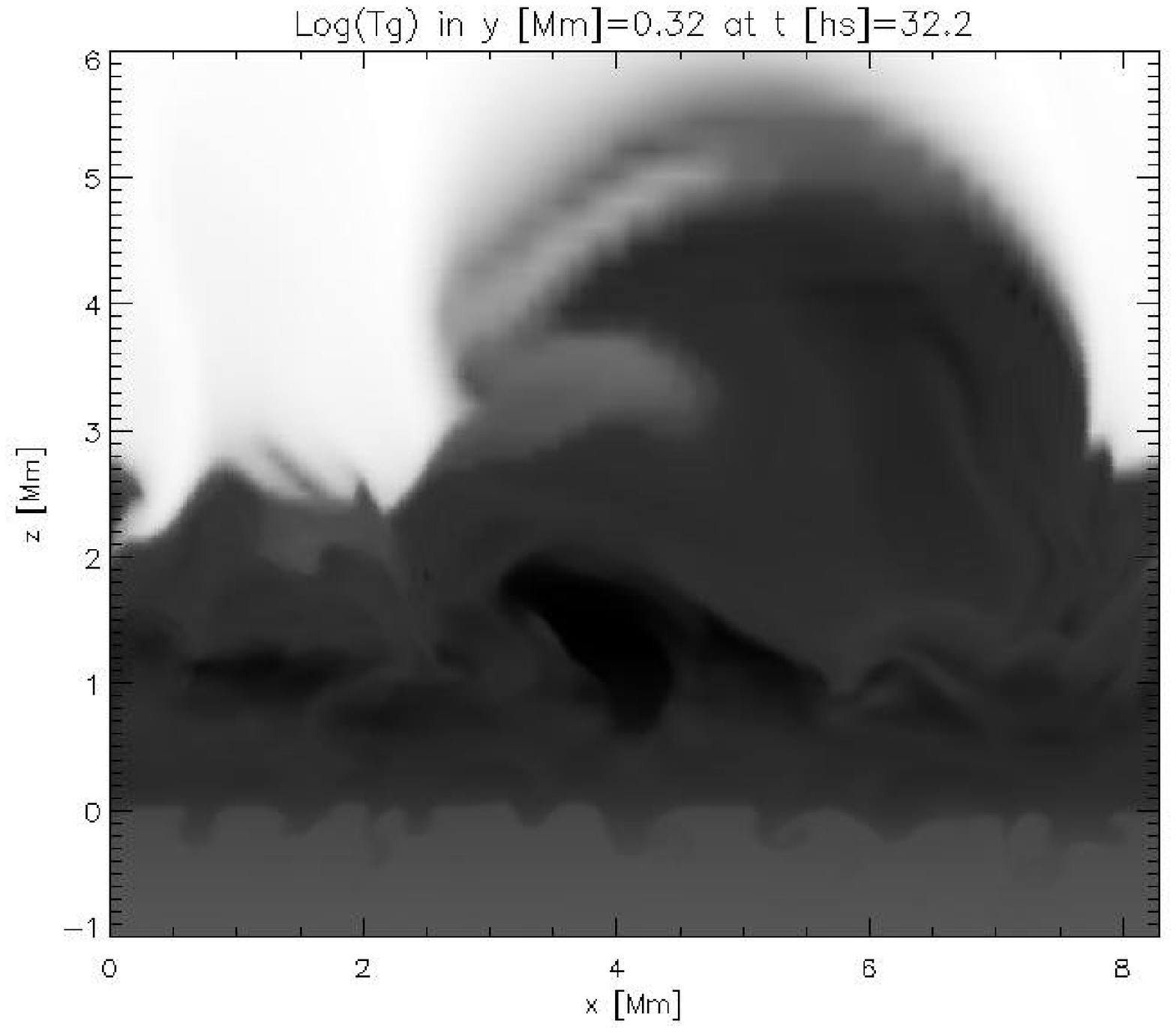}
\includegraphics[width=7.5cm]{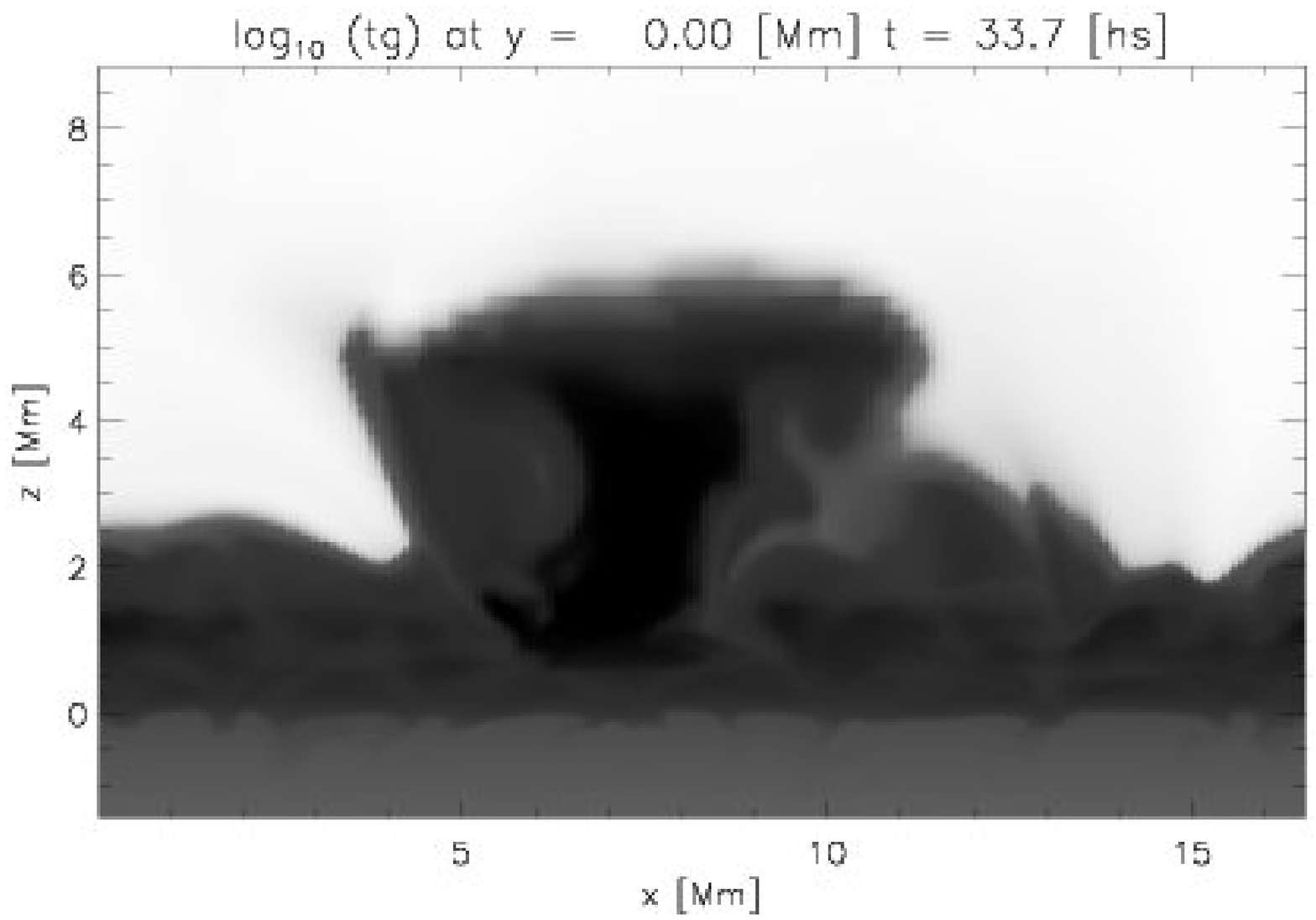}
  \end{center}
  \caption{\label{fig:tgxz} Temperature shown with grey-scale in a
  vertical plane at 3220~s for the simulations A4 (left) and for the simulation
  B1 at time 3370~s (right). Note the different scales.}
\end{figure}

\begin{figure}
\begin{center}
\includegraphics[width=7.5cm]{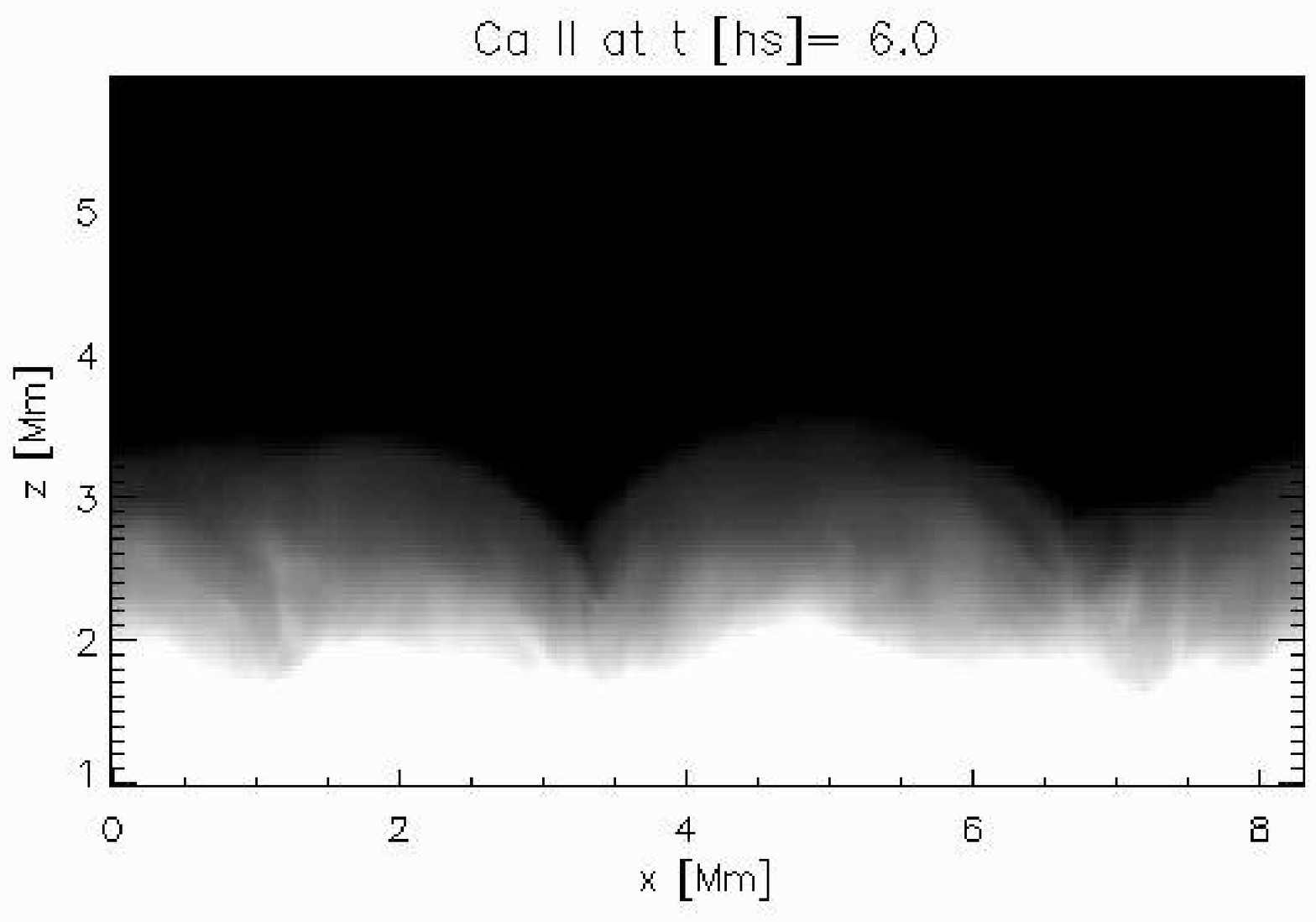}
\includegraphics[width=7.5cm]{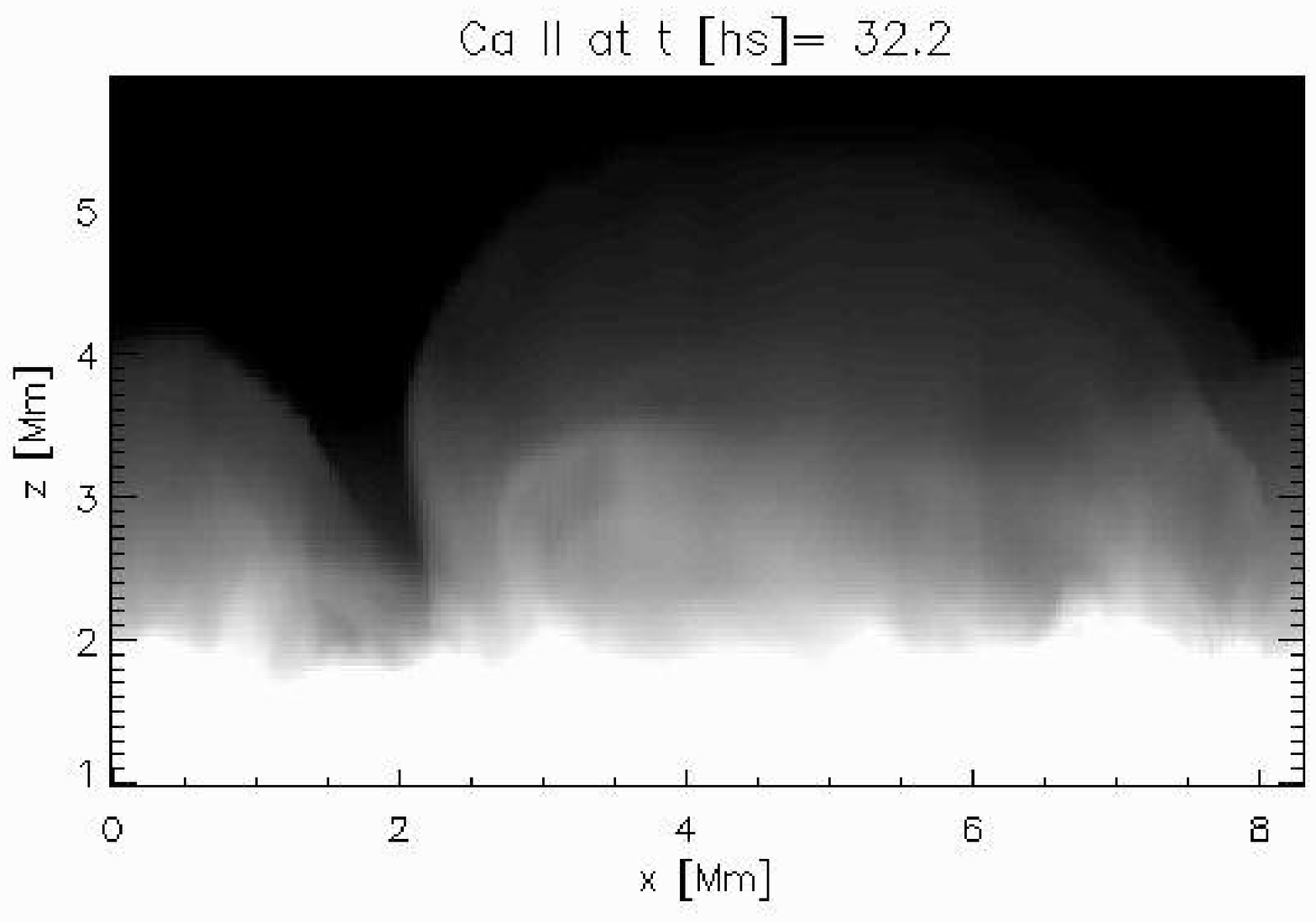}
\includegraphics[width=16cm]{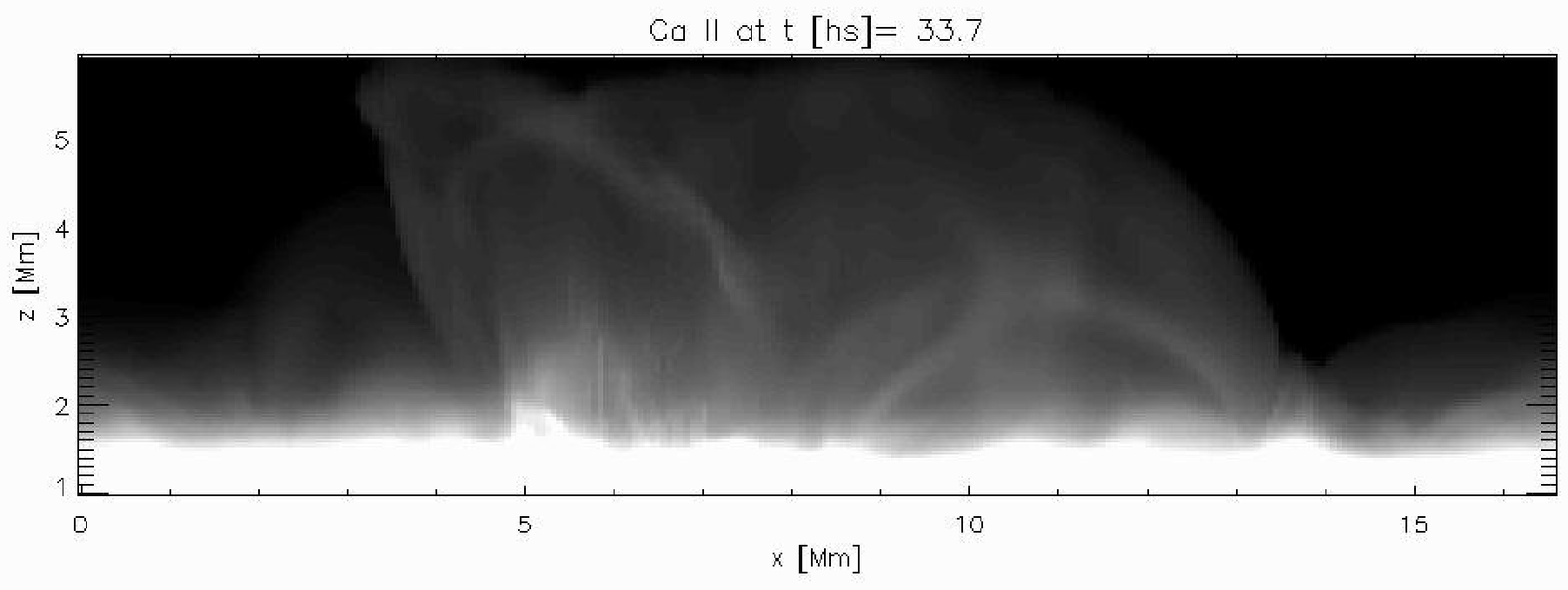}
  \end{center}
  \caption{\label{fig:caxz} The synthetic Ca II H line Hinode intensity 
   to the power of 0.25 (equivalent to a 
   gamma of 0.25 in an image) is plotted to better
  bring out the structure of the bubble from simulation A4 
  at time 600~s, before the flux tube goes through 
  the chromosphere (top left), 
  and at 3220~s when it has crossed the chromosphere (top right), 
  and from the simulation
  B1 at time 3370~s (bottom)}
\end{figure}

That the chromosphere and overlying transition region and lower corona suffer 
large structural changes as a result of flux emergence is driven home by considering a 
vertical slice of the atmosphere as shown in the left panel of figure~\ref{fig:tgxz}.
The transition region has been pushed by the emerging flux from a height that varies 
with chromospheric dynamics from around 2~Mm to almost 6~Mm. This panel also shows that the 
flux tube has grown in horizontal extent and almost fills the computational box. We therefore
also present the equivalent vertical temperature slice from simulation B1 which has a 
computational domain that is twice as large in the horizontal directions. It is evident
also in this model that the rising flux sheet has pushed the transition region and lower corona
aside and has lifted chromospheric plasma to heights of roughly 6~Mm. These cool, 
Lorentz force supported bubbles are long lived, and the simulation that has run the longest, 
B1 to 4500~s, maintains an extended chromosphere for almost 2000~s and is still doing so 
at the end of the simulation. The gas in the extended chromospheric bubble is quite cool, 
down towards a uniform 2000~K, and shows no sign of reheating at heights greater than 1~Mm.
The big bubbles can also be seen in the synthetic Hinode Ca~{\sc ii} H limb images
as shown in figure~\ref{fig:caxz}. These synthetic limb
observations, described in the figure,
show that the chromosphere has expanded and cooled significantly as
the flux tube rises through the chromosphere.
Towards the end of the B1 simulation the emerging magnetic flux shows signs of having
begun fragmentation as it reconnects with the previously present field.
This looks to have interesting implications for transition region and coronal observables, but
the process is not complete in any of our simulations and we will refrain from discussing 
this until the next paper in this series. 

\begin{figure}
\begin{center}
\includegraphics[width=13.5cm]{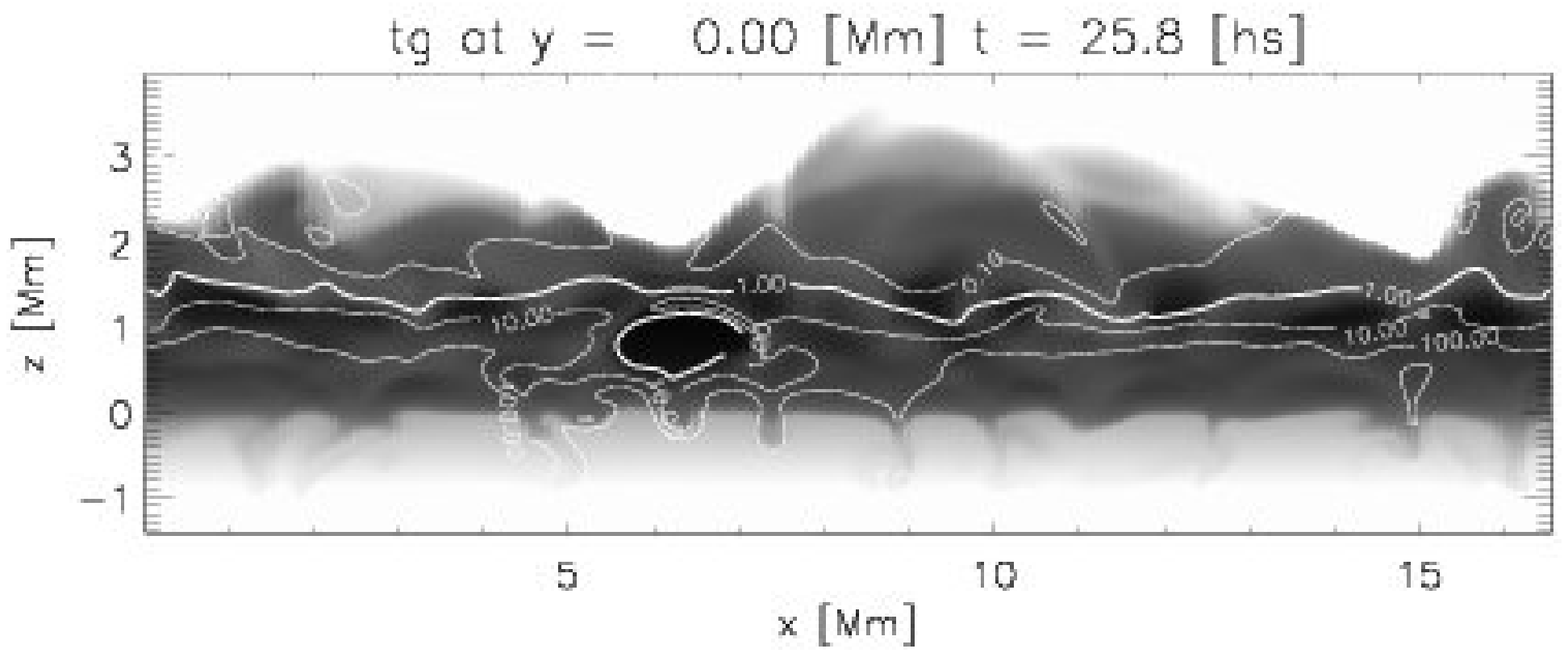}
\includegraphics[width=13.5cm]{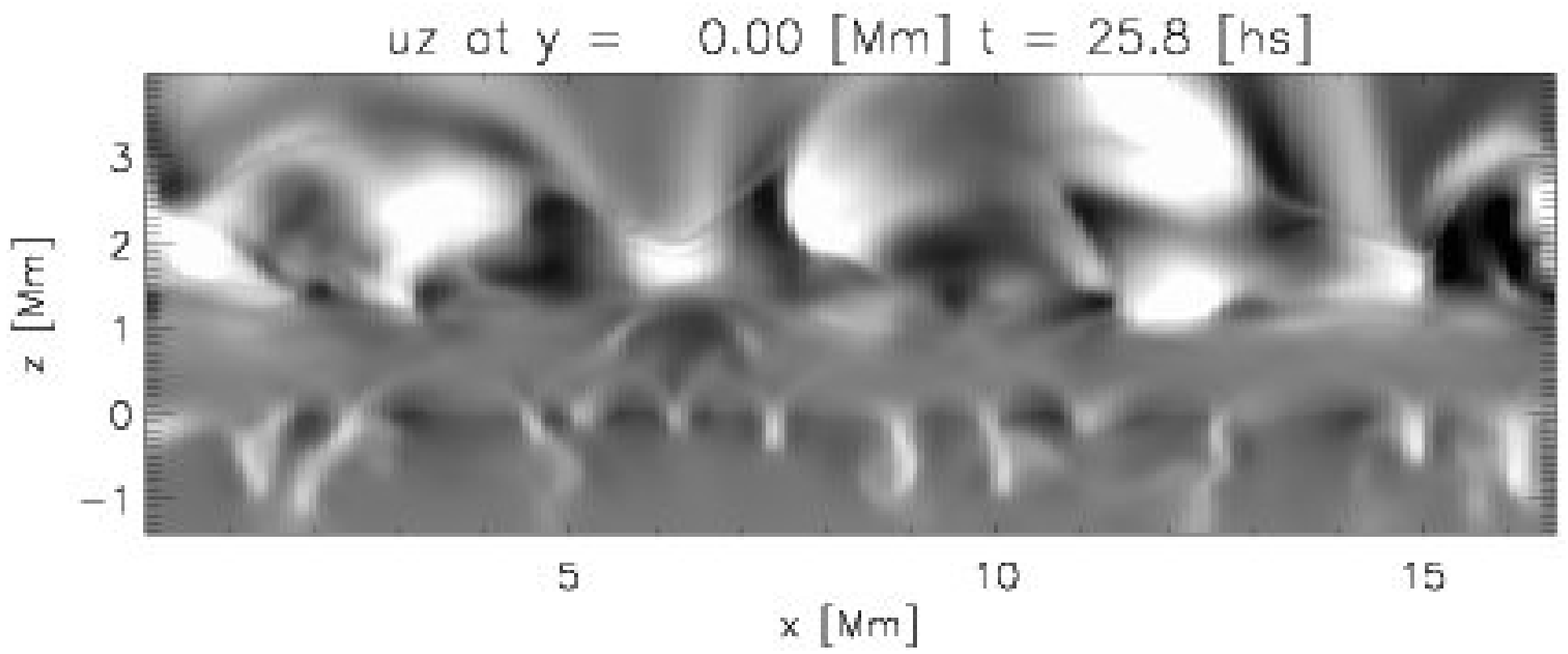}
\includegraphics[width=13.5cm]{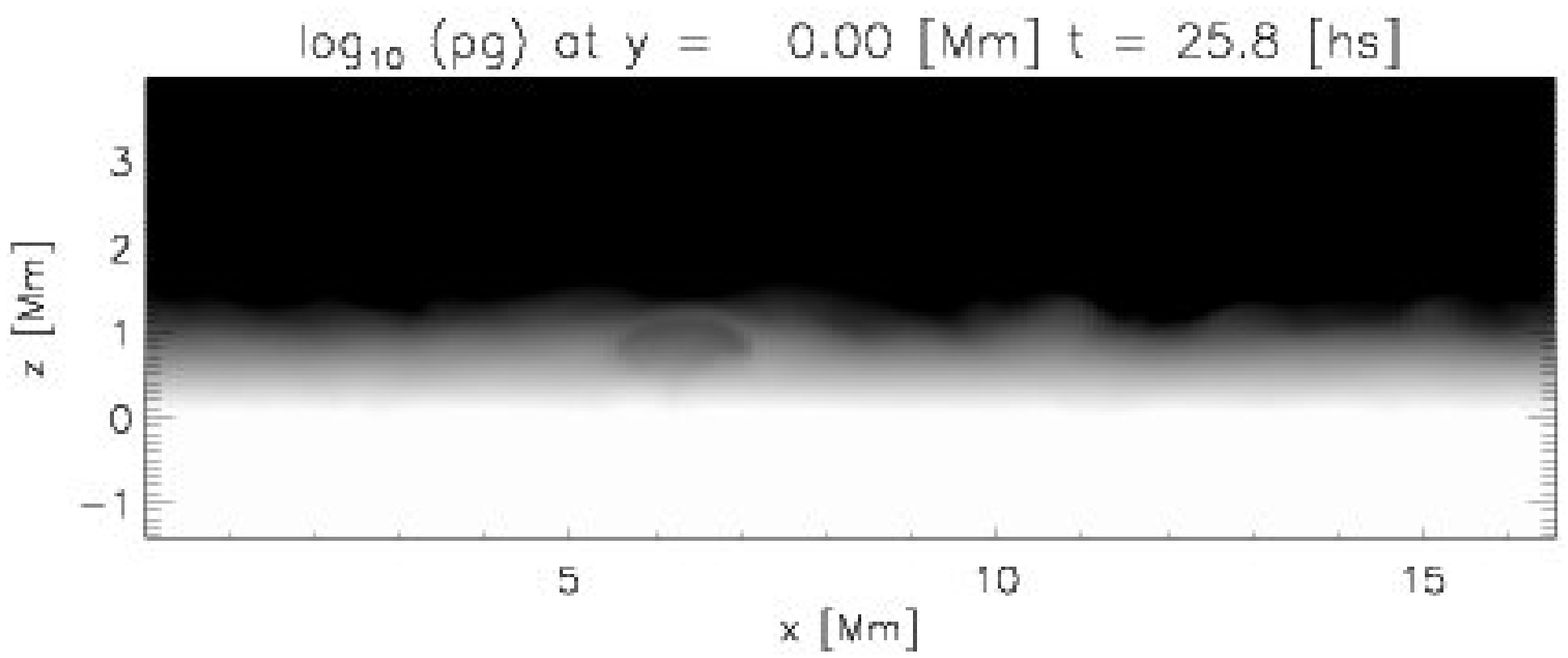}
  \end{center}
  \caption{\label{fig:tg_bubble} Vertical cuts of the logarithm of the temperature (top),
  vertical velocity (middle) and the logarithm of the gas pressure (bottom) from simulation 
  B1 at time $t=2580$~s and $y=0$~Mm. Contours of plasma $\beta$ are overplotted the temperature.}
\end{figure}

Further insight into this phenomena can be found by considering vertical slices of the
B1 simulation at earlier times, as the flux tube bubble rises through the chromosphere,
such as the slices of logarithmic temperature, vertical velocity and logarithmic pressure
we show in figure~\ref{fig:tg_bubble}. We find that a small, low $\beta$ region forms just above the
photosphere, centrally placed over the (partly stalled) rising flux tube. As plasma
$\beta$ falls below one, the bubble becomes buoyant and rises at roughly the sound speed,
expanding rapidly both horizontally and vertically as it does so. The expansion causes 
the plasma to cool and the pressure in the bubble becomes much smaller than in the surrounding 
chromosphere. At some point in the chromosphere the bubble reaches a quasi-equilibrium
with its surroundings and the continued expansion is halted, or at least slows considerably. 

\begin{figure}
\begin{center}
\includegraphics[width=13.5cm]{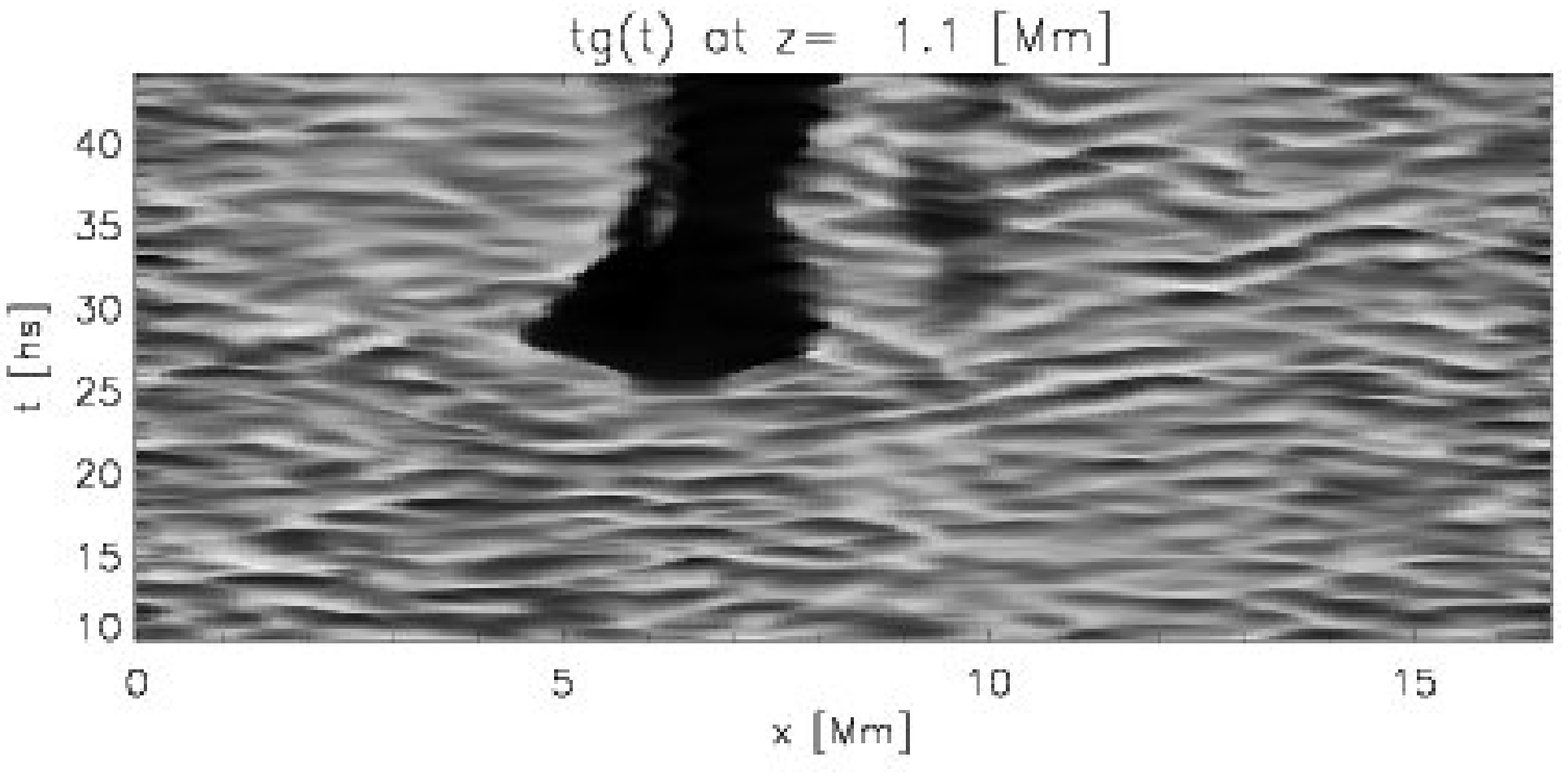}
\includegraphics[width=13.5cm]{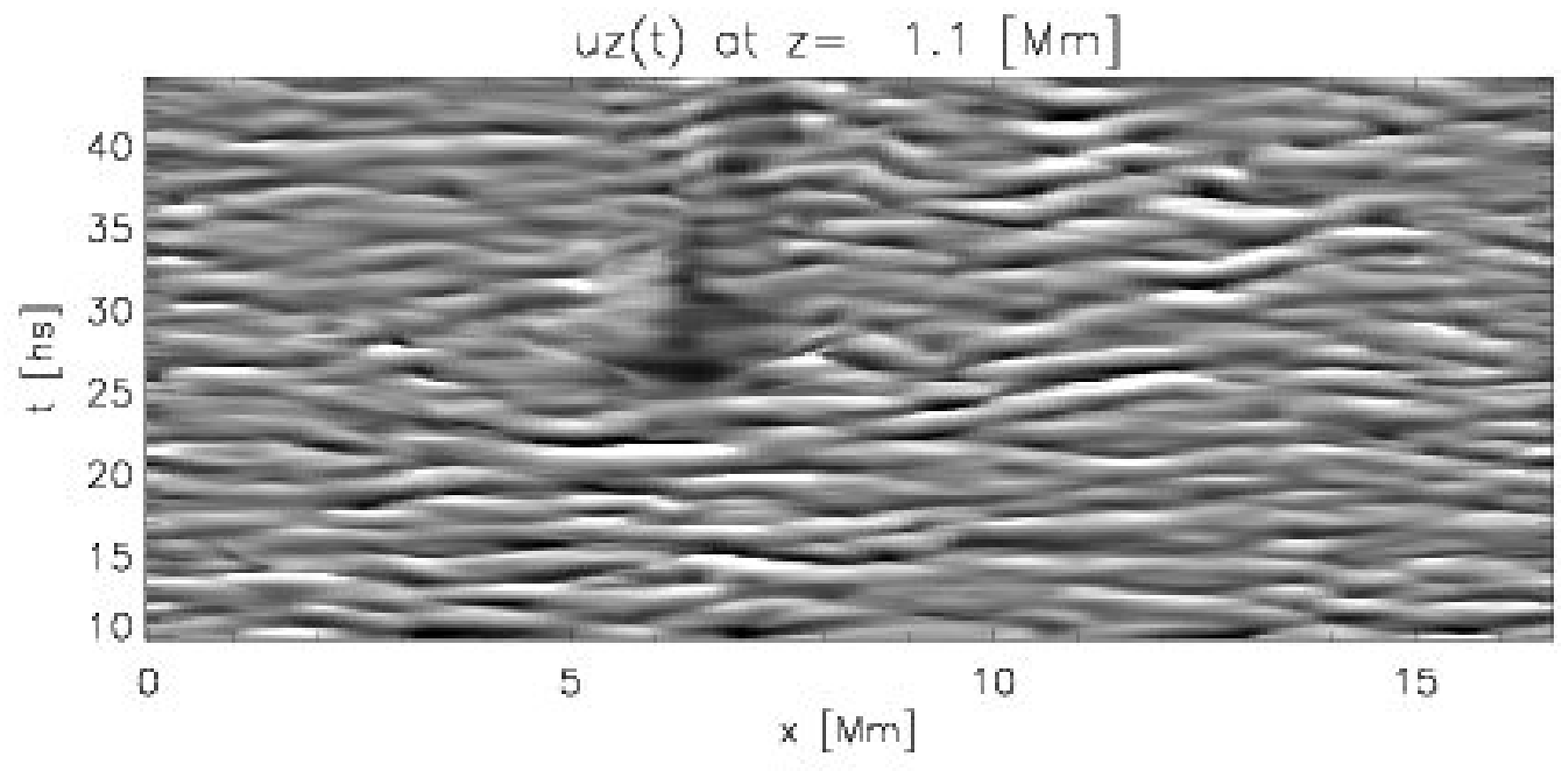}
\includegraphics[width=13.5cm]{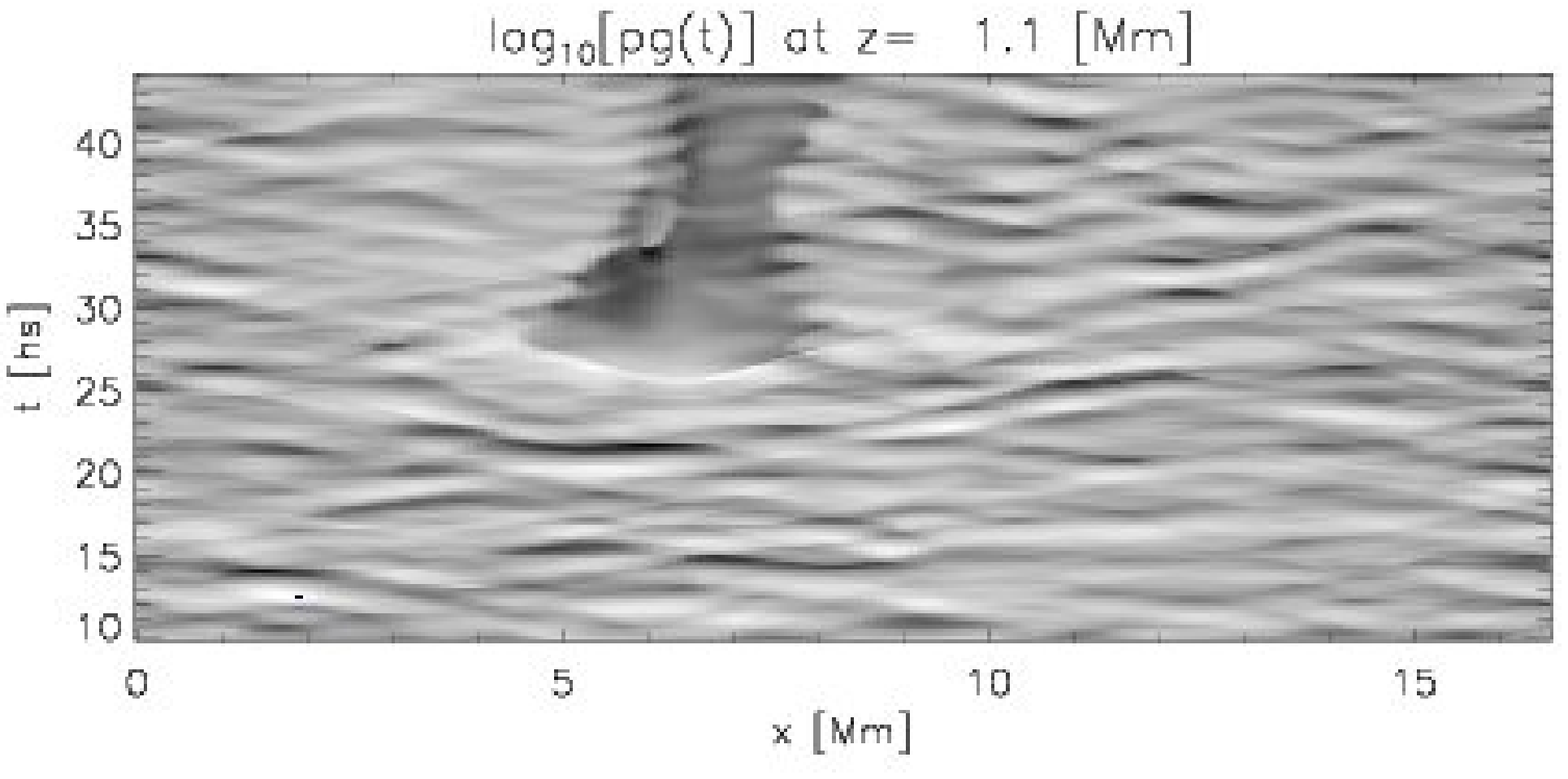}
  \end{center}
  \caption{\label{fig:tgxt} Temperature (top), vertical velocity
  (middle) and logarithm of the pressure (bottom) at a height of
  1.1 Mm as function of position
  in $x$ and time for a vertical cut at $y=0$~Mm for the simulation B1.}
\end{figure}

The appearance of the extended chromospheric bubble and its associated magnetic field has
large repercussions to chromospheric dynamics and energetics. This is abundantly clear
as we consider figure~\ref{fig:tgxt} where we have plotted the variations of the temperature,
velocity, and logarithmic pressure at height 1.1~Mm as a function of time in the B1 simulation.
The region where the bubble is found has very low temperatures and pressures, but also shows 
that the otherwise all pervasive chromospheric oscillations are excluded from the bubble. 
Acoustic waves seem to have difficulty entering this low $\beta$ region - and even when the
bubble oscillates as the rest of the chromosphere at late times the temperature and pressure 
plots show that these oscillation are, at best, only weakly compressive. This expulsion
process appears in every simulation we have run and is important from roughly 800~km above 
the photosphere and extends up to the transition region. 

\section{Conclusions}

The emergence of magnetic flux tubes has been studied in order to
understand not only the dynamics of flux emergence evolution, but also its effects
on the solar atmosphere. One of the major questions we are considering is how flux 
tube emergence changes the energetics and the dynamics of the photosphere,
chromosphere and corona. What contribution to atmospheric energetics 
does the emerging flux represent? Another set of important questions relates to 
the observational signatures of flux emergence throughout the atmosphere.
This paper does not pretend to fully answer these questions, but is rather a 
first attempt to consider a framework for some of the problems and processes 
that arise as one follows the rise of a magnetic field up through the atmosphere.

Specifically, the aim of this work is to simulate a 3D box which
contains a realistic convection zone and the atmospheric elements above;
a photosphere and chromosphere with attendant non-grey radiative losses, 
a transition region and lower corona, including realistic optically 
thin cooling and Spitzer conductivity. The magnetic field injected at the
lower boundary is either a horizontal flux tube with varying degrees of twist 
and magnetic field strengths, or a non-twisted flux sheet. 

Considering first the region spanning the convection zone to photosphere, we find our
results to agree with those reported by \citet{cheung2007} as well as the 
observations of flux emergence of small-scale magnetic loops in the quiet-Sun 
internetwork studied by \citet{Centeno:2007lr}.

We also find that the time evolution of the magnetic flux tube as it rises into the atmospheric
layers above the photosphere is similar in all five simulations we have studied. The flux tube or 
flux sheet crosses the different layers of the atmosphere at roughly the same time in each 
simulation as shown in table \ref{tab:events}.

The results shown in this paper focus on four different
layers; the photosphere at height $10$~km, the region that produces 
reverse granulation at height $234$~km and is represented 
by the bulk of the emission in the Hinode/SOT Ca~{\sc ii} H-line, 
at height $458$~km which is typical for the TRACE 160~nm, and into the chromosphere proper
at a height $906$~km above the photosphere where short wavelength continuum emission 
at 130~nm and 109~nm are observable with SOHO/SUMER. Simulation A4 is the longest running 
simulation that contains a twisted flux tube. In this simulation the flux 
tube rises from the convection zone to the chromosphere at height around 1.1~Mm in 3000~s.
The amount of magnetic flux that reaches each of these regions 
is strongly related to the twist parameter and also to the 
original magnetic field strength injected at the bottom boundary. 
We find that with greater twist and/or greater field strength, 
more flux crosses the photosphere into the chromosphere. The
photosphere is the region where most of the flux is retained. 
The flux that eventually enters the lower chromosphere rises through the various heights we 
have recorded at less than 10~km s$^{-1}$ and does not halt until reaching the transition region/lower 
corona. 

As previously reported in the literature, we find that an 
emerging flux tube produces an expansion of the granulation cells in the 
photosphere. This expansion begins some 400~s before the flux tube itself reaches 
the photosphere.  The simulations show that also the chromospheric plasma expands 
as the emerging flux tube rises. As in the photosphere, we find a delay between 
cell expansion and the arrival of magnetic flux in the lower layers of the chromosphere.
As a result of the expansion the tube plasma is colder than its surroundings,
and emission from the central parts of the flux tube dim successively 
at all heights of the atmosphere according to the schedule outlined in table~\ref{tab:events}.
Such successive dimming has been observed by the SOT and EIS instruments aboard the
Hinode spacecraft \citep{Hansteen+etalSOTEIS2007} for lines up to and including Fe~{\sc xii} 19.5~nm
which is formed at a temperature of some 1.2~MK, but further observations are certainly 
required to search for confirmation or discrepancies with the models described here.

In addition to dimming of the central portion of the flux tube, 
the expansion produces other important effects: For instance,
in the photosphere, the boundary layer of the rising flux tube is a region that 
is conducive in producing granular collapse. 

We also find that chromospheric structure
is substantially altered by the rise and expansion of the flux tube. The chromosphere 
is pushed upward by the magnetic field and cool chromospheric material is found at 
heights up to 6~Mm above the photosphere in contrast to the $2-3$~Mm extent of the 
pre-emergence chromosphere. In expanding, the transition region and corona above 
the flux tube are pushed aside by the chromospheric plasma. However, the pre-existing
coronal magnetic field and transition region and coronal emission are remarkably 
little changed at the times reached by our simulations. Only in simulation B1 do 
we see some initial signs of reconnection between the pre-existing coronal field
and the rising bubble of emerging flux. This reconfiguration, with attendant changes
in transition region and coronal emission, occurs some $30-40$~minutes after the flux
tube initially pierces the photosphere. Since the flux tube also expands dramatically 
in the horizontal direction, reaching widths almost equal to the simulation box
for simulations A1--A4, discussion of transition region and coronal
effects will be postponed to a later paper when simulations contained in a larger 
box are completed. 

While the cooling process in the photosphere and at the lower chromospheric heights
($234, 458$~km) is dominated by adiabatic cooling, we find that at higher layers where
$\beta < 1$ other processes also come into play. The acoustic waves
that are generated as a result of convective overshoot and/or granular dynamics and that 
permeate the non-magnetic chromosphere are expelled from the rising cooling flux tube. 
Thus, the temperature variations that are due to compressive shock waves propagating up 
through the chromosphere are absent in the plasma rising with the flux tube which maintains
a much lower temperature than its surroundings for the duration of our simulations. 
Towards the end of the simulations the magnetic bubble does begin to oscillate with 
the same vertical velocity as the rest of the atmosphere, but the modes penetrating
the bubble are still non-compressive at the time of the completion of the simulation
and the temperature remains fairly constant.

Another interesting effect is the increase of magnetic field strength
in small regions or ``points'' in the atmosphere near the edges of the rising
flux tube. For example, photospheric
bright points are formed in the intergranular lanes edging the flux tube 
after the flux tube has passed through the photosphere. 
These are regions where the magnetic field is concentrated and the field reaches
strengths greater than $10^3$~G. Such bright points have been extensively 
observed in the G-band and other spectral windows previously and 
theoretical and numerical work have explained their appearance: 
The high intensity of bright points is explained by the high magnetic
pressure causing a low gas density. The resulting low opacity means that 
intensity is formed in regions below, where the gas temperature is higher.
However, we also find bright points in the overlying regions formed as 
a result of flux emergence. In the case of the higher lying regions the
bright points are regions of concentrated high magnetic field strength, 
high up- or down-flowing velocity, high velocity convergence and high vorticity.
They are bright in the various spectral bands describing these heights
as a result of high gas temperature due to the compression of the fluid
in these regions.  

There are several interesting and important effects that we have not
reported extensively in this paper, amongst these are the effects of joule 
heating and transition region and coronal diagnostics of flux emergence. This
is mainly due to the finding that the flux emergence process is a quite slow one:
it takes on the order of one hour solar time from the time the magnetic field pierces
the photosphere until it reaches and begins to recombine with the pre-existing 
coronal field. We are currently undertaking simulations that have run long enough 
and that have a large enough computational box to study these phenomena.

\section{Acknowledgements}

This research was supported by the Research Council of Norway through
grant 170935/V30 and through grants of computing time from the Programme for
Supercomputing and by a Marie Curie Early Stage
Research Training Fellowship of the European Community's Sixth Framework
Programme under contract number MESTCT-2005-020395.
Financial support by the European Commission through the SOLAIRE Network (MTRN-CT-2006-035484) is gratefully acknowledged.
To analyze the data we have used IDL and Vapor  (www.vapor.ucar.edu).

\clearpage

\clearpage

\clearpage

\clearpage

\clearpage

\clearpage

\clearpage

\clearpage

\clearpage

\clearpage

\clearpage

\clearpage

\clearpage

\clearpage

\clearpage

\clearpage

\clearpage

\clearpage

\clearpage

\clearpage

\clearpage

\clearpage

\clearpage

\begin{deluxetable}{cccccc}
\tablecaption{\label{tab:runs} Simulation description}
\tablehead{
\colhead{Name} & \colhead{Twist $\lambda$}  & \colhead{B$_0$ [G]} & \colhead{Size} & \colhead{Comment} & \colhead{Time [s]}
}
\startdata
A1 & 0.1 & 4484 & $8\times 4\times 16$~Mm$^{3}$ & Flux tube in $y$ direction & 2200 \\
A2 & 0.2 & 3363 & $8\times 4\times 16$~Mm$^{3}$ & Flux tube in $y$ direction & 2100 \\
A3 & 0.3 & 4484 & $8\times 4\times 16$~Mm$^{3}$ & Flux tube in $y$ direction & 2500 \\
A4 & 0.6 & 4484 & $8\times 4\times 16$~Mm$^{3}$ & Flux tube in $y$ direction & 3200 \\
B1 & 0.0 & 1121 & $16\times 8\times 16$~Mm$^{3}$ & Flux sheet in $y$ direction. & 4500 \\
\enddata
\end{deluxetable}

\clearpage

\begin{deluxetable}{lc}
\tablecaption{\label{tab:events} time evolution}
\tablehead{
\colhead{Process} & \colhead{Time[s]}
}
\startdata
Expansion of granular cells in the photosphere & 900 \\
Expansion of the reverse granulation & 920  \\ 
Cooling of the center of granular cells in the photosphere & 1160 \\
Tube crosses the photosphere & 1300  \\
Tube crosses the chromospheric height forming reverse granulation & 1380 \\ 
Magnetic field moved to the intergranular lanes in the photosphere & 1550 \\
Reverse granulation cooling with expansion& 1560 \\  
Tube crosses the layer $z=450$~km & 1660 \\
The cells return to ``normal'' size in the photosphere & 2100 \\ 
Upper chromosphere vertical expansion and cooling & 2900 \\ 
\enddata
\end{deluxetable}

\end{document}